\documentclass[fleqn,usenatbib]{mnras}

\usepackage{newtxtext,newtxmath}

\usepackage[T1]{fontenc}

\DeclareRobustCommand{\VAN}[3]{#2}
\let\VANthebibliography\thebibliography
\def\thebibliography{\DeclareRobustCommand{\VAN}[3]{##3}\VANthebibliography}

\usepackage{graphicx}	
\usepackage{amsmath}	
\usepackage{float}
\usepackage{natbib}
\usepackage{caption}
\usepackage{gensymb}
\usepackage{bm}
\usepackage[output-decimal-marker={,},exponent-product=\cdot]{siunitx}
\usepackage[utf8]{inputenc}
\usepackage{newtxtext,newtxmath}
\usepackage{hyperref}

\newcommand{\mr}{\mathrm}

\title[AGN winds versus jets]{Winds versus jets: a comparison between black hole feedback modes in simulations of idealized galaxy groups and clusters}

\author[F. Huško et al.]{
Filip Huško$^{1}$\thanks{E-mail: filip.husko@durham.ac.uk},
Cedric G. Lacey,$^{1}$ 
Joop Schaye,$^{2}$ 
Folkert S.J. Nobels,$^{2}$ 
Matthieu Schaller$^{2,3}$ 
\\
$^{1}$ Institute for Computational Cosmology, Department of Physics, University of Durham, South Road, Durham, DH1 3LE, UK\\
$^{2}$ Leiden Observatory, Leiden University, PO Box 9513, NL-2300 RA Leiden,
the Netherlands\\
$^{3}$ Lorentz Institute for Theoretical Physics, Leiden University, PO box 9506, 2300 RA Leiden, the Netherlands
}

\date{Accepted XXX. Received YYY; in original form ZZZ}

\pubyear{2023}

\begin{document}
\label{firstpage}
\pagerange{\pageref{firstpage}--\pageref{lastpage}}
\maketitle

\begin{abstract}
Using the SWIFT simulation code we compare the effects of different forms of active galactic nuclei (AGN) feedback in idealized galaxy groups and clusters. We first present a physically motivated model of black hole (BH) spin evolution and a numerical implementation of thermal isotropic feedback (representing the effects of energy-driven winds) and collimated kinetic jets that they launch at different accretion rates. We find that kinetic jet feedback is more efficient at quenching star formation in the brightest cluster galaxies (BCGs) than thermal isotropic feedback, while simultaneously yielding cooler cores in the intracluster medium (ICM). A hybrid model with both types of AGN feedback yields moderate star formation rates, while having the coolest cores. We then consider a simplified implementation of AGN feedback by fixing the feedback efficiencies and the jet direction, finding that the same general conclusions hold. We vary the feedback energetics (the kick velocity and the heating temperature), the fixed efficiencies and the type of energy (kinetic versus thermal) in both the isotropic and the jet case. The isotropic case is largely insensitive to these variations. On the other hand, jet feedback must be kinetic in order to be efficient at quenching. We also find that it is much more sensitive to the choice of energy per feedback event (the jet velocity), as well as the efficiency. The former indicates that jet velocities need to be carefully chosen in cosmological simulations, while the latter motivates the use of BH spin evolution models.
\end{abstract}






\section{Introduction}

Supermassive black holes (BHs), situated in the central regions of their host galaxies, are often observed to be releasing significant amounts of energy into their environment. In this role they are referred to as active galactic nuclei (AGN). Many AGN have been observed through the radiation they release (often in the form of very luminous quasars), from infrared to X-ray frequencies, from very early in the Universe's history all the way to the present day (e.g.~\citealt{Shen2020}). AGN appear to be affecting their environment on kiloparsec scales through the inflation of lobes of relativistic gas that are visible at radio frequencies, again from the cosmic dawn to today (e.g.~\citealt{Smolcic2017}). This injection of energy by AGN, in various forms, is thought to affect their host galaxies and larger-scale environment -- a process referred to as AGN feedback. Most importantly, AGN feedback appears to be responsible for the quenching of star formation in massive galaxies and could thus explain their current state as ‘red and dead' (e.g.~\citealt{DiMatteo2005}, \citealt{Bower2006}, \citealt{Croton2006}, \citealt{Booth2009}).

BHs can grow through two processes; accretion (of gas, as well as stars and dark matter, albeit the last two are usually ignored when modeling BH growth) and BH-BH mergers. As a consequence of the accreting gas having net angular momentum, an accretion disc is typically formed around an accreting BH. Depending on the accretion rate, different types of discs can form. The classical \cite{ShakuraSunyaev1973} solution (and its general-relativistic counterpart; \citealt{NovikovThorne1973}) describes a geometrically thin and optically thick accretion disc in which gas orbits are almost circular (Keplerian at large distances). As the matter slowly funnels downwards towards the BH, it is heated up by viscous stresses. Around $10$ per cent of the total mass-energy of the matter in this type of accretion disc is radiated outwards through this process, leading to the observed quasars. An alternative solution found by \cite{NarayanYi1994} (see \citealt{PophamGammie1998} for the general-relativistic version) instead describes a geometrically thick and optically thin, advection-dominated accretion flow (ADAF). In this solution, radial gas motions are dominated by advection, and magnetic fields are also thought to be advected inwards (although this is thought to also occur, to a smaller degree, in the thin disc). In combination with the dynamo effect, this leads to the buildup of strong magnetic fields near the BH. These magnetic fields then facilitate the launching of relativistic jets through the \cite{Blandford1977} process, in which energy is extracted from the rotation of the BH.

Radiation from AGN is thought to couple to gas, leading to the launching of winds. These winds are thought to be launched mainly due to radiation or thermal pressure (e.g.~\citealt{Murray1995}). While observations of bright AGN are extremely numerous (e.g.~\citealt{Shen2020}), and winds appearing to emanate from them have also been frequently observed (e.g.~\citealt{Crenshaw2003}, \citealt{Tombesi2010}, \citealt{Feruglio2010}), direct observational evidence of negative feedback on their host galaxies is not conclusive. Observations have found both enhanced and suppressed star formation rates (SFRs) in galaxies hosting AGN (see \citealt{Ellison2016} and references therein). As \cite{Ward2022} show using simulations (see also \citealt{Harrison2017}), this may be explained by highly-accreting BHs (visible as bright AGN) being triggered by large amounts of cold gas and star formation. These simulated galaxies appear quenched only once they are devoid of large amounts of cold gas and star formation, which necessarily means that BHs are accreting at lower rates and the AGN are faint by that point.

Direct evidence of negative AGN feedback is more easily found in galaxy groups and clusters (see reviews by \citealt{Eckert2021} and \citealt{Fabian2012}, respectively). X-ray observations of the circumgalactic/intracluster medium (CGM/ICM) surrounding the central galaxies of these systems (‘brightest group/cluster galaxies', which we refer to simply as BCGs hereafter) have revealed evidence of AGN feedback in the form of cavities in the X-ray emitting gas (\citealt{Gull1973}, \citealt{Boehringer1993}, \citealt{Birzan2004}, \citealt{McNamara2005}, \citealt{Wise2007}). These cavities are often coincident with synchrotron-emitting plasma taking the form of two-sided lobes (\citealt{Biermann1987}, \citealt{Odea1998}, \citealt{Markoff2001}). This plasma originates from jets of relativistic particles launched from close to the BHs (\citealt{Blandford1979}, \citealt{Urry1995}). The power of these jets, inferred from the power required to inflate the cavities, suggests that they inject sufficient energy to shut off the cooling flows that would otherwise develop in the centres of the CGM/ICM (\citealt{Rafferty2006}, \citealt{Fabian2012}, \citealt{Hlavacek-Larrondo2012}, \citealt{Russell2013}, \citealt{Eckert2021}). This feedback mechanism is often referred to as ‘mechanical', ‘maintenance' or ‘radio' mode feedback. We refer to it simply as jet feedback throughout the rest of this paper.

Semi-analytical models of galaxy formation typically employ N-body simulations of cosmic structure formation to populate dark matter haloes with galaxies in post-processing (e.g.~\citealt{Henriques2015}, \citealt{Lacey2016}, \citealt{Lagos2018}). Early versions of such models were successful at reproducing the numbers of massive galaxies, but only if AGN feedback is included (e.g.~\citealt{Bower2006}, \citealt{Croton2006}, \citealt{Lagos2008}). Hydrodynamical cosmological simulations of galaxy formation and evolution also invariably find that AGN feedback is necessary in order to quench star formation in massive galaxies. Most such simulations have implemented AGN feedback as isotropic heating of gas (thermal isotropic feedback), usually intended to represent the effects of radiatively-driven winds from quasars\footnote{While this feedback mode is in principle similar in different simulations, the practical aspects of how it is implemented can lead to significant differences. Most significantly, if the feedback energy is injected in all particles/cells around the BH equally at every time-step (‘thermal dump'), the heated gas can be prone to numerical overcooling, and the feedback is thus not very effective. In contrast, if the feedback energy is held in a reservoir until a sufficient amount of it has been accumulated to heat particles near the BH by some chosen heating temperature $\Delta T$, these problems can be avoided (\protect\citealt{Booth2009}). The feedback is most effective if only a single gas resolution element receives all of the accumulated energy.\label{footnote1}}. Examples of such simulations include Magneticum (\citealt{Hirschmann2014}), EAGLE (\citealt{Schaye2015}), MassiveBlack-II (\citealt{Khandai2015}), Romulus (\citealt{Tremmel2017}) and ASTRID (\citealt{Bird2022}), among others. 

Other simulations have employed somewhat more complicated AGN feedback prescriptions, using different mechanisms of energy injection at low BH accretion rates (alongside thermal isotropic feedback also being used at high accretion rates in all cases). In Illustris (\citealt{Vogelsberger2014}), pairs of thermal bubbles were injected at large distances in haloes (\citealt{Sijacki2015}). This feedback mode represents the late-time effects of relativistic jets that inflate lobes. However, the inflation process itself (which includes strong shocks that may be critical for heating the ICM) was not included, with the bubbles placed ‘by hand', already inflated. IllustrisTNG (\citealt{Nelson2019}) instead uses kinetic isotropic feedback at low accretion rates (\citealt{Weinberger2017b}, \citealt{Weinberger2018}), representing the effects of winds that may be active alongside the jets (e.g.~\citealt{Blandford1999}). For this feedback channel, the critical accretion rate below which it is used is highly dependent on BH mass, leading to effectively no kinetic feedback for low-mass BHs and little thermal feedback for high-mass ones.

The SIMBA simulations (\citealt{Dave2019}) use kinetic jets at low accretion rates (and high BH masses, $M_\mathrm{BH}\geq10^{7.5}$ $\mathrm{M}_\odot$, similar to IllustrisTNG) that are launched in the direction of the angular momentum of the gas surrounding the BH, alongside an additional X-ray feedback mechanism (implemented isotropically, as a mixture of heating and kicking particles), representing the equivalent of the kinetic wind used in IllustrisTNG at low accretion rates. In Horizon-AGN (\citealt{Kaviraj2017}), a similar prescription is used for the jets as in SIMBA (in that the jets are launched in the direction of the angular momentum of the gas close to the BH). Its successor New-Horizon (\citealt{Dubois2021}) uses a more sophisticated prescription based on a model presented in \cite{Dubois2014_spin}, wherein the BH spin is evolved for all BHs using accretion disc models, and the jets are launched along the direction of the BH spin vectors, with spin-dependent efficiencies. In addition to being more realistic, this approach has the benefit (from a numerical perspective) that the BH spin vector is more stable against perturbations compared to the gas angular momentum (in the BH kernel), since the BH spin is a quantity that is integrated over the history of each BH. The radiative efficiency of AGN at high accretion rates also depends on BH spin in this model.

In this work we will focus on modifications to the AGN feedback prescription of the EAGLE galaxy formation model (\citealt{Schaye2015}, \citealt{Crain2015}), which is based on the \cite{Booth2009} AGN feedback scheme developed for the OWLS simulations (\citealt{Schaye2010}). The EAGLE simulations used a fairly simple AGN feedback prescription -- despite this, the model correctly predicts the number of galaxies as a function of mass (as measured through the stellar mass function or the stellar mass-halo mass relation; \citealt{Schaye2015}) and redshift (\citealt{Furlong2015}) as well as many galaxy properties (e.g.~the metallicities and sizes; \citealt{Schaye2015}, molecular gas content; \citealt{Lagos2015}, and colours; \citealt{Trayford2017}). 

The Hydrangea simulations used the EAGLE model to evolve a sample of galaxy clusters (\citealt{Bahe2017}). Despite the EAGLE model working well for the overall population of galaxies, these simulations found that BCGs were too massive, from about a factor of two for low-mass clusters (halo masses of order $10^{14}$ $\mr{M}_\odot$) to a factor of nearly ten for high-mass clusters (halo masses of order $10^{15}$ $\mr{M}_\odot$). The same galaxies were also found to be too highly star-forming compared to observations. This problem possibly originates from overly strong cooling flows in the simulations, which could be a consequence of insufficient heating by thermal isotropic AGN feedback at large radii (e.g.~$>100$ kpc). 

The C-EAGLE project (\citealt{Barnes2017}) also used the EAGLE model to simulate a broadened sample (relative to Hydrangea) of galaxy clusters. Mock X-ray observations (\citealt{Barnes2017}) showed that these clusters appear to have central entropies of the ICM that are too high (a problem confirmed by \citealt{Altamura2023} on a separate sample of galaxy groups and clusters, using an updated version of the EAGLE model). This is also true for the temperature, and the reverse is true for the density. A related problem is in the cool-core (CC) versus non-cool-core (NCC) dichotomy of clusters (e.g.~\citealt{Hudson2010}): simulated clusters are likely too often NCC as compared to observed ones (the fraction of CC clusters is too low), although firm conclusions on this are complicated by varying definitions in the literature of what is a CC versus a NCC cluster (\citealt{Barnes2018}). 

\cite{Nobels2022} studied AGN feedback using an updated version of the EAGLE model in idealized galaxy groups and clusters, and found that thermal isotropic feedback can quench star formation in central galaxies for long times (many Gyr) only in galaxy groups, while in galaxy clusters, the BCGs have recurrent cooling flows. They found that clusters initialized as CC largely remain CC. This indicates that the potential lack of CC clusters, as measured through the central entropy, in realistic, cosmologically simulated samples of clusters (C-EAGLE) may be unrelated to AGN feedback, and could instead be a result of other physical processes in the evolution of these clusters. Alternatively, \cite{Altamura2023} found significant differences in entropy profiles between their clusters and those in the C-EAGLE sample. Their cosmological zoom-in simulations of groups and clusters used a slightly updated EAGLE model with, most significantly, a new hydrodynamics scheme (\citealt{Borrow2022}). They also found substantial differences when turning off artificial conduction in the hydrodynamics solver. These results indicate that the differences between observed and simulated clusters may be partly or wholly due to numerical issues.

If the differences between the observed clusters and ones simulated with EAGLE are not entirely due to numerics, including a more realistic feedback mechanism (representing the effects of relativistic jets) may be helpful, presumably by allowing more effective coupling of the feedback energy to larger radii instead of only to the core of the ICM. A similar modification may be beneficial in the IllustrisTNG model (see e.g.~the results of the MillenniumTNG simulations, \citealt{Pakmor2022}). This is despite that model using kinetic feedback at low accretion rates (alongside thermal isotropic feedback at high accretion rates), and the reason may be that the feedback mechanism is also isotropic. As we will show in this paper, kinetic isotropic and thermal isotropic feedback are fairly similar in their effects, at least in the context of idealized cluster simulations. The potential problems we have discussed may be present even for the SIMBA simulations (\citealt{Dave2019}), which also show somewhat too high entropies, albeit at intermediate radii rather than in the core of the ICM (\citealt{Oppenheimer2021}). While SIMBA includes AGN jets, they are launched in the direction of the angular momentum of the gas near the BH, which may not be very stable (especially in clusters and at low resolutions). As we will show in this work, the jet direction needs to be relatively stable for the jets to lead to significant differences compared to isotropic feedback.


Performing idealized simulations of AGN jets (on kpc scales) is important in order to further our understanding of the effects of these jets, and their precise mechanisms of action, in a more controlled environment than in cosmological simulations (see review by \citealt{Bourne2023}). In \cite{Husko2022_self_similar} we simulated constant-power jet episodes in an idealized ICM: these simulations were the first smoothed particle hydrodynamics (SPH) simulations of their type, i.e.~of idealized episodes of AGN jets. We performed them mainly to validate the numerical method for the jet launching and to confirm that the hydrodynamics of these jets, as well as the lobes they inflate and their interaction with the ambient medium, are correctly simulated with our chosen SPH method (\citealt{Borrow2022}). We found good agreement with theoretical predictions. Surprisingly, jet episodes represented with only $\approx500$ particles per jet were found to be sufficiently resolved in terms of basic properties (e.g.~the sizes of the inflated lobes). In a subsequent paper (\citealt{Husko2022_bubbles}) we studied the evolution of jet-inflated bubbles in an idealized ICM over relatively long time-scales ($\approx$Gyr). We found that heating of the ICM dominates early on (while the jets are active and shortly afterwards), but AGN feedback is done mostly through gas uplift and the reduction of its central density at late times (once buoyancy starts to act on the bubbles). In both papers we found that the jet velocity parameter\footnote{The jet velocity in cosmological simulations, as well as idealized ones such as in this paper, is both a physical and numerical parameter. Typically, real AGN jets are highly relativistic and the lobes they inflate are very light. The mass resolution of the simulations limits the sampling of particles being launched into the jets, and effectively provides a lower limit to the mass of jets and lobes. The spatial resolution of the simulations provides a lower limit to the injection scale on which the AGN jets are launched. For these reasons, jet velocities in these simulations need to be subrelativistic, of order $10^4$ $\mathrm{km}$ $\mathrm{s}^{-1}$.\label{footnote2}} plays a very important role, and thus needs to be carefully chosen. In our latest paper (\citealt{Husko2022_spin_driven}, hereafter Paper I), we studied self-consistent BH accretion and feedback using BH spin evolution and the EAGLE model in simulations of idealized galaxy groups and clusters. We found that jets were successful at preventing cooling flows and quenching star formation in this setting. 

Here we will broaden this analysis and consider isotropic feedback as well -- the AGN feedback mode used in EAGLE and all other large, cosmological hydrodynamical simulations (at least at high BH accretion rates). Our main goal here is to compare these two feedback modes in terms of their impact on the BCGs, their BHs and the ICM. Some previous works focusing on feedback in idealized galaxy clusters have studied different feedback implementations, with many of them comparing thermal and kinetic feedback (e.g.~\citealt{Barai2016}, \citealt{Meece2017}, \citealt{Su2021}, \citealt{Weinberger2022}). Most of these works employed thermal feedback as a ‘thermal dump' (see footnote \ref{footnote1}), meaning that it will likely have been prone to numerical overcooling, unlike the kinetic variety.

The work we are presenting here builds on previous studies by broadening the comparison between AGN feedback modes to include both realistic feedback (with BH spin evolution) and a more simplified implementation (with fixed efficiencies and jet directions). In addition, we compare these feedback implementations for different halo masses, ranging from the galaxy group ($M_{200}=10^{13}$ $\mr{M}_\odot$) to the high-mass galaxy cluster ($M_{200}=10^{15}$ $\mr{M}_\odot$) scales. In our simplified feedback scenario, we systematically vary relevant parameters such as the heating temperatures and kick velocities, as well as feedback efficiencies. Furthermore, for both isotropic and jet feedback, we vary the type of energy being injected (thermal versus kinetic). In terms of results, we focus mostly on SFRs and entropy profiles, as stellar masses of BCGs and entropy profiles of the ICM appear to show the largest or most easily observable discrepancies between observed clusters and those simulated with EAGLE. 

In \S~\ref{sec:sec2} we present our BH spin evolution model and the feedback efficiencies used in the simulations. We focus on the thin, radiatively-efficient disc, with the thick, advection-dominated disc having been presented in Paper I. In \S~\ref{sec:sec3} we discuss the code and galaxy evolution model that we use, alongside the physical set-up. We also list all of the simulations we have performed and discuss how their parameters were chosen. \S~\ref{sec:res_spin} contains our results using the BH spin evolution model, whereas \S~\ref{sec:res_fixed} contains the ones using simpler feedback without BH spin dependencies. In \S~\ref{sec:conclusions} we summarise and conclude. In the Appendices \ref{sec:app1}, \ref{sec:app2}-\ref{sec:app2p2} and \ref{sec:app3} we discuss, in turn: 1) the role of redirection and precession in the jet case, 2) some additional quantities related to our BH spin evolution simulations and 3) the origin of the periodicity in the cooling flows that will be apparent.

\section{Black hole spin evolution and feedback}
\label{sec:sec2}

The dimensionless BH spin, $a$, is a proxy for the angular momentum of the BH, $J_\mr{BH}$, through its definition $a=J_\mr{BH}c/M_\mr{BH}^2G$, where $M_\mr{BH}$ is the mass of the BH, and $c$ and $G$ the speed of light and Newton's constant, respectively. In order to avoid naked singularities, the BH spin is expected to be no larger than $1$ in magnitude (\citealt{Kerr}). We actually limit the upper end of this range to $0.998$ (see \citealt{Thorne1974})\footnote{The emission of radiation by accreting gas and its swallowing by the BH causes a counteracting torque that acts against spinup of the BH, and which is important for $a>0.99$. The difference between a maximal BH spin of 0.998 and 1 may seem negligible, but the radiative efficiency of a thin disc is substantially lower for the former ($32$ per cent) than for the latter ($42$ per cent); see \S~\ref{sec:feedback_effs}.}. The inner accretion disc is expected to be in the equatorial plane due to the effects of \cite{LenseThirring} torques (see \S~\ref{sec:spin_sign}). In this region, the gas may be corotating with the BH (prograde accretion), in which case $a>0$, or it may be counter-rotating (retrograde accretion), with $a<0$. 

For the purpose of evolving the BH spin (and modeling its effects on feedback), we have developed an analytical model and implemented it as a subgrid model in the simulation code we use (see \S~\ref{sec:sec3}). This model is similar to and is inspired by a series of other models that have been used to include BH spin in simulations (\citealt{Volonteri2005}, \citealt{King2008}, \citealt{Fanidakis2011}, \citealt{Dubois2014_spin}, \citealt{Fiacconi2018}, \citealt{Griffin2019a}, \citealt{Dubois2021}). We assume that BHs can be in one of two different accretion states depending on the accretion rate (more precisely, the Eddington ratio -- see the next subsection): 1) the geometrically thick, advection-dominated disc (i.e.~ADAF; advection-dominated accretion flow, \citealt{NarayanYi1994}) at low accretion rates and 2) the geometrically thin, radiatively-efficient disc (\citealt{ShakuraSunyaev1973}) at high accretion rates. We refer the reader to these papers for details on the properties of the discs. For our purpose, it is most important that the thick disc features a turbulent dynamo effect and strong advection -- this includes the advection of magnetic fields, which then build up near the event horizon and lead to the launching of strong jets. The thin disc, on the other hand, releases most of the gravitational binding energy of the gas (as it flows inwards) as radiation, which results in winds that may act as a feedback mechanism on the galaxy scale.

In Paper I we assumed the first of these two accretion states to evolve the BH spin and launch jets. However, for simplicity, we made the unrealistic assumption that the equations describing this accretion flow are valid at all accretion rates, and that the jet efficiencies are also high at all accretion rates. Here we will present an accretion model that includes both accretion states, in which the BH is assumed to be in one of the two modes depending on its current accretion rate. In the simulations we may instead assume that one or the other accretion state is active at all accretion rates, in order to compare simulations with simpler BH spin evolution/feedback prescriptions. Below we give a summary of our method and the assumptions we make for the modelling of the thin disc state. We also refer the reader to Paper I, where we describe the BH spin evolution model in more detail. While that model was presented for the thick disc, a very similar one is used for the thin disc, but with some different assumptions for disc structure. We also modify the thick disc model slightly (as compared to Paper I) by updating the model for the spinup/spindown rates in this accretion state. 

Compared to most other previously published models for BH spin evolution, our model self-consistently tracks the evolution of BHs in the two different accretion regimes. In particular, in the thick disc: 1) accretion is less effective at spinning up the BH than in the thin disc, 2) jet spindown is important and 3) Lense-Thirring torques are much less efficient at aligning/counter-aligning the BH with the surrounding gas (on large scales, beyond the accretion disc). These effects have so far only been included in our model and that presented in \cite{Dubois2021}.

\subsection{Deciding the nature of the accretion state}

The state of the accretion flow is thought to depend on the dimensionless accretion rate (also often referred to as the Eddington ratio), defined as $\dot{m}=\dot{M}_\mr{BH}/\dot{M}_\mr{Edd}$, where the Eddington accretion rate is
\begin{equation}
\dot{M}_\mr{Edd}=\frac{L_\mr{Edd}}{\epsilon_\mr{r,0}c^2}=4\pi\frac{G M_\mr{BH}m_\mr{p}}{\epsilon_\mr{r,0}\sigma_\mr{T}c}.
\label{eq:eq1}
\end{equation}
Here, $L_\mr{Edd}$ is the Eddington luminosity, $m_\mr{p}$ is the proton mass, $\sigma_\mr{T}$ the Thomson cross-section and $\epsilon_\mr{r,0}=0.1$ is a nominal radiative efficiency used only for the definition of $\dot{m}$ in this paper (the actual radiative efficiency is allowed to depend on BH spin, see \S~\ref{sec:feedback_effs}). 

According to numerical calculations by \cite{Narayan1995} done soon after the discovery of the thick disc solution, the thick disc is not always stable and it should transition to being thin once $\dot{m}\gtrsim \alpha^2$. Here, $\alpha$ is a numerical parameter that is related to the kinematic viscosity $\nu$ through $\nu=\alpha c_\mr{s}H$, where $c_\mr{s}$ and $H$ are the sound speed and height of the disc at a given radius, respectively. The factor $\alpha$ is used to encapsulate our ignorance of the detailed behaviour and origin of the kinematic viscosity of accretion discs. It is usually taken to be constant with radius, for simplicity, although it very likely varies with radius and possibly with accretion state.

More recent and detailed calculations suggest that this picture (of a transition between accretion solutions at $\dot{m}\approx\alpha^2$) is somewhat too simple (see the review by \citealt{YuanNarayan2014}). In particular, the properties of the thick disc already begin to change at $\dot{m}=0.2\alpha^2$, and the transition appears to be complete by $\dot{m}=0.7\alpha$. Between these two values, the disc takes on a transition state whose properties are not well understood. For conceivable values of $\alpha$, which may be as low as $0.05$ based on simulations (\citealt{YuanNarayan2014}) and as high as $0.1-0.4$ based on observations (\citealt{King2007}), the transition state may occupy the range $0.001-0.3$ in $\dot{m}$. Observations of both X-ray binary spectra (\citealt{Done2007}) and AGN spectra (\citealt{Noda2018}) find this transition to occupy a narrower range of $\dot{m}=0.01-0.03$. \cite{Russell2013} analysed the radiative and mechanical powers of AGN and found the transition to span the same range. We assume the lower end of this range to be the critical transition rate at which the two accretion states interchange; $\dot{m}_\mathrm{crit}=0.01$. 

Given this choice, we can set a value for the viscosity parameter $\alpha$, which appears in many of the equations describing accretion disc structure that we will discuss. For this purpose we use the finding of numerical calculations that the transition spans the range between $0.2\alpha^2$ and $0.7\alpha$ in $\dot{m}$. We assume that the geometric mean of these two boundaries corresponds to $\dot{m}_\mathrm{crit}=0.01$, which is true for $\alpha\approx0.2$, so we set $\alpha=0.2$ for the remainder of this paper.

\subsection{Feedback efficiencies}
\label{sec:feedback_effs}

For the purpose of simulations such as the ones presented in this paper, the feedback power $P$ is the end-product of interest of any BH spin evolution model (with the jet direction also being of interest). For this reason, we specify here the feedback efficiencies $\epsilon$ used in our two accretion states, before elaborating on how we evolve the BH spin. We define the feedback efficiency $\epsilon$ (both the radiative and jet efficiency) using the relation $P=\epsilon\dot{M}_\mr{BH}c^2$. Thick discs are thought to have low radiative efficiencies and thin discs low jet efficiencies. We therefore assume, for simplicity and as a first-order approximation, that no jets are launched at high accretion rates ($\epsilon_\mr{j}=0$ for $\dot{m}>\dot{m}_\mr{crit}=0.01$) and no radiation is emitted at low accretion rates ($\epsilon_\mr{r}=0$ for $\dot{m}<\dot{m}_\mr{crit}=0.01$). Given some assumed radiative and jet efficiencies, the BH grows at the rate
\begin{equation}
    \dot{M}_\mr{BH}=(1-\epsilon_\mr{r}-\epsilon_\mr{j})\dot{M}_\mr{BH,0},
\label{eq:mdot_real}
\end{equation}
where $\dot{M}_\mr{BH,0}$ is the rest-frame large-scale accretion rate (before radiative or jet losses).

The radiative efficiency of the thin disc, $\epsilon_\mr{r}$, is taken from the general-relativistic solution presented by \cite{NovikovThorne1973}. It is assumed that the radiative efficiency is related to the binding energy of the gas at the innermost stable circular orbit (ISCO). Within this radius, $R_\mr{ISCO}$, orbits are assumed to decay quickly, carrying all of the gas energy into the BH before it can be radiated away. From infinity to the ISCO, a parcel of gas of mass $\Delta M$ loses a fraction $\epsilon_\mr{r,ISCO}$ of its total rest-frame mass-energy $\Delta Mc^2$ to radiation, while a fraction $e_\mr{ISCO} = 1 - \epsilon_\mr{r,ISCO}$ of it is kept. This fraction is the (dimensionless) binding energy. Using the known analytical expression for $e$ as a function of radius from \cite{NovikovThorne1973}, in combination with an analytical expression for the dimensionless radius $r_\mr{ISCO}=R_\mr{ISCO}/R_\mr{G}$ (see e.g.~Online Appendix A of Paper I), where $R_\mr{G}=M_\mr{BH}G/c^2$ is the gravitational radius of the BH, we can express the radiative efficiency of the thin disc as
\begin{equation}
\epsilon_\mr{r}(a) = 1-e_\mr{ISCO}(a)=1-\sqrt{1-\frac{2}{3r_\mr{ISCO}(a)}},
\label{eq:eq4}
\end{equation}
This formula yields an efficiency that grows monotonically as the BH spin is increased from $a=-1$ to $a=1$, due to the ISCO approaching the event horizon with increasing $a$. It grows slowly from $4.5$ per cent at $a=-1$ to $15$ per cent at $a=0.9$. Beyond this, the efficiency grows very steeply to reach a value of $42$ per cent at $a=1$ (or $32$ per cent at $a=0.998$, which is our actual cap).

For the jet efficiency, $\epsilon_\mr{j}$, we take the same approach as in Paper I (Section 2.2), to which we refer the reader for more details. Here we provide a condensed version. We assume that the jets are powered by the \cite{Blandford1977} (BZ) process, i.e.~they are launched by means of the extraction of energy from the rotational ergosphere of the BH. Whilst analytical expressions for jet powers exist for BZ jets, these rely on assuming classical accretion disc solutions and their magnetic fields (e.g.~\citealt{Meier2002}). The magnetic fields are highly uncertain in these solutions. We instead use jet efficiency formulas inferred from general-relativistic magnetohydrodynamical (GRMHD) simulations that have converged onto very similar jet powers (e.g.~\citealt{Tchekhovskoy2010}, \citealt{McKinney2012}, \citealt{Narayan2021}, \citealt{Lowell2023}). These simulations find that magnetic fields are dynamically important in the inner regions of the disc, where they ‘choke' the accretion flow. In this self-regulated and quasi-periodic state (the magnetically arrested disc, i.e.~MAD, see \citealt{Narayan2003}), the magnetic field saturates at some value that depends on the accretion rate and the BH spin. We take the jet efficiency formula presented in \cite{Narayan2021}, which we reproduce in Paper I. The main features of this formula are as follows: 1) at low BH spin, it leads to $\epsilon_\mr{j}\propto a^2$, in agreement with the classical BZ analysis, while at higher BH spin ($a>0.9$) the dependence is steeper ($\epsilon_\mr{j}\propto a^4$ or even $\epsilon_\mr{j}\propto a^6$), 2) the normalization for the thick disc is overall much higher than in the classical accretion disc solution (the efficiency may even be larger than $100\%$, so the BH loses mass as it accretes and launches jets; see Eqn.~\ref{eq:mdot_real}), to the point that jet spindown becomes very important and 3) the efficiency is higher for prograde accretion ($a>0$) than for retrograde accretion ($a<0$).

\subsection{Evolving the magnitude of the black hole spin}
\label{sec:spin_magn}

The evolution of the magnitude of the BH spin can be described by
\begin{equation}
    \frac{\mr{d}a}{\mr{d}M_\mr{BH,0}/M_\mr{BH}}=\ell_\mr{in}-2a e_\mr{in} - s_\mr{j},
\label{eq:da_dlnMSMBH}
\end{equation}
where $\mr{d}M_\mr{BH,0}$ is an increment of mass being accreted at large radii (i.e.~before radiative or jet losses) and $\ell_\mr{in}=cL_\mr{in}/GM_\mr{BH}$ is the dimensionless specific angular momentum, where $L_\mathrm{in}$ is the specific angular momentum at some inner radius $R_\mr{in}$, at which orbits are unstable and at which gas begins to quickly plunge into the BH. The first term in Eqn.~(\ref{eq:da_dlnMSMBH}) is due to gas accretion onto the BH, the second one originates from the definition of the BH spin $a$ through the presence of the BH mass as a factor, while the last term encapsulates spindown from jets\footnote{We should, in principle, add a term representing radiation (\protect\citealt{Thorne1974}) in the thin disc regime. This term causes spindown and is relevant for $a>0.99$. If $a>0.998$, the spindown from this term is stronger than the spinup from accretion, and vice-versa if $a<0.998$. For simplicity we neglect this term and instead simply cap the BH spin to a value of $0.998$.}. The second term includes the specific binding energy $e_\mr{in}$ at $R_\mr{in}$. 

For the thin disc, $R_\mr{in}$ corresponds to the radius of the ISCO. We use an analytical expression for $\ell_\mr{in}=\ell_\mr{ISCO}$, which is given in the Online Appendix A of Paper I. The binding energy $e_\mr{in}=e_\mr{ISCO}$ can be read off from Eqn.~(\ref{eq:eq4}). Since we assume that no jets are launched from the thin disc, we also set $s_\mr{j}=0$. 

For the thick disc, we replace the entire right-hand side of Eqn.~(\ref{eq:da_dlnMSMBH}) with a fitting formula for the spinup/spindown rates provided by \cite{Narayan2021}, who confirmed the results obtained by \cite{Tchekhovskoy2010}, and many authors since, on the jet production mechanisms and its dependence on BH spin in GRMHD simulations. Note that this is different from Paper I, where we used a mixture of numerical and analytical expressions that were not motivated by these simulations. Since we use the jet powers from GRMHD simulations (see \S~\ref{sec:feedback_effs}), using the spinup(down) rates from the same simulations is more consistent. The fitting formula used in the present paper is given by
\begin{equation}
    \frac{\mr{d}a}{\mr{d}M_\mr{BH,0}/M_\mr{BH}}=0.45 - 12.53a - 7.8a^2 +9.44a^3 + 5.71a^4 -4.03a^5.
\label{eq:da_dlnMSMBH_th}
\end{equation}
The right-hand-side of this equation is positive for $a<0.05$, leading to spinup, while it is negative for $a>0.05$, leading to spindown. Thus, $a_\mr{eq}\approx0.05$ is an equilibrium BH spin value at which accretion and jet launching balance each other in terms of angular momentum flux into/out of the BH (a result recently also confirmed by \citealt{Lowell2023} with even more sophisticated simulations, albeit with a slightly different value $a_\mr{eq}\approx0.07$). This GRMHD-derived value of the equilibrium spin is significantly lower than $a_\mr{eq}\approx0.25$, the value we obtained by using our analytical prescription in Paper I. The spindown is so much stronger (for positive spins) when using the results from the GRMHD simulations because of two separate reasons: 1) accretion provides even less angular momentum than is typically assumed from analytical calculations and 2) jets tap angular momentum (at a fixed power) from the rotation of the BH even more efficiently. 

\subsection{Deciding the sign and direction of the black hole spin}
\label{sec:spin_sign}

Eqn.~(\ref{eq:da_dlnMSMBH}) for the evolution of the BH spin depends only on how much matter is being accreted and the current BH spin, including its sign (direction). The sign of the BH spin encapsulates whether gas accretion is prograde or retrograde relative to the BH spin vector. In the inner accretion disc, the BH's angular momentum always dominates, so the accretion disc becomes either aligned or counter-aligned with the BH's spin vector through \cite{LenseThirring} torques. In the case of counter-alignment, we consider accretion to be retrograde and the BH spin negative. 

The thin disc develops a warp due to \cite{LenseThirring} torques and is aligned or counteraligned with the BH within a warp radius $R_\mr{warp}$, which is the radius out to which the ‘communication' of the BH and the disc is effective (\citealt{BardeenPetterson}), in terms of torques. Outside this radius, the accretion disc is undisturbed and aligned with the large-scale accretion flow (see \citealt{Fanidakis2011} and \citealt{Griffin2019a} for a detailed discussion of the structure of the disc in this case). For the thick disc, the assumption of exact (counter-)alignment is invalid. Instead, the disc precesses about the BH spin vector. This precession occurs on very short time-scales, much shorter than the ones we are simulating. For this reason we may also assume (counter-)alignment of the thick disc, in a time-averaged sense. Thus, in our model, the two accretion states are treated equally in this regard (but with different assumptions about the properties and structure of the accretion disc, which affects the size of the aligned or precessing region).

The sign of the BH spin (i.e.~whether the disc aligns or counteraligns) is decided based on the \cite{King2005} criterion (see Paper I for a detailed discussion). In this prescription, the BH and the inner accretion disc are assumed to come into (counter-)alignment in such a way that the magnitude of the BH spin does not change, and that the total angular momentum (of the BH + inner accretion disc) is conserved. The condition for counteralignment (and for spin to be negative) in this approach can be stated as follows:
\begin{equation}
\cos \theta<-\frac{J_{\mathrm{warp}}}{2 J_{\mathrm{BH}}},
\label{eq:counteralignment}
\end{equation}
where $\cos \theta=\bf{\hat{J}_\mr{BH}}\cdot\bf{\hat{J}_\mr{d}}$ is the initial misalignment between the BH and the (large-scale) angular momentum of the disc, whose direction is $\bf{\hat{J}_\mr{d}}$. $J_{\mathrm{warp}}$ is the total angular momentum of the inner accretion disc out to $R_\mathrm{warp}$. We describe how both are calculated in \S~\ref{sec:disc_structure}. 

Eqn.~(\ref{eq:counteralignment}) implies that if $\cos \theta>0$ (i.e.~if the angle $\theta$ between the BH spin vector and the angular momentum of the outer accretion disc is smaller than $90\degree$), the inner accretion disc is always aligned with the BH spin vector. On the other hand, if $\cos \theta<0$ (the BH spin vector and the angular momentum of the outer accretion disc are misaligned by more than $90\degree$), the warp angular momentum has to be at most similar in magnitude to $J_{\mathrm{BH}}$ for counteralignment to be possible. If $J_{\mathrm{warp}}$ is much larger than $J_{\mathrm{BH}}$, counteralignment cannot occur (even in the case of complete misalignment between the BH spin vector and the angular momentum of the outer accretion disc), which can be understood as a consequence of \cite{LenseThirring} torques being incapable of overpowering the large amount of angular momentum in the inner accretion disc compared to that of the BH.

From a numerical standpoint, the direction of the BH spin is evolved in the following way. For each increment of mass $M_\mr{warp}$ consumed by the BH, the BH-inner accretion disc system is assumed to come into equilibrium (with the inner accretion disc aligned or counter-aligned with the BH), so that the direction of the angular momentum of both the BH and the inner accretion disc is parallel with the direction of the total angular momentum $\bf{J_\mr{tot}}=\bf{J_\mr{BH}}+\bf{J_\mr{warp}}$. Here, $\mathbf{J_\mr{warp}}=J_\mr{warp}\mathbf{\hat{J}_\mr{d}}$ is the angular momentum of a single warp increment, which is assumed to be directed along the angular momentum of the outer accretion disc (i.e.~the large-scale accretion flow, which we calculate directly from the simulation). For a more detailed description of this process and the motivation for this implementation, we again refer the reader to Paper I.

\subsection{The structure of the accretion disc}
\label{sec:disc_structure}

In order to calculate the warp angular momentum, $J_\mathrm{warp}$, we have to: 1) know the size (radius) of the warp, $R_\mathrm{warp}$, 2) assume some accretion disc solution, which yields a surface density profile, $\Sigma(R)$, and 3) assume the specific angular momentum as a function of radius, $L(R)$. For the thick disc, we refer the reader to Paper I for all three of these. 

For the thin disc, the radius $R_\mr{warp}$, which separates the inner and outer accretion disc, can be calculated by equating the Lense-Thirring precession time-scale ($t_\mr{p}=2\pi/\Omega_\mr{p}$, with $\Omega_\mr{p}=2GJ_\mr{BH}/c^2R^3$ the precession rate) and the vertical warp propagation time-scale ($t_\mr{warp}=R^2/\nu_2$, with $\nu_2$ the kinematic viscosity in the vertical direction) (e.g.~\citealt{Pringle1992}, \citealt{Martin2007}, \citealt{Cielo2014}). The vertical kinematic viscosity $\nu_2$ can be related to the horizontal one, $\nu_1$, by $\nu_2=\xi\nu_1$, with $\xi$ a numerical factor (e.g.~\citealt{Lodato2010}). We use the relation $\dot{M}=3\pi\nu_1 \Sigma$ (for $R\gg R_\mr{ISCO}$, e.g.~\citealt{Fiacconi2018}) to calculate $\nu_1$, and therefore $\nu_2$. 

The warp radius depends on which regime of the thin disc we assume, with each having its own expression for $\Sigma$. The \cite{ShakuraSunyaev1973} solution of the thin disc describes three regions: a) an inner one where radiation pressure dominates, which is often unstable and usually does not extend far out, b) a middle one where gas dominates the pressure and electron-electron scattering dominate the opacity and c) an outer one where gas also dominates the pressure, but the opacity is dominated by free-free absorption. We ignore region a) (because the mass and angular momentum associated with that region is relatively small for our purpose) and assume, for simplicity, that the entire accretion disc, at least out to $R_\mr{warp}$, can be described by either region b) or c). We have tested both assumptions and they appear to have little effect. However, we keep both choices as options in our model and specify them both here for clarity and completeness. For the remainder of the paper, we assume the disc to be described by region b).

In region b), the surface density can be expressed as
\begin{equation}
\Sigma_\mr{TD,b}=6.84 \times 10^{5} \mathrm{~g} \mathrm{~cm}^{-2} \alpha^{-4 / 5} \dot{m}^{3 / 5}\left(\frac{M_{\mathrm{BH}}}{10^{8} M_{\odot}}\right)^{1 / 8}\left(\frac{R}{R_{\mathrm{S}}}\right)^{-3 / 5}
\label{eq:sigma_2}
\end{equation}
(\citealt{Collin}) whereas in region c)
\begin{equation}
\Sigma_\mr{TD,c}=3.41 \times 10^{4} \mathrm{~g} \mathrm{~cm}^{-2} \alpha^{-4 / 5} \dot{m}^{7/10}\left(\frac{M_{\mathrm{BH}}}{10^{8} M_{\odot}}\right)^{1 / 20}\left(\frac{R}{R_{\mathrm{S}}}\right)^{-3 / 4}
\label{eq:sigma_3}
\end{equation}
(see appendix in \citealt{Fiacconi2018}). Here, $R_\mr{S}=2R_\mr{G}$ is the Schwarzschild radius. Using these surface densities, the warp radii can be calculated as
\begin{equation}
R_{\text {warp,TD,b}}=3410 R_{S} a^{5 / 8} \xi^{-5/8}\alpha^{-1 / 2} \dot{m}^{-1 / 4}\left(\frac{M_{\mathrm{BH}}}{10^{8} M_{\odot}}\right)^{1 / 8}
\label{eq:Rwarp_2}
\end{equation}
for region b) (\citealt{Griffin2019a}) and 
\begin{equation}
R_\mathrm{warp,TD,c}=2629R_\mathrm{S}a^{4/7}\xi^{-4/7}\alpha^{-16/35}\dot{m}^{-6/35}\bigg(\frac{M_\mathrm{BH}}{10^8\hspace{0.5mm}\mathrm{M}_\odot}  \bigg)^{4/35},
\label{eq:Rwarp_3}
\end{equation}
for region c). The latter is equivalent to equation A8 from \cite{Fiacconi2018} (but with a different definition of $\xi$; we use $\xi=\nu_2/\nu_1$, whereas they use $\xi=(\nu_2/\nu_1)2\alpha^2$).

The ratio of the vertical and horizontal viscosity, $\xi$, is a constant parameter, often also expressed in the form $\alpha_2/\alpha$. Early theoretical calculations predicted $\alpha_2=1/2\alpha$ for small $\alpha$ (\citealt{Papaloizou1983}), which has also been confirmed by simulations (\citealt{Lodato2007}). Later simulations have found that higher-order corrections to this prediction may need to be included for realistic values of $\alpha$ (e.g.~\citealt{Lodato2010}), such as $\alpha=0.2$, as assumed in this paper. These numerical results agree with the theoretical prediction by \cite{Ogilvie1999}, which we assume here:
\begin{equation}
    \xi=\frac{\nu_2}{\nu_1}=\frac{\alpha_2}{\alpha}=\frac{2}{\alpha^2}\frac{1+7\alpha^2}{4+\alpha^2},
\label{eq:alpha_2}
\end{equation}
which reduces to $1/2\alpha^2$ for small $\alpha$. For our assumed value of $\alpha$ ($\alpha=0.2$) we obtain $\xi=15.84$ using the full expression, as opposed to $\xi=12.5$ when using the approximation. We use the former value.

We are finally able to calculate the warp angular momentum by using the expression
\begin{equation}
J_\mr{warp}(R_\mr{warp})=2\pi\int_0^{R_\mr{warp}}L(R)\Sigma(R)R\mr{d}R,
\label{eq:J_warp}
\end{equation}
where $L(R)$ is the specific angular momentum at a distance $R$ from the BH. A similar integral (without the $L(R)$ factor) is used to calculate the warp mass $M_\mr{warp}$. For the thin disc, we assume Keplerian orbital velocities, i.e.~$L(R)=\sqrt{M_\mr{BH}GR}$, and the surface densities are given by equations (\ref{eq:sigma_2}) and (\ref{eq:sigma_3}) for the two cases. 

Thin accretion discs can extend to large enough radii that they are prone to the effects of self-gravity (see \citealt{Lodato2008} for a review). At these distances, the gravity due to the disc locally becomes comparable to that due to the BH. The stability of the disc can be described using the Toomre instability parameter, $Q=\Omega c_{\mathrm{s}} /\pi G \Sigma$. For $Q<1$, the disc is prone to local gravitational instabilities and it likely undergoes gravitational collapse/fragmentation and star formation. We thus assume that the disc extends out to a radius $R_\mr{sg}$ where the Toomre instability parameter is equal to the critical value of $Q=1$. This equation, the Toomre instability criterion, can be solved to obtain
\begin{equation}
R_{\text {sg,TD,b}}=6460 R_{S} \alpha^{28/51} \dot{m}^{-18/51}\left(\frac{M_{\mathrm{BH}}}{10^{8} M_{\odot}}\right)^{-49/51}
\label{eq:Rsg_2}
\end{equation}
for region b) and 
\begin{equation}
R_\mathrm{sg,TD,c}=2456 R_{S} \alpha^{28/45} \dot{m}^{-22/45}\left(\frac{M_{\mathrm{BH}}}{10^{8} M_{\odot}}\right)^{-52/45}
\label{eq:Rsg_3}
\end{equation}
for region c) (\citealt{Fiacconi2018}). In the case that $R_\mr{sg}<R_\mr{warp}$, we simply assume that the entire accretion disc is (counter-)aligned and use $R_\mr{sg}$ instead of $R_\mr{warp}$ in all equations where $R_\mr{warp}$ makes an appearance.

\section{Simulations, methods and set-up}
\label{sec:sec3}

In this section we will describe the code, subgrid galaxy evolution model and physical set-up used to perform the simulations presented in this paper, the details of which we will also describe here. The simulations of idealized galaxy groups and clusters discussed in this paper are the same in substance as the ones presented in Paper I. For this reason, we will provide only a summary of the methods we use. For an even more detailed description than in Paper I, we refer the reader to \cite{Nobels2022}, where the physical set-up of these idealized galaxy groups and clusters is discussed in great detail. 

\subsection{Numerical code and subgrid physics model}

We use the SWIFT\footnote{\href{https://swiftsim.com}{https://swiftsim.com}} hydrodynamics and gravity code (\citealt{Schaller2023}) and its SPH method SPHENIX (\citealt{Borrow2022})\footnote{We use the quartic spline kernel with a resolution $\eta=1.2$, leading to $\approx60$ neighbours in each gas particle's kernel, on average (\protect\citealt{Dehnan2012}). We allow a minimum smoothing length of 0.1 the gravitational softening length (see Table \ref{tab:tab0} for the different values we use).}. SWIFT includes various subgrid physical processes, including our BH spin evolution model presented in \S~\ref{sec:sec2} for the thin, radiatively efficient disc. Additionally, it includes the BH spin evolution model for the thick, advection-dominated disc described in Paper I. In a future paper, such a model will also be presented for the slim, super-Eddington disc (e.g.~\citealt{Abramowicz1988}, \citealt{Wang1999}). SWIFT includes a thermal isotropic AGN feedback mode (\citealt{Booth2009}) that we use in the thin and slim disc, as well as kinetic AGN jets that we use in the thick and slim disc. We describe these feedback modes in \S~\ref{sec:agn_feedback_vars} (the kinetic jet mode is described in more detail in Paper I), alongside other feedback variations that we test.

In addition to AGN feedback, we include subgrid physics in the form of radiative gas cooling, an entropy floor and star formation. We do not include stellar feedback (nor stellar enrichment) in order to simplify the interpretation of the results. We have, however, performed test runs with both stellar feedback and enrichment included, the results of which we do not include here for the sake of brevity. From these runs we find that stellar enrichment and feedback can affect the time evolution of various quantities (e.g. feedback powers and star formation rates), but their effects are minor in a time-averaged sense.

We use essentially the same model for additional subgrid processes (other than AGN feedback) as in the EAGLE galaxy formation model (\citealt{Schaye2015}). We again refer the reader to that paper, \cite{Nobels2022} or Paper I for details. We use a slightly updated version of the EAGLE model with new cooling tables (\citealt{Ploeckinger2020}). The large-scale accretion rate $\dot{M}_\mr{BH,0}$ is set equal to the Bondi-Hoyle-Lyttleton rate (\citealt{BondiHoyle1944}):
\begin{equation}
\dot{M}_\mr{B}=4\pi\frac{G^2M_\mr{BH}^2\rho}{(c_\mr{s}^2+v^2)^{3/2}},
\label{eq:bondi}
\end{equation}
where $\rho$, $c_\mr{s}$ and $v$ are the kernel-weighted density, isothermal sound speed and velocity (relative to the SMBH) of the gas, respectively.

Our usage of the Bondi-Hoyle-Lyttleton rate also differs slightly from the EAGLE model, where that rate was suppressed by an additional factor related to the angular momentum of the gas near the BH. For simplicity we do not suppress the Bondi-Hoyle-Lyttleton rate for the effects of gas turbulence nor vorticity (unlike in \citealt{Nobels2022}, where the suppression due to both effects was accounted for). We also do not boost it, which would account for unresolved high gas densities that the BH would sometimes be embedded in (\citealt{Booth2009}). We do not implement this boost since it is largely used to ensure that BHs grow sufficiently in cosmological simulations of galaxy formation, but in these simulations we place BHs with a given mass by hand. Furthermore, their growth is self-regulated by their own feedback, and should thus be less sensitive to resolution.

Alongside being used to calculate the accretion rate, we also use the gas in the BH smoothing kernel to calculate the \textit{direction} of its angular momentum. We then assume that this determines the direction of the angular momentum of the outer regions of the subgrid accretion disc $\bf{\hat{J}_\mr{d}}$ (\S~\ref{sec:spin_sign}). This is a strong assumption: the direction of the angular momentum of the gas may change significantly as the gas moves down from the scales we are simulating ($\sim100-1000$ pc), to the scales of the accretion disc ($<1$ pc) (see section 2.6 of Paper I for a detailed discussion of this assumption).

\subsection{Implementation of AGN feedback}
\label{sec:agn_feedback_vars}

When implementing any feedback mechanism, several choices must be made: 1) how the energy is directed, 2) what is the feedback power, 3) how much energy is imparted per each feedback event and 4) what form the energy takes. In this paper we compare two different forms of AGN feedback in terms of how it is directed: isotropic and jet feedback (the former of which is done as in \citealt{Booth2009}, at least for the thermal case). For both of these options, we thoroughly compare different choices related to points 2-4 above. 

In the isotropic case, energy is imparted to the closest particle in the BH smoothing kernel. Note, however that this implementation is not precisely isotropic, since isotropic feedback would entail choosing random angles and imparting energy to the particles closest to those chosen angles. \cite{Chaikin2022} compared different numerical implementations of kinetic feedback (albeit stellar, but this makes no difference for the following argument), including ‘Min distance' and ‘Isotropic'. In the former, the closest particles to the BH are heated (corresponding to what we do here), while in the latter, particles were chosen in pairs along rays (that do not generally pass through the central star or BH that is injecting the energy) to ensure conservation of not only energy, but also linear and angular momentum. They found the results to be very similar in the two kicking schemes. Throughout the rest of this paper, for simplicity we refer to the scheme we use (‘Min distance' from \citealt{Chaikin2022}) as ‘isotropic', since we use it to represent the effects of isotropic winds, and since it is much different from jet feedback regardless of the details of how it is implemented.

In the jet case, energy is always imparted to two particles instead of one, and the same criterion is used to choose the particles as in our isotropic feedback (the ones closest to the BH; see Paper I for other choices and their effects on jet feedback). In order to find a pair of particles to kick in roughly opposite directions, we define two hemispheres within the BH smoothing kernel. The equatorial plane separating them is perpendicular to the vector that defines the launching direction of the jets (the $z-$axis or the BH spin vector).

Several parameters can be tuned to affect the behaviour of these feedback mechanisms (as described by points 2-4 in the beginning of this section). The first of these is the feedback efficiency $\epsilon$, which controls how much feedback energy is injected given some amount of BH accretion.  We use variable feedback efficiencies in the case where the BH spin and its evolution are used (\S~\ref{sec:res_spin}), but also values fixed throughout the duration of a given simulation in a simplified model (\S~\ref{sec:res_fixed}). 

The feedback power is funneled to a reservoir of energy. Once the reservoir exceeds some threshold value $\Delta E$, a feedback event occurs (either one particle receiving energy in the isotropic case, or a pair in the jet case). The energy $\Delta E$ is imparted in either thermal, kinetic or mixed form (in the latter case, half of the energy is injected as thermal and half as kinetic). Thus, there are three choices to make in both the isotropic and jet case: 1) the feedback efficiency $\epsilon$, 2) the energy threshold $\Delta E$ and 3) the type of energy being received. In all of our isotropic cases, we use large enough values of $\Delta E$ that the feedback is energy-dominated, rather than momentum-dominated (see e.g.~\cite{Faucher2012},~\citealt{Costa2014}). We thus expect no additional radiative cooling (of a physical or numerical nature) in the regions immediately ahead or behind the outflows associated with feedback, as seen for momentum-driven outflows that appear if low velocities are used for kinetic feedback.

In the thin, radiatively efficient disc (used at high accretion rates), we use the thermal isotropic variant of AGN feedback to represent the effects of radiation-driven winds\footnote{In the thermal isotropic case, we generally refer to the total feedback efficiency $\epsilon$, which is different from the radiative efficiency $\epsilon_\mr{r}$ for the following reason. The BH radiates at a rate $\epsilon_\mr{r}\dot{M}_\mr{B}c^2$, but only a fraction $\epsilon_\mathrm{f}$ (the coupling, or feedback, efficiency) of that actually couples with the gas in the simulation. The total feedback efficiency is therefore $\epsilon=\epsilon_\mr{f}\epsilon_\mr{r}$. This distinction has a small effect in the simulations in that the BH accretes only $(1-\epsilon_\mr{r})$ of the total accretion rate, rather than a fraction $(1-\epsilon)$ of it. We fix $\epsilon_\mr{f}=0.1$ and vary $\epsilon_\mr{r}$ in our simulations. For the jets we assume $\epsilon_\mr{f}=1$ and drop the factor hereafter.}. This assumption is valid if the radiatively-driven winds shock and deposit their energy on small scales (e.g.~$1-100$ pc) that we do not resolve in these simulations, leading to hot gas that expands on account of thermal pressure (e.g.~\citealt{Faucher2012}). For the thick, advection-dominated disc, we use kinetic jets to represent the effects of relativistic jets launched in this accretion regime. In both cases, our BH spin evolution model is used to evolve the radiative and jet feedback efficiencies (\S~\ref{sec:sec2} for the former, and Paper I for the latter), when we allow them to vary. For the jet case, this also results in a variation of the jet direction, which is assumed to be aligned with the BH spin axis.

In the case that particles are being isotropically heated, we refer to (and vary) the heating temperature $\Delta T$ instead of $\Delta E$; the two are related by $\Delta E = (3m_\mathrm{g}/2\mathrm{\mu}m_\mathrm{p})k_\mathrm{B}\Delta T$, where $m_\mathrm{g}$ is the gas particle mass, $\mu=0.62$ the mean molecular weight of ionized gas and $k_\mathrm{B}$ the Boltzmann constant. 

In the kinetic jet case, we express the energy being received by the particles through the jet velocity $v_\mathrm{j}$ as $\Delta E=2\times m_\mathrm{g}v_\mathrm{j}^2/2$, where the multiplication by two is present since we always kick in pairs. We do not kick particles perfectly along the jet direction, but instead implement a finite half-opening angle of $\theta_\mr{j}=10\degree$. This is accomplished by assigning a new kick direction every time a kick event occurs; this direction is given by a unit vector $\mathbf{n}_\mr{j}$ that is drawn randomly and uniformly in solid angle within a cone with a half-opening angle $\theta_\mr{j}$ directed along the chosen jet direction (either aligned with the BH spin vector or the $z-$axis). Since we always kick in pairs, the above procedure is done for one particle in the ‘positive' direction (along the jet direction) and for another particle in the ‘negative' direction (counteraligned with the jet direction).

We kick particles by increasing their velocity (in the frame of the BH) by $\Delta \mathbf{v} = \Delta v \mathbf{n}_\mr{j}$. The magnitude of the velocity increase $\Delta v$ is chosen in such a way that the kinetic energy of each particle increases exactly by $\Delta E/2$. Conservation of kinetic energy gives
\begin{equation}
\frac{1}{2}m_\mr{g}(\mathbf{v}_i+\Delta\mathbf{v})^2 - \frac{1}{2}m_\mr{g}\mathbf{v}_i^2 = \frac{\Delta E}{2},
\label{eq:conservation}
\end{equation}
where $\mathbf{v}_i$ is the initial velocity. This equation can be solved for the magnitude of the velocity increase $\Delta v$, yielding
\begin{equation}
\Delta v = \sqrt{v_\mr{i,j}^2 + v_\mr{j}^2} - v_\mr{i,j},
\label{eq:conservation2}
\end{equation}
where $v_\mr{i,j}=\mathbf{v}_i \cdot \mathbf{n}_\mr{j}$ is the initial velocity projected onto the kick direction. This equation implies that the change in the particle velocity is always smaller than the target velocity, i.e.~$\Delta v < v_\mr{j}$, if the initial velocity is non-zero. However, we use fairly large values of $v_\mr{j}$ that are at least a factor of $10$ larger than the initial particle velocities, so in practice $\Delta v \approx v_\mr{j}$.

Heated and kicked particles can have much shorter time steps than their neighbours that may make up the ambient medium. We have thus used a time step limiter that ensures that particles never differ by more than a factor of four in the size of their time steps (within a given particle's kernel). We have also ensured, with simple tests of individual particle kicks, that this time step limiter effectively `wakes up' particles ahead of an incoming kicked particle.

\subsection{Physical set-up}
\label{sec:setup}


In order to test the different implementations of AGN feedback introduced above, we simulate idealized galaxy groups and clusters. The initial set-up of these systems follows \cite{Nobels2022}. We focus on three halo masses, which correspond to a galaxy group ($M_{200}=10^{13}$ $\mr{M}_\odot$), a low-mass galaxy cluster ($M_{200}=10^{14}$ $\mr{M}_\odot$) and a high-mass galaxy cluster ($M_{200}=10^{15}$ $\mr{M}_\odot$). Here, the halo masses are defined as the masses within the radius $R_{200}$, the radius of a sphere within which the mean density is 200 times the critical density at $z=0$.

We use a \cite{NFW} (NFW) gravitational potential to represent the dark matter. The value of the concentration parameter, $c_\mr{NFW}$, is chosen for each halo to be in line with the mass-concentration relation found by \cite{Correa2015}. A \cite{Hernquist1990} profile is used to represent the stellar population of the BCG (for which we use live particles), given some total stellar mass $M_*$ and a stellar half-light radius $a_*$ (i.e.~the scale length of the \citealt{Hernquist1990} profile). Using the NFW and Hernquist potentials, a gaseous halo representing the ICM is generated in such a way that it is in hydrostatic equilibrium, and that the baryonic mass fraction (ratio of enclosed baryonic and total masses) within radius $R_{500}$ (defined in a similar way as $R_{200}$, but using an overdensity factor of $500$) is equal to some value $f_\mathrm{b,500}$. These values are calibrated on the BAHAMAS simulations (\citealt{McCarthy2017}). In the central regions of the ICM, the gaseous halo is modified such that its temperature approaches some value $T_0$, which controls how cool-core (CC) or non-cool-core (NCC) the halo is. At the centre of the halo we place a BH, which is fixed there throughout the simulation. We assume some initial BH mass and spin, the latter of which is directed along the $z-$axis.

All of the above parameters vary with halo mass. For the following parameters we assume values based on general expectations and scaling relations between these quantities and halo masses: halo concentration, baryonic fraction, stellar mass and half-light radius, and BH mass. Our assumed values for each halo mass are listed in Table \ref{tab:tab0}. These parameters do not vary in any of our simulations, other than with halo mass, as shown in the table.

The initial central temperature of the gas $T_0$ has a strong impact on the simulations (see \citealt{Nobels2022} for the thermal isotropic feedback case, and \citealt{Husko2022_spin_driven} for kinetic jet case). For this study we choose relatively low values that lead to significant cooling and feedback on the Gyr time-scales of the simulations we are performing here. In other words, these choices of $T_0$ correspond to a relatively CC set-up, rather than NCC (the majority of observed groups and clusters do not have appreciable amounts of ongoing cooling or AGN feedback). While this choice may not be representative of the entire population of galaxy clusters, we make it for a few reasons: 1) it leads to more AGN activity, allowing us to compare different AGN feedback schemes more easily, 2) the cooling flows are stronger, so the potential of various AGN feedback schemes to shut them off is tested to a stronger degree, 3) the BH accretion rates are higher, leading to the accretion regime more often corresponding to the thin, radiatively-efficient disc (the regime for which we have developed a model that we wish to test in detail using this setting). The BH spins we choose are relatively low; in galaxy groups and clusters we do not expect fully spun-up BHs due to spindown from jets and BH-BH merger activity.

Other choices also have to be made in setting up the ICM, although they are independent of halo mass (for our study). We assume a constant gas metallicity (as found in at least some observations, e.g.~\citealt{Werner2013} and \citealt{McDonald2016}) of $0.3Z_\odot$ (with the solar metallicity chosen as $Z_\odot=0.0134$; \citealt{Asplund2009}). We also assume rotation of the ICM about the $z-$axis (see \citealt{Nobels2022} for details on how this is set up) with a spin parameter of $\lambda=0.05$ (\citealt{Bullock2000}), which is slightly larger than that of the DM (\citealt{Oppenheimer2018}). 

\captionsetup[table]{skip=0pt} 
\begin{table*}
\begin{center}
\caption{List of parameters for the initial conditions (first eight columns) and numerical resolution (last two columns) of our idealized galaxy group and cluster simulations. These are, in order: 1) $M_\mr{200}$ - halo mass, 2) $r_\mr{200}$ - halo virial radius, 3) NFW halo concentration parameter $c_\mr{NFW}$, 4) baryonic mass fraction within $R_{500}$, $f_\mathrm{b,500}$, 5) central gas temperature $T_\mathrm{0}$, 6) stellar mass of the BCG, 7) Hernquist scale length (half-light radius) of the galaxy, 8) mass of the central BH, 9) spin of the BH, 10) central gas mass resolution, 11) gravitational softening length.}
\label{tab:tab0}
\end{center}

\begin{tabular*}{1.\textwidth}{@{\extracolsep{\fill}}llllllcrrrrrr}
  \hline
  $M_\mr{200}$ $[\mr{M}_\odot]$ & $R_\mr{200}$ $[\mathrm{kpc}]$ & $c_\mathrm{NFW}$ & $f_\mr{b,500}$ & $T_\mr{0}$ [K] & $M_{*}$ $[10^{11}\hspace{0.3mm}\mr{M}_\odot]$ & $a_*$ [$\mr{kpc}$] & $M_\mr{BH}$ $[10^{9}\hspace{0.3mm}\mr{M}_\odot]$ & $a_0$ & $m_\mr{g,0}$ $[10^5\hspace{0.3mm}\mr{M}_\odot]$ & $\epsilon_\mr{g}$ [$\mr{kpc}$]   \\
  \hline 
  $10^{13}$ & $442.7$ & $7.2$  & $0.05$ & $10^6$ & $1$ & $10$ & $0.25$ & $0.2$ & $1$ & $0.25$ \\
  $10^{14}$ & $953.8$ & $5.6$ & $0.10$ & $10^{6.75}$ & $2.5$ & $20$ & $0.5$ & $0.2$ & $8$ & $0.5$  \\
  $10^{15}$ & $2054.8$ & $4$ & $0.15$ & $10^{7.5}$ & $6$ & $30$ & $6.5$ & $0.4$  & $64$ & $1$ \\
  \hline
\end{tabular*}

\end{table*}

We assume that the ICM extends out to $3R_\mr{200}$. In the central 500 kpc of the ICM, we use a fixed gas particle mass resolution $m_\mr{g,0}$, while outside $500$ kpc, the particle masses (in the initial conditions) increase as $m_\mr{g,0}(r/500\hspace{0.3mm}\mr{kpc})^2$. This drop in resolution allows us to perform relatively higher-resolution simulations (in terms of how well the central regions of the halo are resolved). The central mass resolution $m_\mr{g,0}$ is chosen to increase with the halo mass (by factors of 8), since more massive objects require more computational resources to be simulated. This means that we resolve more massive haloes more poorly in terms of spatial scales, but these haloes also contain more gas, both hot and cold (when star formation is ongoing), and have stronger feedback episodes with more power and mass flux. As a result, on average, we actually resolve all phases (hot gas in the ICM, cold star-forming gas, as well as gas making up the outflows associated with feedback) with more resolution elements in more massive haloes, despite the decrease in the gas mass resolution. Furthermore, \cite{Nobels2022} demonstrated convergence of simulations with thermal isotropic feedback down to the resolutions we use, and we have done the same in Paper I for simulations with kinetic jet feedback. The values we have chosen for our gas mass resolution are given in Table \ref{tab:tab0}, alongside the gravitational softening lengths $\epsilon_\mr{g}$. We run all of our simulations for a duration of 8 Gyr.

\subsection{Simulations}
\label{sec:simulations}

We perform a total of nine simulations using the BH spin evolution model presented in Paper I and \S~\ref{sec:sec2}; three for each halo mass. The three for each case use different variations of BH spin evolution and feedback: 1) one simulation using the thin, radiatively efficient disc and thermal isotropic feedback, 2) one using the thick, advection-dominated disc with kinetic jets and 3) one with hybrid accretion and feedback modes, with the thin disc mode used at high accretion rates ($\dot{m}>\dot{m}_\mr{crit}$) and the thick disc one at low accretion rates ($\dot{m}<\dot{m}_\mr{crit}$). The details of these simulations are given in
Table \ref{tab:tab1}. This last model represents the most realistic one and should thus replicate the behaviour of BHs in the real Universe most closely.

In this work we use heating temperatures, $\Delta T$, of order $10^9$ K as motivated by many previous works (e.g.~\citealt{Schaye2015}). For jet velocities, $v_\mathrm{j}$, we choose values of order $10^4$ $\mathrm{km}$ $\mathrm{s}^{-1}$, instead of relativistic ones, mainly due to limitations related to resolution (see footnote \ref{footnote2}). We increase the heating temperatures and jet velocities with halo mass, in order to sample feedback at a similar level (using the same values would result in the sampling of feedback being significantly better as halo mass is increased, which might thus lead to artificial numerical differences between the three simulated haloes). The increase in jet velocity with halo mass is also motivated by previous simulations we have done (e.g.~\citealt{Husko2022_self_similar} and \citealt{Husko2022_spin_driven}), where we found that jets need to be highly supersonic relative to the external medium (by a factor $M=v_\mathrm{j}/c_\mathrm{s,ICM}\geq10$) in order to inflate lobes. As the ICM temperature increases with increasing halo mass, this implies that an increase in jet velocity is well-motivated.

\captionsetup[table]{skip=0pt} 
\begin{table*}
\begin{center}
\caption{List of simulations performed with the BH spin evolution model (see Paper I and \S~\ref{sec:sec2}). For each halo mass, three simulations were performed with different feedback and accretion modes. In the hybrid mode, the thin disc and thermal isotropic feedback is used when the BH is accreting with $\dot{m}>\dot{m}_\mr{crit}=0.01$, while the thick disc and kinetic jets are used otherwise. The details of these simulations are given below.}
\label{tab:tab1}
\end{center}

\begin{tabular*}{1.\textwidth}{@{\extracolsep{\fill}}cccccc}
  \hline
  $M_\mr{200}$ $[\mr{M}_\odot]$ & Accretion disc & Feedback mode & Feedback efficiency $\epsilon$ & Heating temperature $\Delta T$ [K] & Jet velocity [$10^4\hspace{0.5mm}\mathrm{km}\hspace{0.3mm}\mathrm{s}^{-1}$]  \\
  \hline
  $10^{13}$ & Thin & Thermal isotropic & $\epsilon_\mr{f}\epsilon_\mr{r}(a)$ & $10^{8.5}$ & -- \\
  $10^{13}$ & Thick & Kinetic jets & $\epsilon_\mr{j}(a)$ & -- & $0.5$ \\
  $10^{13}$ & Hybrid & Hybrid & $\epsilon_\mr{f}\epsilon_\mr{r}(a)$ or $\epsilon_\mr{j}(a)$ & $10^{8.5}$ & $0.5$ \\
  \hline
  $10^{14}$ & Thin & Thermal isotropic & $\epsilon_\mr{f}\epsilon_\mr{r}(a)$ & $10^{9}$ & -- \\
  $10^{14}$ & Thick & Kinetic jets & $\epsilon_\mr{j}(a)$ & -- & $1.5$ \\
  $10^{14}$ & Hybrid & Hybrid & $\epsilon_\mr{f}\epsilon_\mr{r}(a)$ or $\epsilon_\mr{j}(a)$ & $10^{9}$ & $1.5$ \\
  \hline
  $10^{15}$ & Thin & Thermal isotropic & $\epsilon_\mr{f}\epsilon_\mr{r}(a)$ & $10^{9.5}$ & -- \\
  $10^{15}$ & Thick & Kinetic jets & $\epsilon_\mr{j}(a)$ & -- & $3$ \\
  $10^{15}$ & Hybrid & Hybrid & $\epsilon_\mr{f}\epsilon_\mr{r}(a)$ or $\epsilon_\mr{j}(a)$ & $10^{9.5}$ & $3$ \\
  \hline
\end{tabular*}

\end{table*}

We also perform simulations with simplified feedback prescriptions. For these we fix the feedback efficiencies to constant values, as well as fixing the jet directions to be along the $z-$axis. The details of these simulations are given in Table \ref{tab:tab2}. We perform these simulations only for the galaxy clusters ($M_{200}=10^{14}$ $\mr{M}_\odot$ and $M_{200}=10^{15}$ $\mr{M}_\odot$) since these simulations show more interesting (or variable) behaviour than the galaxy group ones. The motivation for these simulations is to provide a comparison of different feedback modes by removing any differences due to variations in the feedback efficiency. To this end we include runs where we fix the efficiency to $\epsilon=0.01$ in both the thermal isotropic and kinetic jet cases. For the kinetic jet case we test two options: 1) using jet velocities that are $\geq10$ times higher than the sound speed of the ICM and 2) using lower velocities (by a factor $\approx3$ relative to option 1) that, however, lead to the energy per feedback event $\Delta E$ being the same as in the equivalent thermal isotropic simulations. We consider option 1) our fiducial choice, for the reasons laid out in the paragraph above.

For the low-mass galaxy cluster ($M_{200}=10^{14}$ $\mr{M}_\odot$) case, we also perform a series of simulations whose parameters are specified in the last two rows Table \ref{tab:tab2}. The purpose of these simulations is to vary all parameters of interest: the feedback efficiency, the energy per feedback event and the type of energy being injected. These variations were done for both the isotropic and jet cases. For the jet case, we also tested the importance of the jet direction by manually redirecting the jets in random directions with a given periodicity, and also by precessing them with varying opening angles and periods. These simulations, and their results, are discussed in detail in Appendix \ref{sec:app1}. We found the jet direction to be largely unimportant for the type of simulations being performed here.

\captionsetup[table]{skip=0pt} 
\begin{table*}
\begin{center}
\caption{List of simulations performed with simplified feedback prescriptions (without BH spin evolution, i.e.~with fixed feedback efficiencies and jets in the direction of the $z-$axis). Three simulations were performed for the low- and high-mass galaxy clusters. For the low-mass galaxy cluster, further variations of all the parameters were performed in a total of 16 simulations. The parameters of these simulations are given in the last two rows. $^*$ Th. - thermal, Mix. - mixed (half thermal, half kinetic), Kin. - kinetic.}
\label{tab:tab2}
\end{center}

\begin{tabular*}{1.\textwidth}{@{\extracolsep{\fill}}ccccccccccccc}
  \hline
  $M_\mr{200}$ $[\mr{M}_\odot]$ & Feedback mode & Energy type & Feedback efficiency $\epsilon$ & Heating temperature $\Delta T$ [K] & Jet velocity [$10^4\hspace{0.5mm}\mathrm{km}\hspace{0.3mm}\mathrm{s}^{-1}$]  \\
  \hline
  $10^{14}$ & Isotropic & Thermal & $0.01$ & $10^{9}$ & -- \\
  $10^{14}$ & Jet & Kinetic & $0.01$ & -- & $0.65$ \\
  $10^{14}$ & Jet & Kinetic & $0.01$ & -- & $1.5$ \\
  \hline
  $10^{15}$ & Isotropic & Thermal & $0.01$ & $10^{9.5}$ & -- \\
  $10^{15}$ & Jet & Kinetic & $0.01$ & -- & $1.15$ \\
  $10^{15}$ & Jet & Kinetic & $0.01$ & -- & $3$ \\
  \hline
  $10^{14}$ & Isotropic & Th., Mix., Kin.$^*$ & $0.01-1$ & $10^{8-9.5}$ & -- \\
  $10^{14}$ & Jet & Th., Mix., Kin.$^*$ & $0.01-1$ & -- & $0.47-2.66$ \\
  \hline
\end{tabular*}

\end{table*}

\subsection{Observational sample of entropy profiles}
\label{sec:obs}

In this work, we mainly focus on the gas entropy when discussing the impact of feedback on the ICM. For this purpose we define the entropy as $K=k_\mr{B}T/n_\mr{e}^{2/3}$, where $k_\mr{B}$ is the Boltzmann constant and $T$ and $n_\mr{e}$ are the gas temperature and electron number density, respectively. We will compare our simulated entropy profiles of the ICM (as a function of radius) to observed ones inferred from X-ray observations. 

For high-mass clusters there are plentiful such samples due to the hot ICM gas falling well into the range observable by X-ray observatories such as \textit{Chandra}, and since these clusters are easier to observe due to a larger intrinsic brightness. We compare the simulated high-mass galaxy cluster ($M_{200}=10^{15}$ $\mr{M}_\odot$) with the observed ones from \cite{Pratt2010}, who studied 31 nearby clusters using $\textit{XMM-Newton}$, as well as those from \cite{Ghirardini2019} using the same telescope, but with a different sample of 12 galaxy clusters. We also compare with \textit{Chandra} observations by \cite{Cavagnolo2009}, who provide entropy profiles for a large sample of 239 high-mass galaxy clusters ($M_\mr{500}\approx10^{15}$ $\mr{M}_\odot$, where $M_\mr{500}$ is the halo mass using a virial overdensity of factor 500 relative to the critical density). They also split their sample into CC and NCC clusters based on whether the central entropy is below or above $50$ keV$\hspace{0.3mm}\mr{cm}^2$.

For galaxy groups and low-mass clusters ($M_\mr{500}\leq10^{14}$ $\mr{M}_\odot$), such observations are inherently difficult (e.g.~\citealt{Werner2019},~\citealt{Eckert2021},~\citealt{Lovisari2021},~\citealt{Oppenheimer2021}). The sample sizes tend to be small and/or they span a large range in halo mass. The halo masses of these galaxies cannot currently be measured through X-ray observations, since their ICM/CGM may not be in hydrostatic equilibrium, nor is the X-ray emission typically measured up to the virial radius (or an appreciable fraction of it so that one may extrapolate the pressure profile). The samples may also be biased towards CC (low-entropy) ones since such X-ray atmospheres are more likely to be bright and therefore observed. Finally, it is also likely that many of these observed X-ray atmospheres surround satellite galaxies rather than being the central ones of primary haloes. Tidal stripping may be affecting many such galaxies, or it may also be biasing the samples towards the X-ray bright ones.

Notwithstanding those currently unavoidable shortcomings, we compare the entropy profiles of our galaxy group ($M_{200}=10^{13}$ $\mr{M}_\odot$) and low-mass cluster ($M_{200}=10^{14}$ $\mr{M}_\odot$) simulations with a set of different observational papers. We use the data based on 28 and 43 observed galaxy groups and clusters by \cite{Johnson2009} (using \textit{XMM-Newton}) and \cite{Sun2009} (using \textit{Chandra}), which provide useful constraints on the entropy profiles between roughly $r=30$ kpc and $r=1$ Mpc. At relatively small radii ($r<100$ kpc) we compare with data from \cite{Babyk2018}, who compiled observed profiles of 40 galaxies/groups and 110 galaxy clusters, all observed with \textit{Chandra}. For all these systems, \cite{Babyk2018} find a universal median entropy profile, which they fit with $K\propto r^{2/3}$ at small radii and $K\propto r^{1.1}$ at large ones. Finally, we compare with \cite{Lakhchaura2018}, who presented entropy profiles of 49 bright elliptical galaxies observed with \textit{Chandra}. These data are largely consistent with the \cite{Babyk2018} ones, although they tend to follow a single slope with radius.

\section{Results I: Feedback with black hole spin evolution}
\label{sec:res_spin}

\begin{figure*}
\includegraphics[width=0.98\textwidth, trim = 0 10 0 0]{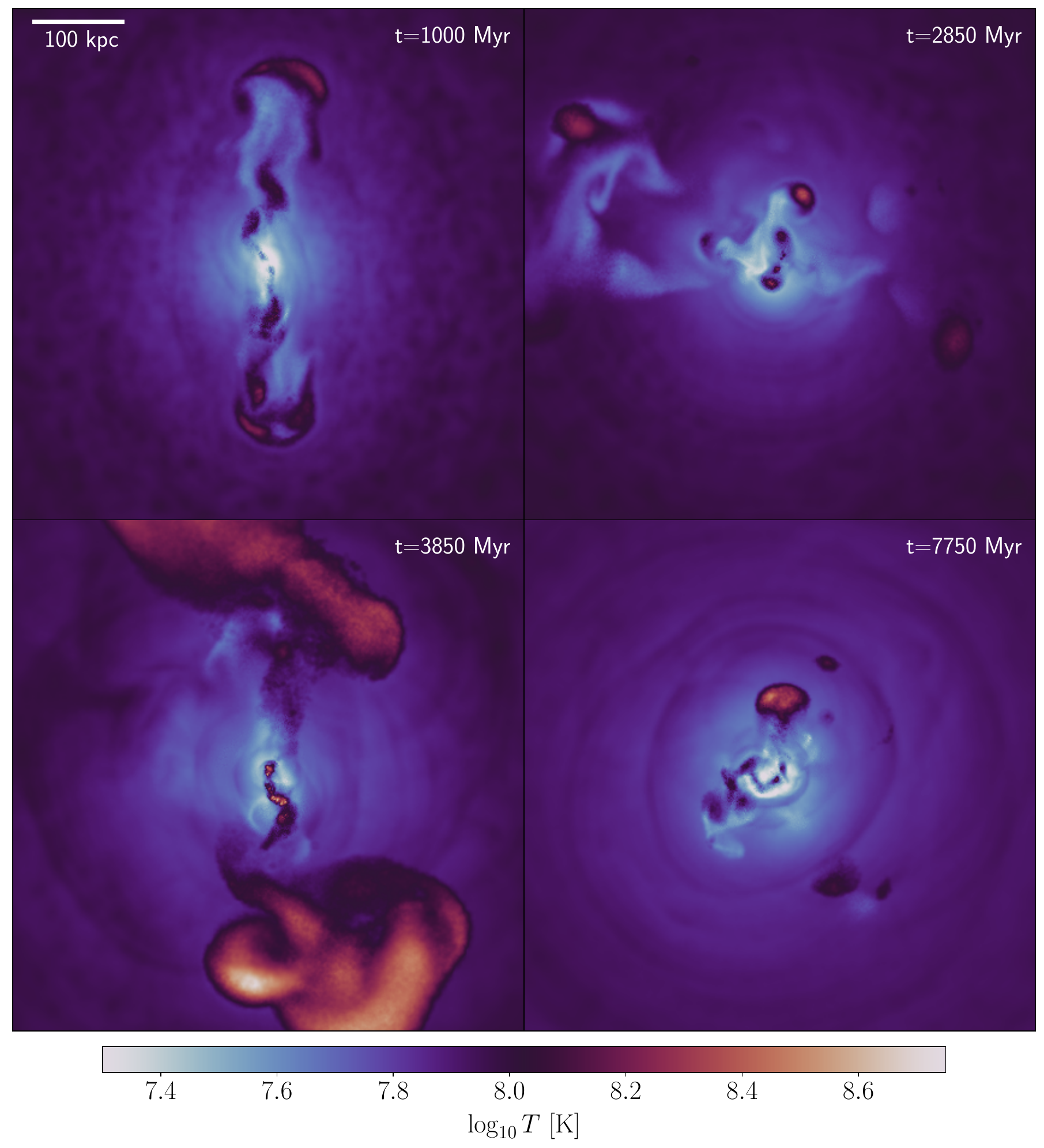}
\caption{A visualization of the gas temperature at four representative times in our hybrid simulation (with both kinetic jets and thermal isotropic feedback, interchanging at an Eddington ratio $\dot{m}_\mathrm{crit}=0.01$) of the high-mass galaxy cluster ($M_\mathrm{200}=10^{15}$ $\mathrm{M}_\odot$). The colours indicate the projected, mass-weighted gas temperature, as indicated by the colour bar, and we include all gas in a $50$ kpc-deep slice. The left-hand panels show times when kinetic jet feedback dominates, while the right-hand panels show times when thermal isotropic feedback is dominant. The bottom two panels show that both types of feedback lead to spherical shock waves. At all times shown here, ambient gas uplifted by feedback-induced outflows is visible in the form of cool filamentary structures.
}
\label{fig:fig0}
\end{figure*}%

We first consider the results of using the BH spin evolution model for all three of the halo masses, from the galaxy group ($M_\mathrm{200}=10^{13}$ $\mathrm{M}_\odot$) to the high-mass cluster ($M_\mathrm{200}=10^{15}$ $\mathrm{M}_\odot$) scale. For each of the halo masses, we performed three simulations: 1) using the thin, radiatively-efficient disc and thermal isotropic feedback, 2) using the thick, advection-dominated disc and kinetic jet feedback and 3) a hybrid case where the two accretion and feedback modes interchange at $\dot{m}=\dot{m}_\mr{crit}=0.01$. The details of these simulations are given in \S~\ref{sec:setup} and Table \ref{tab:tab0} (in terms of physical set-up and halo mass) as well as \S~\ref{sec:simulations} and Table \ref{tab:tab1} (in terms of feedback implementation).

In Fig.~\ref{fig:fig0} we show visualizations of the gas temperature in our hybrid simulation of the high-mass cluster. These show the qualitative behaviour of the feedback and cooling cycle, which we consider to be representative of all our simulations. These visualizations highlight the rich variety of structures we find, with many of them similar to features observed in the ICM of real galaxy clusters. The bulk of the ICM on the spatial scales shown in Fig.~\ref{fig:fig0} has a temperature of order $T\approx10^{7.5}-10^8$ K (light-blue to dark-purple colours), varying with radius. Black colours indicate gas that is slightly hotter (mostly due to shock waves), while orange-to-white colours indicate gas that is a factor of several times hotter than the ambient medium (the gas launched as part of feedback or entrained in the same process). These visualizations also show gas that is strongly cooling (white colours).

The two left-hand panels show simulation times when the kinetic jet activity is peaking, while the two right-hand panels show the same for thermal isotropic feedback. From the two left-hand panels, we see that jet feedback can lead to asymmetrical large-scale outflows, as a result of several processes, some of which are: 1) jet redirection and/or precession, 2) variability in the jet power and 3) the complex structure of the ICM in the jets' path (including uplifted low-entropy gas due to previous feedback episodes; we discuss this below). From the right-hand panels, we see that thermal isotropic feedback generally does \textit{not} lead to isotropic outflows. This is partly a result of how it is implemented in our simulations: gas is heated to large temperatures ($\Delta T=10^{9.5}$ K in this case). This hot gas tends to not expand isotropically, but rather in the 'path of least resistance' away from the BCG. The first few heating events in a given feedback episode create a channel that represents the preferred direction in which the subsequently heated gas will expand. 

For both thermal isotropic and kinetic jet feedback, we see that the typical temperature of the hot gas outflows and bubbles is not similar to the temperature associated with the launching events ($\Delta T=10^{9.5}$ K and $\Delta T_\mathrm{j}\approx10^{10}$ K\footnote{This temperature represents the typical temperature of hot gas making up the jet-inflated lobes if one assumes that all of the kinetic energy of a single jet kicking event, with a velocity of $v_\mathrm{j}=3\times10^4$ km s$^{-1}$ in this case, is transformed to thermal energy through shocks, as well as that none of it is transferred to the ambient medium through the shocks, and that no ambient ICM is entrained. This typical temperature, obtained through $(3/2)k_\mathrm{B}\Delta T_\mathrm{j}=(1/2)\mu m_\mathrm{p}v_\mathrm{j}^2$, is expected to be an overestimate for the aforementioned reasons, but it is a useful order-of-magnitude estimate, especially when comparing to thermal feedback.}, respectively). It is instead a factor of $10$ or so lower in temperature, which is likely on account of several processes, including the transferal of energy from the outflows to the ambient medium (through shocks or other processes), as well as adiabatic expansion and entrainment of ambient gas.

From the bottom two panels we see that both kinetic jets and thermal isotropic feedback lead to the generation of roughly spherical shock waves, which is one of the ways in which AGN feedback can heat the ambient medium (e.g.~\citealt{Li2017}, see also review by \citealt{Fabian2012}). From all four panels we see that the ICM has a generally very complex structure, with actively cooling gas draping and trailing the outflows and bubbles associated with feedback (to distances as large as 300 kpc). Particularly noticeable are filamentary structures that arise from the feedback-induced uplift of the low-entropy ICM from the core of the ICM to larger radii (see discussion in Paper I and \citealt{Husko2022_bubbles} for a detailed study of AGN feedback-induced uplift). The process of gas uplifting is one of the ways feedback is done, by reducing the central gas density and therefore delaying radiative cooling. However, the uplifted gas rises to some radius where the thermal pressure is lower, so its thermal pressure also reduces. The gas cools adiabatically and it thus may be more prone to further radiative cooling. It is thus possible that a positive AGN feedback loop exists, at least to some degree (not necessarily dominant over the negative feedback), in the systems we are simulating.

\subsection{Feedback powers}

\begin{figure*}
\includegraphics[width=0.99\textwidth, trim = 0 10 0 0]{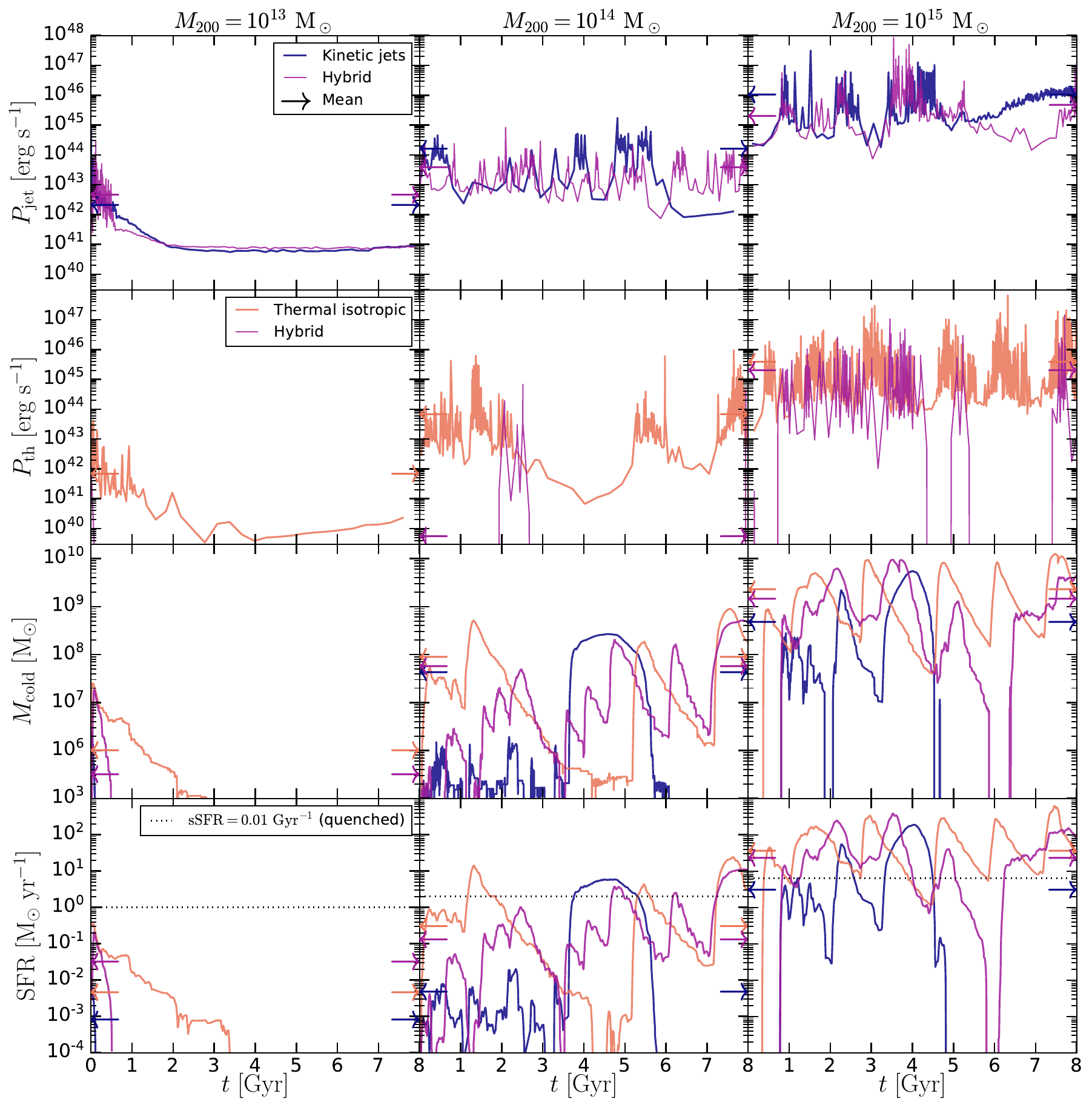}
\caption{Comparison of the cooling and feedback cycle in simulations using the BH spin evolution model for our idealized galaxy group ($M_\mathrm{200}=10^{13}$ $\mathrm{M}_\odot$), low-mass cluster ($M_\mathrm{200}=10^{14}$ $\mathrm{M}_\odot$) and high-mass cluster ($M_\mathrm{200}=10^{15}$ $\mathrm{M}_\odot$), from left to right. From top to bottom we show the kinetic jet power, the thermal heating power, the cold gas masses ($T<2\times10^4$ K) and the star formation rates. The details of these simulations are given in \S~\ref{sec:setup} and Table \ref{tab:tab0} (in terms of physical set-up and halo mass) as well as \S~\ref{sec:simulations} and Table \ref{tab:tab1} (in terms of the feedback implementation). The model uses the thin, radiatively-efficient accretion disc with thermal isotropic feedback (orange lines) and/or the thick, advection-dominated accretion disc with kinetic jet feedback (blue lines). The purple lines show cases with hybrid feedback, in which the feedback and accretion modes interchange at an Eddington ratio $\dot{m}_\mr{crit}=0.01$ (below this value kinetic jets are used, above it thermal isotropic feedback). The feedback powers are calculated using adaptive time bins such that during each bin, 10 feedback events (heating or kicking particles) occurred. The cold gas masses and star formation rates are calculated as moving averages in $5$ Myr-wide bins. The arrows indicate averages over the 8 Gyr simulation run time. }
\label{fig:fig1}
\end{figure*}%

We begin our quantitative comparison of the different simulations with BH spin evolution by considering the variation of feedback powers with time. In Paper I we showed the power to be high when the central regions of these simulated haloes are strongly cooling, i.e.~undergoing a cooling flow that leads to significant amounts of cool gas (which we define to be $T<2\times10^4$ K for the purposes of this paper) and a non-zero SFR (see \citealt{Nobels2022} for a discussion of the small delay between cooling and feedback). On the other hand, if the central regions of these haloes have been sufficiently heated by feedback, or if gas has been transported outwards through feedback-induced uplift, the feedback powers are low since the BHs are accreting directly from the hot halo, rather than the cold gas. As a result, the feedback power serves as a good tracer of the overall behaviour of the cooling and feedback cycle of these haloes.

In the top two rows of Fig.~\ref{fig:fig1} we show the feedback power as a function of time in simulations with different feedback prescriptions, for all three of our studied halo masses. The top two left-hand panels show the feedback powers for the galaxy group. In all cases there is an initial feedback episode, after which the feedback power settles down to much lower, roughly constant values for the rest of the simulations. This constant value is around 5 times lower for the thermal isotropic case (bottom panel) than the kinetic jet case. The difference can be explained by considering the feedback efficiencies in these simulations, which are set by the spins of the BHs (see \S~\ref{sec:feedback_effs} and Fig.~\ref{fig:fig3} for a detailed discussion of the evolution of the BHs in these simulations). In the jet case, the feedback efficiency is $\epsilon_\mathrm{j}\approx0.025$, whereas the radiative efficiency in the thermal case is $\epsilon_\mathrm{r}\approx0.06$. The thermal isotropic feedback power is 10 times lower than that due to a coupling efficiency factor of $\epsilon_\mathrm{f}=0.1$, so the total thermal isotropic efficiency is $\epsilon=\epsilon_\mathrm{f}\epsilon_\mathrm{r}\approx0.006$. This value is around $5$ times lower than the jet one, $\epsilon_\mathrm{j}\approx0.025$, leading to a 5 times lower feedback power at late times (given similar accretion rates, see Fig.~\ref{fig:fig3}). The thermal isotropic power is also (on average) lower at all except very early times ($t>100$ Myr). This indicates that these haloes go through a very similar thermodynamic state as the feedback is in the process of quenching them. In other words, the system is not self-regulated. Instead, any feedback mechanism is sufficient to quench the cooling flow in the centre very quickly, and any residual feedback is merely an ‘after-effect'. While this is not easily visible from the plot, the thermal power is higher than the jet power at very early times. This could be either due to the lower feedback efficiency in that case or due to thermal isotropic feedback generally being less effective at quenching cooling flows than kinetic jets even with the same efficiency (as we show in \S~\ref{sec:feedback_variations}), so a stronger initial cooling flow develops. As a consequence, there is more cold gas in the centre of the halo (visible in the third row) at these times, feeding the BH more strongly. The feedback power is also more variable in that case as compared to the jet case. This difference is a result of isotropic feedback regularly blowing away clumps of cool gas from the centre of the halo, which eventually fall back and periodically feed the BH. 

In the same two panels (top two left-hand ones) we show the feedback powers in a simulation with hybrid feedback and interchanging accretion modes. We find that there is only a small amount of thermal feedback in the very beginning in this simulation, with jets dominating at all other times (because the accretion rate in terms of the Eddington ratio is generally $\dot{m}<\dot{m}_\mr{crit}=0.01$; see Fig.~\ref{fig:fig3}). The jet power in this case is very similar to the jet-only case, although it appears to be more variable, possibly as a result of more cold gas being present (see third row).

We now turn to the more massive, galaxy cluster-size haloes ($M_{200}\geq10^{14}$ $\mr{M}_\odot$), which quickly become self-regulated. We begin with the low-mass cluster ($M_{200}=10^{14}$ $\mr{M}_\odot$), the results for which are shown in the top two middle panels of Fig.~\ref{fig:fig1}. In all cases we see multiple cycles of cooling and feedback. The peaks in the feedback powers do not occur at the same times for the different simulations, for two reasons: 1) the feedback implementation is inherently different and 2) these simulations are chaotic (see Appendix A of \citealt{Nobels2022}). The feedback powers averaged over the 8 Gyr simulation run times, shown with arrows on the plot, are $P\approx10^{44}$ $\mr{erg}$ $\mr{s}^{-1}$ in both the kinetic-only and thermal-only cases (slightly above that value for the former and slightly below it for the latter). The hybrid case again shows very little thermal feedback, except some activity at $t=2-3$ Gyr. The mean jet power in this simulation is, however, roughly a factor of two lower than in the kinetic-only one. 

Despite the overall similarity in the mean feedback powers between these three simulations, there are differences in their variability. The kinetic-only case has 3-7 distinct episodes of feedback (depending on how one counts them)\footnote{The shape of the curves in these plots is dependent on how we calculate the feedback power and on the binning. However, episodes of activity can, regardless of how the power is calculated, be gleaned as features taking the form of a clear increase from the global minimum power, peak and subsequent decrease (with possible variability in between) to the minimum power. This is what we mean when we refer to feedback episodes.} with some activity at all times except at the end of the simulation. The thermal-only one has $3$-$4$ episodes (depending on whether the first bout of activity, between $t=0$ Gyr and $t=2$ Gyr, is considered as one or two episodes) with very clear quiescent periods at $t\approx4$ Gyr and $t\approx7$ Gyr. This difference is likely a result of jet feedback being able to react more quickly to the formation of a cooling flow, possibly due to the higher feedback efficiency (see \S~\ref{sec:feedback_effs} and Fig.~\ref{fig:fig3}), which allows a cooling flow to be shut off before it becomes overly strong. In the hybrid case, jet feedback appears yet more variable. Instead of multiple coherent episodes being discernible in the variability of the jet power, we see relatively frequent variations around a jet power of $P\approx3\times10^{43}$ $\mr{erg}$ $\mr{s}^{-1}$. This difference is likely caused by the higher jet efficiency in this case, since in the jet-only mode the BH can be and does become spun down to very low BH spins (see \S~\ref{sec:spin_magn}, \S~\ref{sec:bh_props} and Fig.~\ref{fig:fig3}).

We now move to our most massive galaxy cluster, with $M_{200}=10^{15}$ $\mr{M}_\odot$, the results for which are shown in the top two right-hand panels of Fig.~\ref{fig:fig1}. Similar to the low-mass cluster, the feedback powers show multiple cycles of activity, with the thermal-only case this time showing significant variability, while the jet-only case has a few distinct episodes of activity. From the hybrid case we see that that thermal feedback is often active. While it may appear that thermal isotropic and kinetic jet feedback are often active at the same time, this is merely a consequence of the feedback modes interchanging more frequently than our sampling of the feedback powers (which are in this case plotted using adaptive bin widths containing 10 feedback events), as well as other quantities (e.g. Eddington ratios shown in Fig.~\ref{fig:fig3})

Comparing the jet powers in the jet-only and hybrid case, we find that they are overall similar (even in the positions of the peaks), but there is a difference towards the end of the simulations. The jet-only one has a jet power that increases towards $P_\mathrm{j}\approx10^{46}$ $\mr{erg}$ $\mr{s}^{-1}$ by the end -- this is a result of jet-induced spindown leading to a very low BH spin and therefore low jet efficiency, which in turn leads to a very high (unrealistically so) BH mass (see \S~\ref{sec:bh_props} and Fig.~\ref{fig:fig3}). With such a high mass, the BH is able to launch strong jets by accreting from the hot gas halo, leaving the system fully quenched (see bottom panels). From the thermal power, we see that thermal feedback is active more often in this case than in the low-mass galaxy cluster. This is a result of the massive galaxy cluster having significant amounts of gas cooling and star formation, which is connected to the accretion rate of the BH, which is also higher (similar as in Paper I; see bottom panels and Fig.~\ref{fig:fig3}).

We can also compare the mean feedback powers in all of the simulations we have discussed thus far (see arrows in Fig.~\ref{fig:fig1}). We find that the kinetic jet power is higher than the thermal power in all cases. We interpret this as a result of a larger fraction of the energy related to jet feedback reaching larger radii (regions that do not ‘need' to be heated, since they already have long cooling times) than in the thermal isotropic case, which generally has more central heating. A larger fraction of feedback energy coupling to larger radii thus leads to overall more energy needing to be injected to shut off the central cooling flows.

\subsection{Impact of feedback on galaxy growth}


We will now discuss quantities related to the BCGs and their growth in our simulations with BH spin evolution, which are shown in the bottom two panels of Fig.~\ref{fig:fig1}. These are the cold gas masses (defined as the total masses of all gas with $T<2\times 10^4$ K, at all radii) and SFRs. We consider galaxies as quenched if their specific SFR (sSFR), i.e.~the SFR divided by $M_*$, is below $0.01$ Gyr$^{-1}$ (e.g.~\citealt{Weinmann2006}). We find that our results are largely insensitive to this exact choice. We calculate both the SFR and $M_\mathrm{cold}$ as moving averages in $5$ Myr wide time bins.

We again begin with the left-hand panels, showing the results for the lowest-mass simulations ($M_{200}=10^{13}$ $\mr{M}_\odot$). In the kinetic-only case, there is barely any cold gas and star formation, and then only at the very beginning (hardly discernible in the plot). The thermal-only and hybrid cases show a similar amount of cold gas ($M_\mr{cold}\approx3\times10^7$ $\mr{M}_\odot$) and star formation ($\mr{SFR}=0.1$ $\mr{M}_\odot$ $\mr{yr}^{-1}$) at the peak, although the hybrid case more quickly reaches a state of no cold gas being present and therefore no star formation. In all three cases, the system is considered quenched at all times.

We now move to the low-mass galaxy cluster case with $M_{200}=10^{14}$ $\mr{M}_\odot$ (middle panels). The cold gas and SFR exhibit multiple episodes that generally coincide with the peaks in the feedback powers (see top rows). The cold gas mass has peak values close to $M_\mr{cold}=10^{9}$ $\mr{M}_\odot$, with the peaks being slightly lower in the cases with jet feedback (see also the mean values, indicated on the plot with arrows). The SFR peaks at $\approx10$ $\mr{M}_\odot$ $\mr{yr}^{-1}$, which is sufficient to consider the galaxies non-quenched at these rare times. The kinetic jet case exhibits very little cold gas or star formation at early times (before $t=3.5$ Gyr). This indicates that hot halo accretion is sufficient to keep the halo quenched with this feedback mode for quite a long time. By $t=3.5$ Gyr, a strong cooling flow develops, and it lasts $\approx2$ Gyr. During this time, the BH experiences a significant amount of growth. Since it was spun down to a very low value of the BH spin even earlier (see \S~\ref{sec:bh_props} and Fig.~\ref{fig:fig3}), it means that the BH cannot quickly react to the development of a cooling flow. As a result, a strong cooling flow develops, to the degree that it results in feedback strong enough to heat the ICM at large radii, thus preventing any cooling flows from occurring in at least the next $2.5$ Gyr (until the end of the simulation). The thermal isotropic case has the largest amounts of cold gas and star formation, and its first cooling flow develops in the very beginning of the simulation (whereas jet feedback, in both the jet-only and hybrid simulations, is able to delay the initial cooling flow). The hybrid case has a moderate amount of cool gas and star formation. The shape of each peak is similar to the thermal-only case. Whereas the jet-only case has sharp declines in the cold gas mass and SFR after every peak, these two cases have gradual declines that can last up to $2$ Gyr. We interpret this as possibly being due to thermal feedback blowing away clumps of cold gas, which thus take a longer time to be consumed through SF, and in the meantime they are not feeding the BH and producing feedback.

Finally, we discuss the massive galaxy cluster case ($M_{200}=10^{15}$ $\mr{M}_\odot$). The cold gas mass reaches peaks of up to $10^{10}$ $\mr{M}_\odot$ in all three cases, with the SFR reaching several hundred $\mr{M}_\odot$ $\mr{yr}^{-1}$. The hybrid case has only a mildly lower mean cold gas mass and SFR than the thermal-only case, since the operating feedback mode is quite often thermal (Fig.~\ref{fig:fig1}). The jet-only simulation has an appreciably lower cold gas masses and SFR, and is fully quenched at around $t=4.5$ Gyr.

In Appendix \ref{sec:app2} we discuss whether BH growth and feedback interferes with star formation directly or indirectly. We probe this by considering the ratio $M_\mr{i}/M_\mr{*,formed}$ as a function of time, where $M_\mr{i}$ is the total mass accreted, launched into the jets or heated by the BH, and $M_\mr{*,formed}$ is the total mass of all stars formed. We find that this ratio is often comparable to or larger than unity, suggesting that BH growth and feedback do indeed directly interfere with star formation in our simulations, by depriving it of its fuel (cold gas) through direct processes (algorithmically choosing it to be heated or kicked), rather than, for example, through entrainment. 

The implications of this finding for realistic, cosmological simulations may not be problematic for BH accretion, as long as we assume that BH growth is not excessive in these simulations. However, the high mass flux of particles associated with feedback may be more problematic, especially since these fluxes are also typically higher than those associated with BH accretion. The rate at which the BH is heating or kicking gas particles depends not only on the feedback powers, but also on the heating temperature $\Delta T$ and jet velocity $v_\mr{j}$. Both of these parameters are at least partially numerical in nature. Decreasing their values (at a fixed feedback power) increases the mass flux of particles being heated/kicked. If too low values are chosen, the mass flux of particles associated with feedback may be close in magnitude to the SFR, which we sometimes find to be the case in our simulations. One would ideally want to avoid this situation, and ensure that the mass flux of particles being heated/kicked is always much smaller than the rate at which the gas is being converted into stars. In practice, this limit may be hard to avoid, at least at low resolutions, since decreasing the mass flux of the particles being heated/kicked also decreases how well sampled the feedback is, which then means that feedback is resolved more poorly. We do not propose a particular solution here, but merely point out that the mass flux in question is probably quite large (close to the SFR) in most implementations of AGN feedback in cosmological simulations.

\subsection{Evolution of black hole properties}
\label{sec:bh_props}

\begin{figure*}
\includegraphics[width=0.95\textwidth, trim = 0 20 0 0]{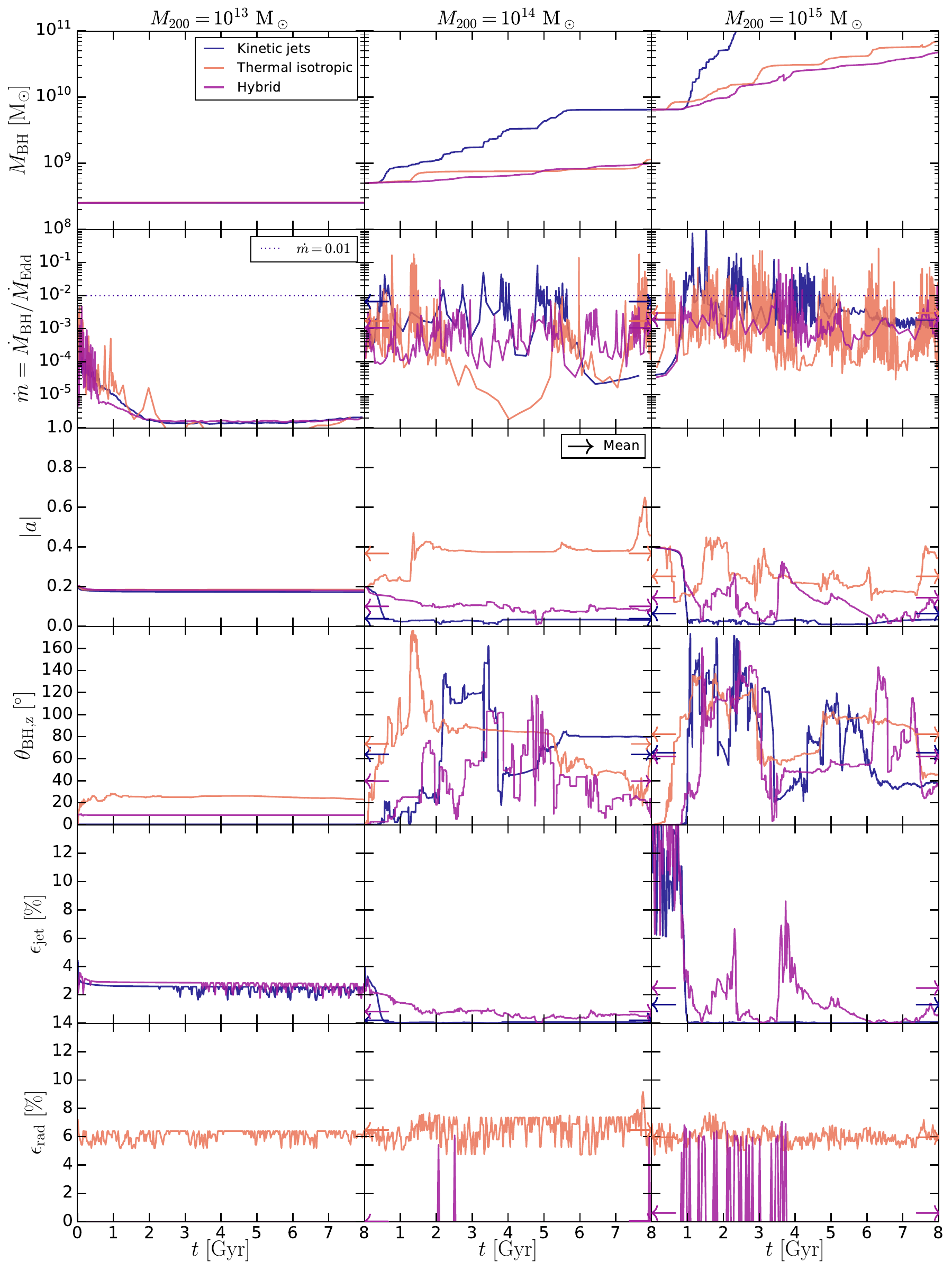}
\caption{The properties of the BHs and their evolution in simulations using the BH spin evolution model. From top to bottom, we show the BH masses, the Eddington fractions, the BH spin magnitudes, the angles between the BH spin vectors and the $z-$axis, and the jet and radiative efficiencies. All quantities except the feedback efficiencies are sampled every 5 Myr. The feedback efficiencies are sampled every 1 Myr and then a moving average in a 5 Myr-wide bin is calculated using only times when the BH is in the appropriate accretion state (the thick disc for the jet efficiency and the thin disc for the radiative efficiency), and they are weighted by the accretion rate. The arrows indicate averages over the 8 Gyr simulation run time. Everything else is the same as in Fig.~\ref{fig:fig1}.}
\label{fig:fig3}
\end{figure*}%

In Fig.~\ref{fig:fig3} we show the evolution of various BH-related properties, including from top to bottom the BH mass (both the subgrid and the dynamical mass, which are the same in this case), Eddington-normalised mass accretion rate, the BH spin magnitude, the angle between the BH spin vector and the z-axis, and finally the jet and radiative efficiencies. We discuss each of these quantities in turn.

The BH mass (first row of Fig.~\ref{fig:fig3}) remains unchanged in the galaxy group case, whereas in the galaxy cluster cases, there is always some appreciable growth. The low-mass cluster exhibits BH growth by more than a factor of ten in the kinetic-only case, partly as a result of low efficiencies (due to jet-induced spindown). The thermal-only and hybrid cases show much less growth -- about a factor of two for both cases. The results are similar for the high-mass galaxy cluster, but these simulations show even more growth. The kinetic-only one shows BH growth beyond $M_\mr{BH}=10^{11}$ $\mr{M}_\odot$ by $t=2.5$ Gyr (the final BH mass reached in this simulation, which we do not show here, is $\approx5\times10^{11}$ $\mathrm{M}_\odot$, which highlights the unrealistic nature of using the jet mode in isolation, at least with the strong jet spindown rates we have assumed). The other two cases both show growth by a factor of $5$-$10$, with the hybrid one, interestingly, showing the least amount of growth (which may be interpreted as the hybrid feedback being most effective at self-regulation of BH growth).

The Eddington-normalized accretion rates (second row of Fig.~\ref{fig:fig3}) peak near $\dot{m}_\mr{crit}=0.01$ for all three galaxy group simulations. They also all reach $\dot{m}=10^{-5}$ by the end, although the thermal-only case takes the longest time to reach that value. The two higher-mass cases show much more variability in the accretion rate, with it often peaking above $\dot{m}_\mr{crit}$ (which is why the hybrid cases have the feedback modes often interchanging). Interestingly, the hybrid simulations do not feature high values of the accretion rate (e.g.~$\dot{m}=0.1$ or approaching $\dot{m}=1$) as often as the two other simulations, which again may be related to more effective self-regulation.

From the evolution of the Eddington ratios, it is clear that BHs sometimes have an accretion rate high enough for the accretion mode to correspond to the thin disc, instead of the thick disc, and feedback to be thermal isotropic ($\dot{m}>\dot{m}_\mathrm{crit}=0.01$), at least in the galaxy cluster cases ($M_{200}\geq10^{14}$ $\mathrm{M}_\odot$). However, it is not clear from these plots how much growth actually occurs in which accretion regime. While the Eddington ratio appears to be $\dot{m}<0.01$ most of the time for all 9 simulations shown in Fig.~\ref{fig:fig3}, it is possible for most of the growth to occur at $\dot{m}>0.01$ due to the accretion rates being higher. 

In Appendix \ref{sec:app2p1} we discuss the cumulative mass fractions of growth that occur at low versus high Eddington ratios. We find that neither regime is negligible in terms of growth. Perhaps surprisingly, we find that most growth occurs when $\dot{m}>0.01$ in the galaxy cluster cases, despite the accretion rate satisfying $\dot{m}<0.01$ most of the time. This implies that radiatively-efficient accretion and its associated ‘quasar feedback' should not be ignored for galaxy clusters (at least not CC ones), despite its rarity. This finding is likely a consequence of cooling flows becoming progressively stronger for more massive clusters. The picture of ‘maintenance-mode' feedback (by relativistic jets) that keeps BCGs quenched is thus probably an oversimplification for relatively CC clusters, such as the ones we are simulating here. This is in agreement with the analysis of a wide range of observations done by \cite{McDonald2018}, who found that the systems with the largest cooling flows (and star formation rates) tend to have the highest star formation efficiencies, which could be explained by the central BH more often being in the radiative versus mechanical feedback mode (the former of which is less efficient as a feedback mechanism, as we have found already in this section, and which we also confirm more robustly in \S~\ref{sec:res_fixed}).

The BH spin magnitude (third row of Fig.~\ref{fig:fig3}) exhibits very little evolution in the galaxy group case (except a small amount of spindown at the very beginning), which is a direct result of little-to-no BH mass growth. In the other two cases there is significant BH spin evolution. The low-mass cluster shows spindown in the kinetic-only simulation (down to values $\vert a\vert<0.05$, as a result of using a GRMHD spindown formula, see \S~\ref{sec:spin_magn}), as well as in the hybrid one, where larger values of the BH spin are reached (although still very low ones, $\vert a\vert \approx0.05-0.1$). The thermal-only case instead shows occasional spin-up to values around $\vert a\vert=0.4$. In the massive galaxy cluster case, all three simulations have a median BH spin that is below the initial value ($a_0=0.4$). The kinetic-only one behaves similar to the low-mass cluster case, although the mean BH spin is even lower. The hybrid one has the BH spin varying between $a=0$ and $a=0.3$ -- higher values are achieved than in the low-mass cluster case due to more spinup, as the BH spends more time in the thin disc regime due to high accretion rates. The thermal-only case, on the other hand, shows a lower mean BH spin than in the low-mass cluster simulation. This could be due to the cold gas being more chaotic in terms of its angular momentum (or due to the high-mass simulation having poorer resolution), which would lead to more frequent retrograde accretion of the BH, and therefore more frequent spindown.

The angle between the BH spin vector and the $z$-axis (fourth row of Fig.~\ref{fig:fig3}) contains information on how much redirection the BH spin vector has experienced. In the galaxy group case, there is some redirection that occurs in the very beginning of both the thermal-only and hybrid simulations. In the two higher-mass cases, there is much more redirection -- these plots highlight the chaotic nature of the angular momentum of the accreting gas in these simulations. Our results are in qualitative agreement with the ‘chaotic cold accretion' (CCA) scenario presented by \cite{Gaspari2013}. This is despite the fact that we use Bondi accretion, which is often portrayed as being mutually exclusive with CCA. We argue instead that a version of CCA naturally emerges if cold gas is included in the Bondi formula (instead of restricting it to hot, X-ray emitting gas). This mixed approach can reproduce the main features of CCA, including the chaotic nature of the cold gas that is accreting onto the BH and the boosting of the BH accretion rate (relative to the Bondi rate inferred from hot gas). 

In the final two rows of Fig.~\ref{fig:fig3} we show the feedback efficiencies in these simulations. These are calculated as moving averages over $5$ Myr wide bins, but we only include times when the BH is in the appropriate accretion state (the thick disc for the jet efficiency and the thin disc for the radiative efficiency), and they are also weighted by the accretion rate at every time-step. In the galaxy group case, the efficiencies show some variability--this can occur despite the magnitude of the BH spin not evolving because the feedback efficiencies also depend on the sign of the BH spin, which itself depends on the angular momentum direction of the gas in the BH smoothing kernel. The jet efficiency quickly drops to per cent-level values for the kinetic-only case in both galaxy cluster simulations. In the hybrid cases, the jet efficiencies depend highly on the current BH spin; in the low-mass case it is below $2$ per cent, while in the high-mass case it sometimes grows to several per cent. The radiative efficiency in the thermal-only simulations is in all cases between $4$ and $8$ per cent. This lack of strong variability is a result of the radiative efficiency being weakly dependent on BH spin, except at $a>0.9$. 


\subsection{Entropy profiles}
\label{sec:entropy_profiles_spin}

\begin{figure*}
\includegraphics[width=1.01\textwidth, trim = 0 15 0 0]{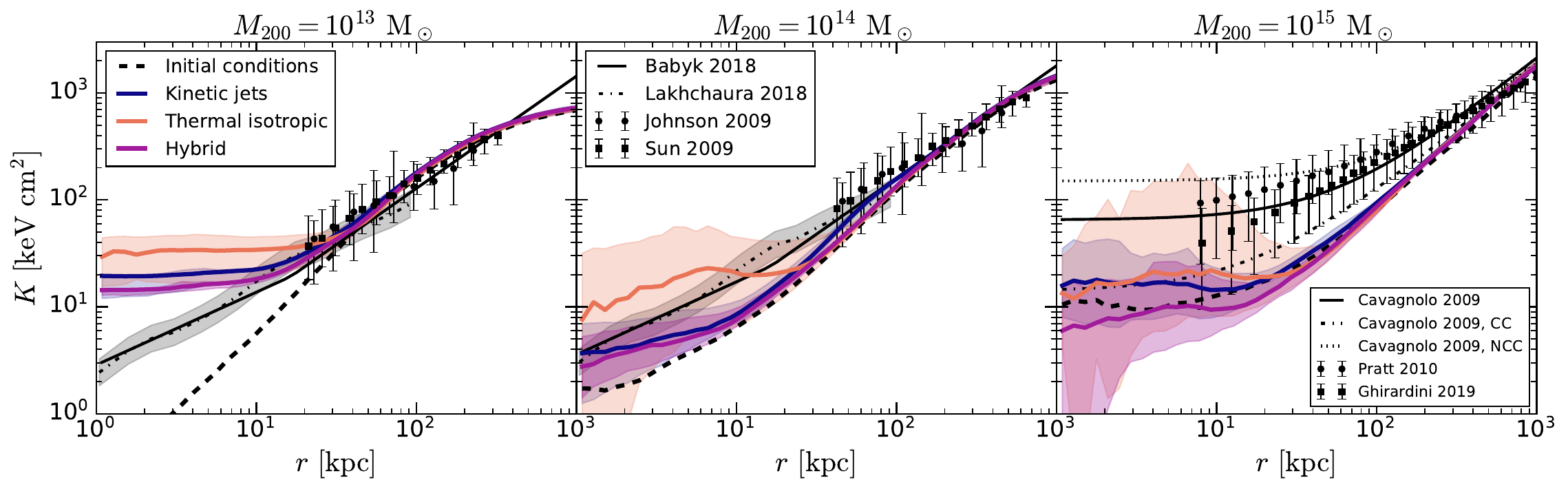}
\caption{The radial gas entropy profiles (volume-weighted) of the ICM using the BH spin evolution model, from the same set of simulations as in Figs. \ref{fig:fig1} and \ref{fig:fig3}. The solid lines show medians calculated using 160 snapshots between $t=0$ and $t=8$ Gyr, while the shadings indicate the $16^\mathrm{th}-84^\mathrm{th}$ percentile ranges. The profiles in the initial conditions are shown with the dashed lines. The observational data sets are described in \S~\ref{sec:obs}. The profiles from \protect\cite{Johnson2009} and \protect\cite{Sun2009} are converted from their dimensionless form to physical form for the two left-hand panels separately, which is why these data differ in the two panels. The sample from \protect\cite{Cavagnolo2009} is split onto cool-core (CC) and non-cool-core (NCC) clusters.}
\label{fig:fig4}
\end{figure*}%

We now turn to the entropy profiles in these simulations, which are shown in Fig.~\ref{fig:fig4}. These profiles are compared to the observed ones, which are described in \S~\ref{sec:obs}. We do not expect the simulated profiles to agree perfectly with the observed ones, for several reasons. Firstly, our simulations represent only single realizations in terms of how CC they are, i.e.~we could vary the initial central ICM temperature parameter $T_0$ to obtain different profiles. We chose low values (relatively CC systems) for reasons laid out in \S~\ref{sec:setup}. Secondly, real clusters undergo merging activity, which is not included in our idealized simulations. Thirdly, observed profiles are deprojected given some assumed model. Fourthly, the profiles are not volume-weighted, but X-ray emission-weighted. Finally, for the galaxy group and low-mass cluster simulations, the observed profiles we are comparing these simulations with span the mass range $M_{200}=10^{13}$-$10^{14}$ $\mr{M}_\odot$, i.e.~they are centred on a median halo mass of $M_{200}\approx10^{13.5}$ $\mr{M}_\odot$. Additional shortcomings of these observations (in the context of applying them here for purposes of comparison) are discussed in \S~\ref{sec:obs}. For these reasons, care should be taken when comparing the galaxy group ($M_{200}\approx10^{13}$ $\mr{M}_\odot$) and low-mass cluster ($M_{200}\approx10^{14}$ $\mr{M}_\odot$) simulations with these observations (the observed profiles shown in the left-hand and middle panels of Fig.~\ref{fig:fig4} are the same). Furthermore, this sample of observed galaxies does not only include centrals, but also satellites. Given these considerations, we use the observed profiles as a baseline to compare the shapes of the profiles (and their differences between models), rather than seeking full agreement.

Before discussing the cases individually (and comparing with the observed profiles), we first comment on some common features seen in all three cases. From Fig.~\ref{fig:fig4} we see that the hybrid simulations have the lowest central entropy, even lower than the kinetic-only simulations. This is potentially caused by a combination of two effects whose impacts on the central entropy are opposite. Firstly, jets are able to heat the halo at larger distances than thermal isotropic feedback, and they deposit less of their energy in the very central regions (see also Fig.~\ref{fig:fig6}). This means that the cores in simulations with jets should be cooler. Secondly, the haloes are more effectively quenched by jets than by thermal isotropic feedback, which means that the central ICM undergoes strong cooling flows less frequently, or they are weaker and/or shorter-lived. This in turn means that the central entropy should, on average, be higher if jets are used. When comparing our kinetic-only and thermal-only simulations, it appears that the first of these two effects dominates, at least for the two lower-mass cases. Allowing the two feedback modes to interchange, as in the hybrid simulations, leads to the lowest central entropies for all three halo masses since these simulations exhibit both strong cooling flows and less central heating. Another common feature between all three feedback cases is the difference in scatter. The thermal-only simulations show the largest scatter because they have both central ICM heating and strong cooling flows, whereas the opposite is true for the kinetic-only simulations. The hybrid ones have an intermediate amount of scatter.

For the galaxy group case, shown in the left-hand panel of Fig.~\ref{fig:fig4}, all three simulations (with differing feedback implementations) fail to reproduce the shape (slope) of the observed entropy profiles within $10$ kpc, but agree with them at larger distances (by construction). Within $10$ kpc, the observed profiles behave as $K\propto r^{2/3}$, whereas our simulated entropy profiles all have cores. This is unlikely to be affected strongly by the $T_0$ parameter (\citealt{Nobels2022}, Paper I). Instead we interpret this disagreement as showing that lower jet velocities may need to be used (the velocity used for these simulations was $v_\mr{j}=5\times10^3$ km $\mr{s}^{-1}$). As we show in Fig.~\ref{fig:fig9}, lower velocities lead to lower central entropies and more sloped profiles. Alternatively, as already explained, it is possible (if not likely) that the observations used for comparison here are biased towards brighter groups that therefore have lower central entropies. If the satellites were removed from the observational samples, it is likely that the disagreement would be worse, since satellites are less likely to be undergoing cooling flows, due to stripping of their CGM.


For the low-mass cluster simulations (middle panel of Fig.~\ref{fig:fig4}), we again find agreement with the observed profiles at large distances (in this case $r>50$ kpc). In the central regions, the thermal-only case appears to have the correct slope at small distances ($r<10$ kpc), but it has a flat section extending from $r=10$ kpc to $r=30$ kpc -- a feature not visible in the observed entropy profiles. Our kinetic-only and hybrid simulations show similar slopes as the observed profiles, with perhaps a slightly too shallow slope in the very centre. This could be mitigated by a different choice of $T_0$ or a slightly lower jet velocity.

Finally, we move to the high-mass galaxy cluster case, shown in the right-hand panel of Fig.~\ref{fig:fig4}. In the inner regions, all of our entropy profiles are lower than those in observations of \cite{Pratt2010}, \cite{Cavagnolo2009} and \cite{Ghirardini2019} (although this could have been prevented by choosing a higher $T_0$, but we instead chose a highly CC setup to maximize differences between the AGN feedback implementations). They also show a central entropy core, but signs of such a core appear to be present in observations as well. All three of the simulations are consistent with being CC at most times (in agreement with the CC sample of \citealt{Cavagnolo2009}, as well as with the lower end of the scatter from \citealt{Pratt2010} and \citealt{Ghirardini2019}). Out of the three simulations, only the thermal feedback case sometimes has a central entropy approaching the median entropy of the NCC sample from \cite{Cavagnolo2009}. However, NCC clusters may also be explained as a result of mergers (e.g.~\citealt{Poole2008}, \citealt{Hudson2010}).

We note that changing the implementation of AGN feedback is not the only way of affecting simulated entropy profiles (e.g.~\citealt{Altamura2023}). The details of the hydrodynamics scheme appear to be at least as important (e.g.~\citealt{Borrow2022}, \citealt{Altamura2023}). In particular, turning off artificial conduction in the SPH solver appears to lead to significantly more sloped entropy profiles.

Entropy profiles are often compared in a rescaled form, such that instead of plotting $K$ versus $r$, one plots $K/K_{500}$ versus $r/r_{500}$, where $K_{500}$ is a typical entropy that depends only on the halo mass. We discuss such profiles in Appendix \ref{sec:app2p2}. We find that they are fairly similar between the different simulated haloes, and all of them lie below the median observed entropies, likely because we simulate relatively CC systems.

\section{Results II: Simplified feedback}
\label{sec:res_fixed}

In the previous section we presented results of our model with BH spin evolution, for both thermal isotropic and kinetic jet feedback. Here we will simplify our implementation by instead fixing the efficiencies, as well as the direction for the jets (along the $z-$axis). In Appendix \ref{sec:app1} we show that the latter is justified as long as jet redirection occurs less frequently than $\approx1000$ Myr, and if jets precess with small opening angles ($\leq15\degree$) and long time-scales ($\Delta t\geq20$ Myr). The typical redirection time-scale, if redirection is allowed, is indeed typically longer than this (see the hybrid feedback cases in Fig.~\ref{fig:fig3}), since it occurs only a few times during an $8$ Gyr long simulation (if we define ‘redirection' as a change in direction that is larger than a few dozen degrees).

These simplifications are motivated by a desire to isolate the effects of varying efficiencies, as well as to make the simulations with different feedback implementations as similar as possible. To this end we fix the efficiencies to a value $\epsilon=0.01$ for both the thermal isotropic and kinetic jet feedback. We do not test hybrid cases in these simplified simulations, and instead assume the feedback to be either thermal isotropic or kinetic jets regardless of accretion rate. We test the case where the feedback energies per heating/kicking event are the same for thermal isotropic and kinetic jet feedback, but this is not our fiducial choice. We instead typically use much higher jet velocities, since they are required in order to lead to inflation of lobes that turn into bubbles and create cavities in X-ray emitting gas, as seen in observations. We present results for the low- and high-mass galaxy clusters here ($M_{200}=10^{14}$ $\mr{M}_\odot$ and $M_{200}=10^{15}$ $\mr{M}_\odot$, respectively). We then vary the feedback efficiency, heating/kicking energy and energy type (thermal versus kinetic, as well as mixed) for both isotropic and jet feedback. These variations are intended to show the effects of choosing a particular implementation of feedback. For simplicity we vary these only for the low-mass galaxy cluster.

\subsection{General results}

\begin{figure*}
\includegraphics[width=1.01\textwidth, trim = 0 10 0 0]{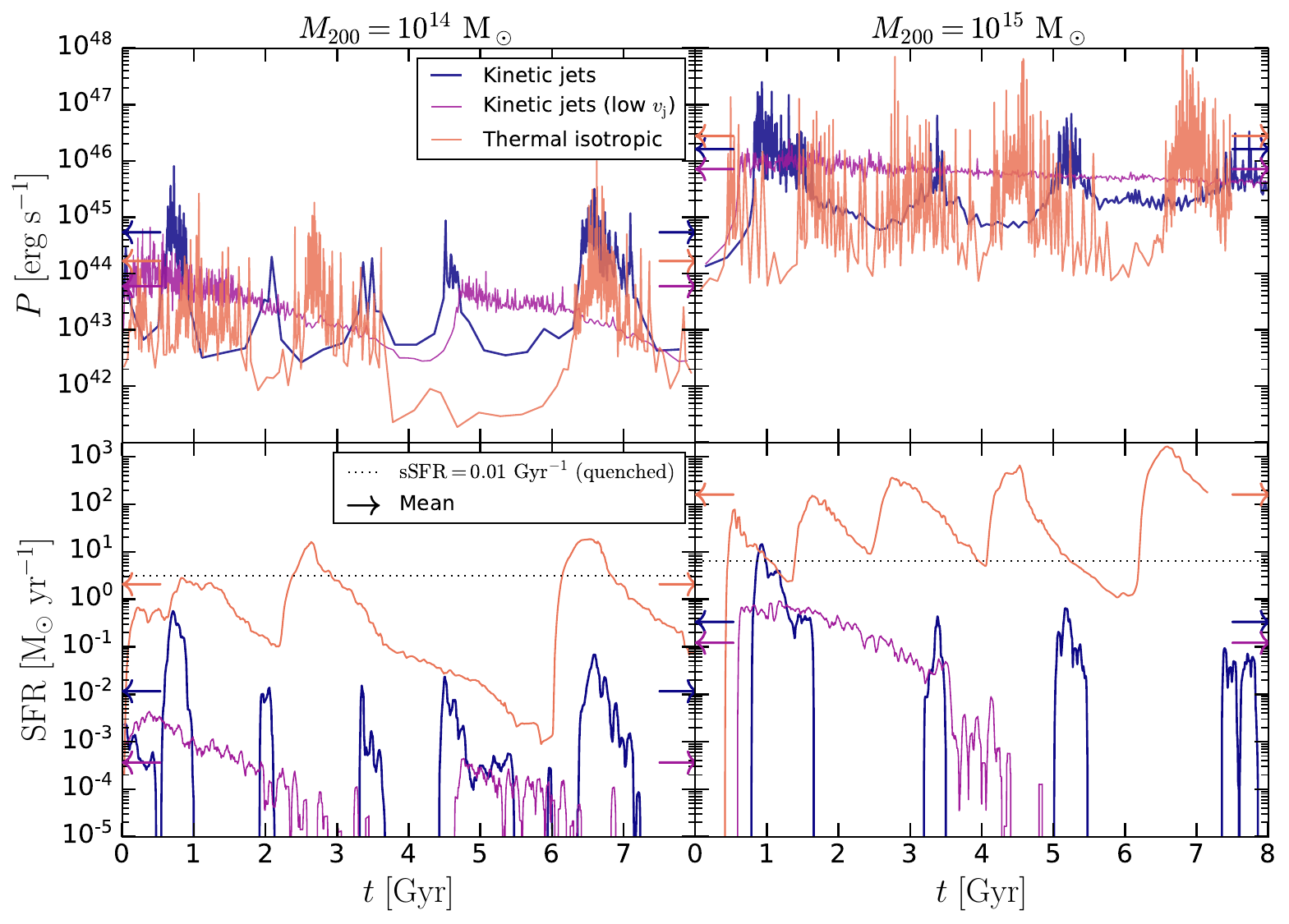}
\caption{The feedback cycle for our simulations with fixed feedback efficiencies (as well as fixed jet directions in the jet feedback case, where they are launched in the $z-$direction). The left-hand panels show results for the low-mass galaxy cluster case ($M_\mathrm{200}=10^{14}$ $\mathrm{M}_\odot$), while the right-hand ones show the results for the high-mass galaxy cluster ($M_\mathrm{200}=10^{15}$ $\mathrm{M}_\odot$). In the top panels we show the feedback powers, while the bottom panels show the SFRs. For the jet case, we perform fiducial simulations that have velocities high enough to lead to the inflation of hot lobes of gas (blue lines), as well as lower-velocity ones (purple lines) that instead have an equal energy per kicking event as the thermal isotropic simulations (orange lines). The feedback powers are calculated using adaptive time bins such that during each bin, 10 feedback events (heating or kicking particles) occurred, while the star formation rates are calculated as moving averages using $5$ Myr-wide bins. The arrows indicate averages over the 8 Gyr simulation run time. Further details of the simulations are given in \S~\ref{sec:simulations} and Table \ref{tab:tab2}.}
\label{fig:fig5}
\end{figure*}%

In Fig.~\ref{fig:fig5} we show the feedback powers and SFRs in the galaxy cluster simulations with the simplified feedback prescriptions. We find that the thermal isotropic simulations are quite similar to those presented in \S~\ref{sec:res_spin}, i.e.~with BH spin evolution. This is likely due to the radiative efficiencies being near-constant in the case with BH spin evolution (Fig.~\ref{fig:fig3}). The kinetic jet simulations (with fiducial jet velocities, $1.5\times10^4$ km $\mr{s}^{-1}$ and $3\times10^4$ km $\mr{s}^{-1}$ for the two halo masses) are somewhat different from the BH spin evolution case. This is largely due to the jet efficiency not dropping below $1$ per cent (unlike in the BH spin evolution case; Fig.~\ref{fig:fig3}), which means that very strong cooling and BH growth are prevented. As in Paper I, we find that fixing the jet direction to be along the $z-$axis does not prevent efficient feedback. 

Comparing the thermal isotropic and kinetic jet simulations, we find that the former reach lower minima of the feedback power, despite the fact that the same constant efficiency is used. This means that the BH reaches lower accretion rates in the thermal isotropic case (same as in the cases with BH spin evolution, see Fig.~\ref{fig:fig3}). This is caused by the thermal feedback simulations often featuring a significant presence of hot gas near the BH (originating from the feedback itself), which reduces its accretion rate. We find that using a constant efficiency leads to periodicity between cooling flow episodes, which seems more pronounced in the high-velocity jet cases. In these cases, we see periods of $\approx1.5$ Gyr in the low-mass cluster and $\approx2$ Gyr in the high-mass cluster. This periodicity occurs because AGN feedback effectively heats all gas out to a radius at which the ratio of the cooling time, $t_\mathrm{cool}$, and the dynamical time, $t_\mathrm{dyn}$, is $t_\mathrm{cool}/t_\mathrm{dyn}\approx10$. The period between cooling flows is then roughly equal to the cooling time at that radius. These findings are illustrated in Appendix \ref{sec:app3} and are supported by previous works (e.g.~\citealt{Nobels2022}, see also discussion therein).

In Fig.~\ref{fig:fig5} we also show the results of using low jet velocities ($6.5\times10^3$ km $\mr{s}^{-1}$ and $1.15\times10^4$ km $\mr{s}^{-1}$ for the low-mass and high-mass cluster, respectively), which are supersonic by only a factor of a few relative to the ICM. These velocities are chosen such that the energy per kicking event is equal to the heating energies used in the corresponding thermal isotropic simulations ($\Delta T=10^{9}$ K and $\Delta T=10^{9.5}$ K for the low-mass and high-mass cluster, respectively). We find that such low velocities lead to the period between cooling flow episodes increasing (to the point that the high-mass cluster shows no periodicity in this case, at least within 8 Gyr), and the SFR reaching smaller peaks, as well as being lower on average.

\begin{figure*}
\includegraphics[width=1.01\textwidth, trim = 0 10 0 0]{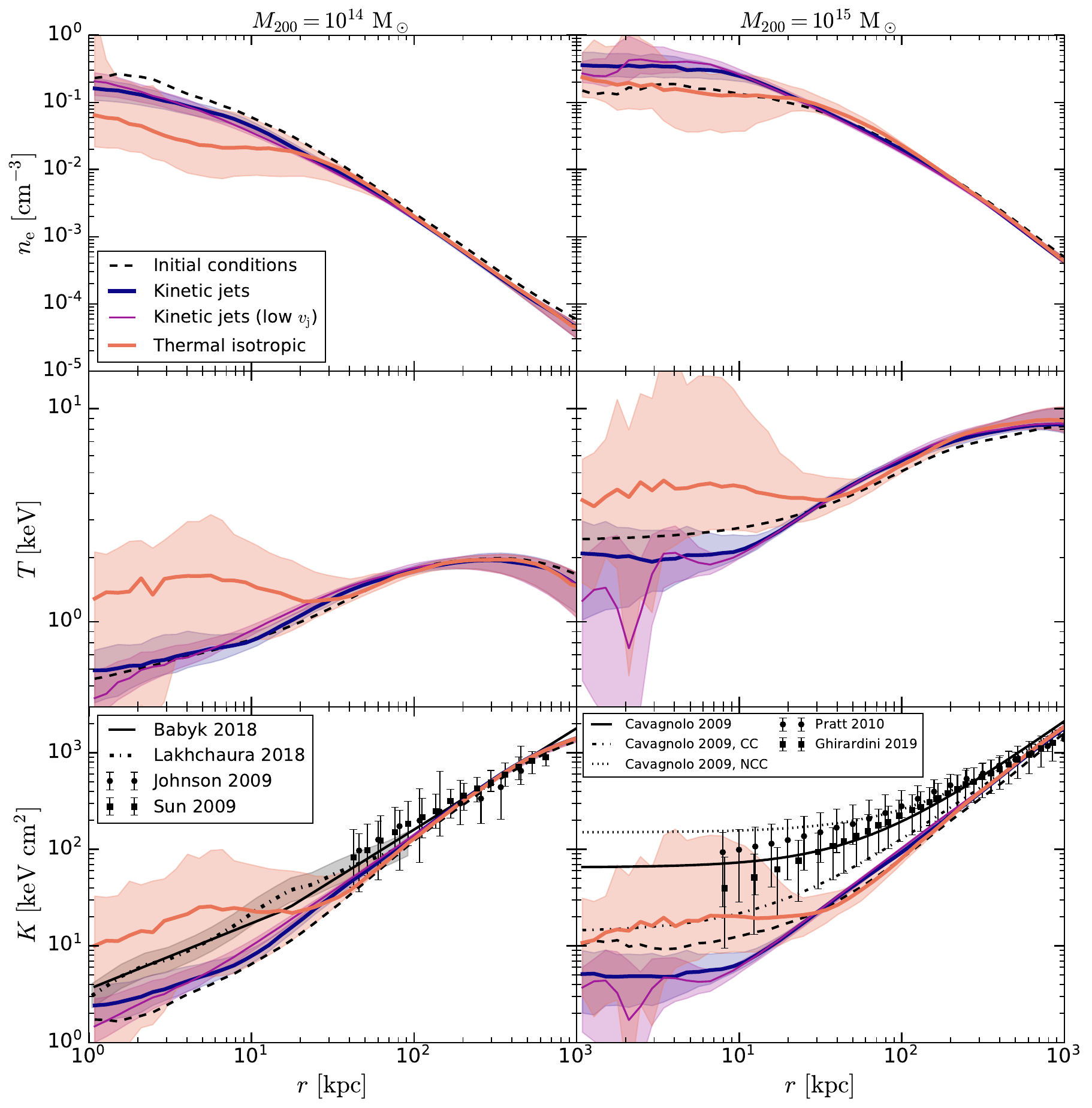}
\caption{Radial gas profiles (volume-weighted) of the ICM for the same set of simulations as shown in Fig.~\ref{fig:fig5}. From top to bottom we show the electron number density, the temperature and the entropy. The solid lines are medians calculated using 160 snapshots between $t=0$ and $t=8$ Gyr, while the shadings indicate the $16^\mathrm{th}-84^\mathrm{th}$ percentile ranges. The profiles in the initial conditions are shown with dashed lines. The observational data sets are described in \S~\ref{sec:obs}. The sample from \protect\cite{Cavagnolo2009} is split onto cool-core (CC) and non-cool-core (NCC) clusters.}
\label{fig:fig6}
\end{figure*}%

In Fig.~\ref{fig:fig6} we show radial profiles of the ICM density, temperature and entropy for these same simulations. From the top panels we see that using jets leads to higher central densities (within $20-40$ kpc), by a factor of a few. There is only a small difference between the fiducial and low-velocity jet cases. In the middle panels we compare the temperatures. The jet cases have lower central temperatures (within the same radii as for the densities) than the thermal isotropic ones, by up to a factor of two. At intermediate radii (up to $r=100$ kpc), the jet cases have higher temperatures, indicating that more of the feedback energy couples to larger radii in the jet cases. In the bottom panels we compare the entropy profiles. Due to a combination of higher central densities and lower central temperatures, the central entropies in the jet cases are lower by a factor of $\approx5$ and $\approx2$ for the two halo masses, respectively. In the low-mass case, the low-velocity simulation appears to have the same slope as the observed profiles, which are also shown in the figure. This potentially indicates that lower velocities should be used (rather than highly supersonic ones with Mach numbers $\geq10$, which we find to be required for the inflation of hot lobes and for X-ray cavities to be present), at least in these lower-mass systems. For the high-mass case, we again find that using thermal isotropic or kinetic jet feedback leads to similarly-shaped profiles as the observed ones (the same conclusion as found from Fig.~\ref{fig:fig4}, showing the BH spin evolution case). The entropies are visibly higher for jet feedback at $r=30-100$ kpc and $r=40-300$ kpc in the two mass cases, respectively. This supports the interpretation that kinetic jets are able to heat at larger radii than thermal isotropic feedback. In all of the presented profiles we see less scatter in the kinetic jet cases than with thermal isotropic feedback -- this is a result of fewer or weaker cooling flows, and less violent central heating. The thermal isotropic form of feedback leads to very similar results in terms of the entropy profiles as the cases with BH spin evolution (Fig.~\ref{fig:fig4}). This is likely due to very similar feedback efficiencies (Fig.~\ref{fig:fig3}), although as we show in the next section, the entropy profiles are also largely insensitive to a much larger variation of the efficiency.

\subsection{Varying the implementation of feedback}
\label{sec:feedback_variations}

\subsubsection{Visual differences}

\begin{figure*}
\includegraphics[width=1\textwidth, trim = 0 10 0 0]{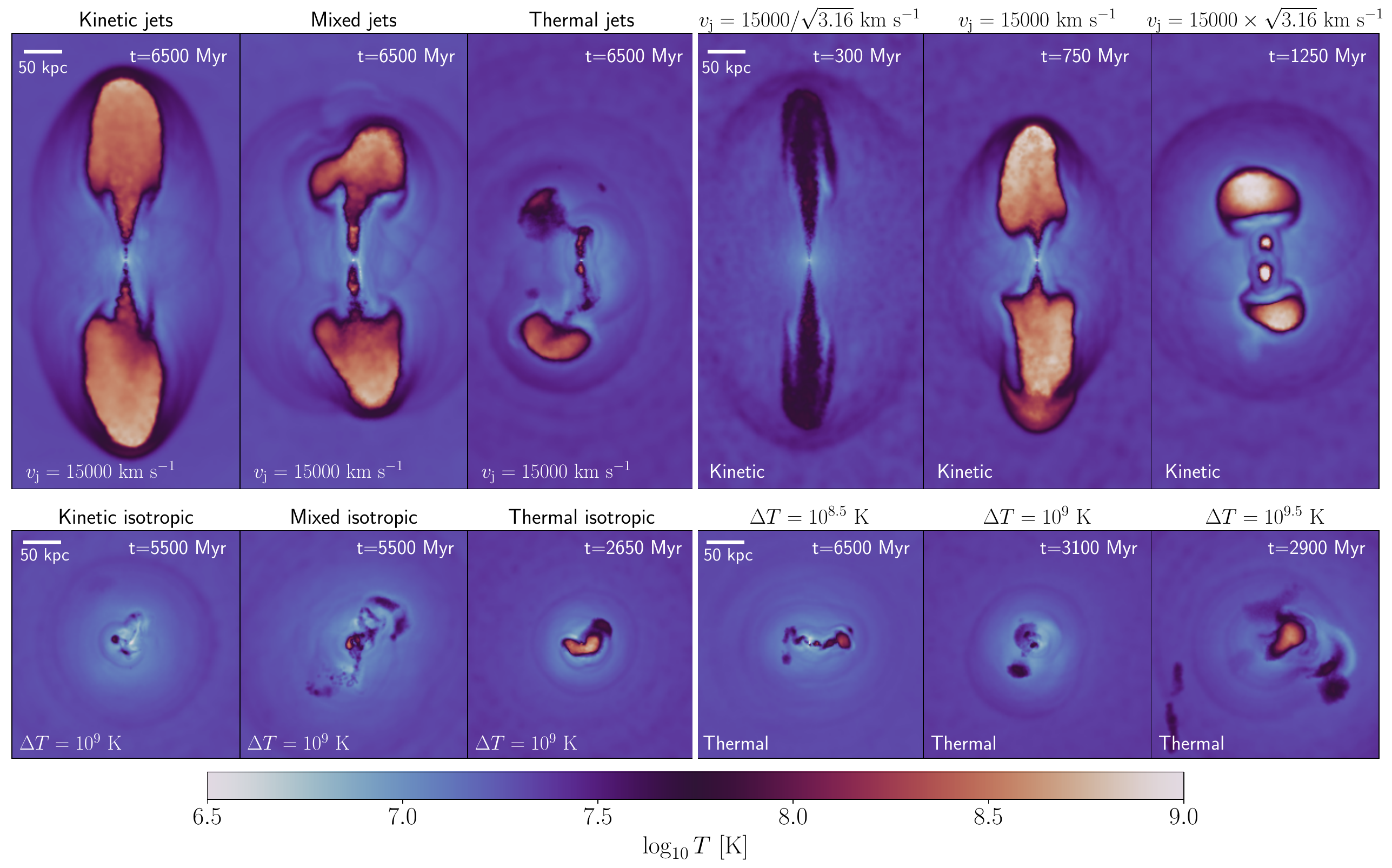}
\caption{A mosaic of different jet (top row) and isotropic (bottom row) feedback simulations with varying feedback parameters, for the low-mass galaxy cluster ($M_\mathrm{200}=10^{14}$ $\mathrm{M}_\odot$). The colours indicate the gas temperature, as shown by the colour bar, and we include all gas in a $50$ kpc-deep slice. The panels all show the same spatial scales. The simulation times shown here are generally different since the timing of the feedback activity is highly chaotic. For both the jets and isotropic feedback we vary: 1) the fraction of energy injected as thermal as opposed to kinetic (values of $0$, $0.5$ and $1$), shown on the left-hand side, and 2) the energy increment received by the particles (by factors of $\sqrt{10}\approx3.16$), shown on the right-hand side. The variations of the jet energy type use a jet velocity of $1.5\times10^4$ km s$^{-1}$ (or its corresponding heating temperature if feedback is mixed or purely thermal), while the corresponding isotropic variations use a corresponding temperature increase of $\Delta T=10^9$ K. The variations of the energetics are done for kinetic feedback in the jet case and thermal feedback in the isotropic case. 
}
\label{fig:fig7}
\end{figure*}%

We now turn to variations on the cases presented above. We vary the efficiencies of both types of feedback (isotropic and jet),  energy per each feedback event and the type of energy used for feedback -- thermal\footnote{In the thermal jet variant, particles are preferentially heated along a particular direction (the $z-$axis in this case). No kinetic energy is imparted to the gas, but it can still form outflows in the form of jets.}, mixed or kinetic. We performed all of these for the low-mass galaxy cluster ($M_{200}=10^{14}$ $\mr{M}_\odot$).

In Fig.~\ref{fig:fig7} we show visualizations of some of these simulations. In particular, we show jets with different energy types and velocities (top row, left- and right-hand sides, respectively), and the same for isotropic feedback (bottom row, with the latter variation corresponding to the heating temperature). These are shown on the same spatial scales for purposes of comparison, but we find that isotropic feedback is generally more confined to the central regions than jet feedback. We also note that these visualizations are generally \textit{not} shown for the same simulation time. Doing so would result in very little visible activity in some of the cases, since all of these simulations peak in feedback activity at different times. We have therefore attempted to show these visualizations at representative times for each of the cases.

We begin our comparison of different types of feedback with variations of energy type for jet feedback (left-hand side of the top row in Fig.~\ref{fig:fig7}). Kinetic jets inflate well-defined ellipsoidal lobes, and they also create strong bow shocks. Using mixed jets also leads to fairly symmetrical lobes that create bow shocks, although they appear to be weaker (judging by the typical temperature in the shock fronts). Thermal jets do lead to biconical outflows, but these are asymmetrical since they are much more susceptible to perturbations. Relatively weak shocks are visible in this case.

In the right-hand side of the top row of Fig.~\ref{fig:fig7} we show variations of the jet velocity in the kinetic jet case. The lowest-velocity case ($v_\mr{j}\approx8500$ km $\mr{s}^{-1}$) does not appear to feature hot, ellipsoidal lobes. Instead, the outflows resemble \cite{Fanaroff} type I (conical) jets. Increasing the jet velocity leads to the inflation of lobes and stronger generation of spherical shocks, and this activity tends to be concentrated to smaller radii. The highest-velocity case shows lobes that appear similar to observed X-ray cavities, although we caution that this may be merely a consequence of low resolution (increasing the jet velocity at fixed power decreases the number of jet particles inside the lobes/bubbles, making them more spherical).

In the left-hand side of the bottom row of Fig.~\ref{fig:fig7} we show results of varying the type of energy in the isotropic case. Using less kinetic energy leads to weaker spherical shocks, but typically hotter outflows. In the last row of Fig.~\ref{fig:fig7} we show the results of varying the heating temperature in the thermal isotropic case. These simulations all appear similar, and we do not find that increasing the heating temperature leads to hotter outflows, as one might have expected. From the visualizations shown here, it is also apparent that thermal isotropic feedback can sometimes lead to the emergence of biconical outflows -- this is typically a result of a cold gas disc forming in the centre and feeding the BH. The feedback then results in the launching of biconical outflows that are perpendicular to the disc (see also \citealt{Nobels2022}), since the heated gas tends to expand along the ‘path of least resistance'.

\subsubsection{Differences in feedback powers and SFRs}

\begin{figure*}
\includegraphics[width=1\textwidth, trim = 0 10 0 0]{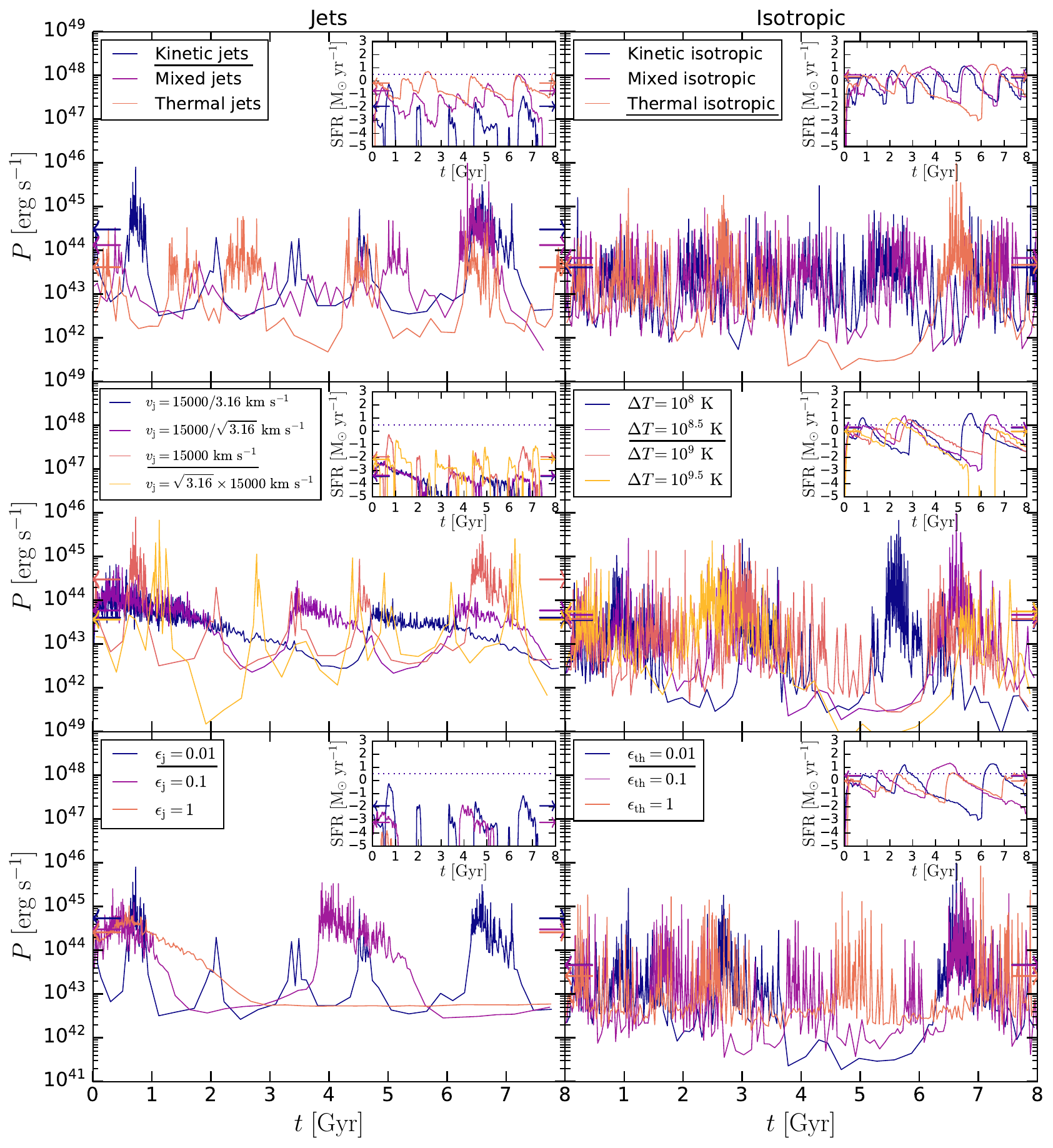}
\caption{Impact of parameter variations on the feedback cycle for both jet (left) and isotropic (right) feedback as measured through the feedback power and the star formation rate (insets in each panel), for the low-mass galaxy cluster ($M_\mathrm{200}=10^{14}$ $\mathrm{M}_\odot$ halo). The feedback powers are calculated using adaptive time bins such that during each bin, 10 feedback events (heating or kicking particles) occurred, while the star formation rates are calculated as averages in bins with a fixed width of 5 Myr.  The arrows indicate averages over the 8 Gyr simulation run time. The dotted horizontal lines in the inset panels show the SFR at which the BCG is classified as marginally quenched ($\mathrm{sSFR}=0.01$ $\mathrm{Gyr}^{-1}$). The details of the simulations are given in \S~\ref{sec:simulations} and Table \ref{tab:tab2}. The parameters that are varied are shown in the legend of each panel, and they are: 1) the fraction of energy injected in thermal as opposed to kinetic form (top row, values of $0$, $0.5$ and $1$), 2) the kicking/heating energy increments, given by the choice of jet velocity and heating temperature (middle row, by factors of $10^{1/4}\approx1.78$ and $\sqrt{10}\approx3.16$, which corresponds to logarithmic intervals of 0.25 and 0.5 dex, respectively) and 3) the feedback efficiency (bottom row, by factors of 10). The fiducial cases are those with thermal isotropic and kinetic jet feedback, using efficiencies of $\epsilon=0.01$, heating temperatures of $\Delta T=10^{8.5}$ K and jet velocities of $v_\mathrm{j}=1.5\times10^4$ km s$^{-1}$. These parameters are underlined in the legend of each panel.
}
\label{fig:fig8}
\end{figure*}%

In Fig.~\ref{fig:fig8} we show the feedback power and SFR for all of the cases discussed above, as well as cases with varying feedback efficiency. We begin by discussing the jet cases (left-hand column), and then the isotropic ones (right-hand column). We find that varying the type of jet energy (top-left panel) does not lead to very large differences in the jet powers. The mean jet power does increase slightly, however, by making it more kinetic rather than thermal. In addition, thermal jets lead to lower minima in the jet power, similar as in the thermal isotropic case, due to the gas near the BH often being hotter (which leads to lower accretion rates). From the SFR plot we see that kinetic jets are the most efficient at quenching, with the purely thermal ones quite similar to isotropic feedback (discussed below), and the mixed ones somewhere in between.

In the middle-left panel of Fig.~\ref{fig:fig8} we show results of varying the jet velocity of kinetic jets. We already showed a variation of this kind in Figs. \ref{fig:fig5} and \ref{fig:fig6}, although we do it here more systemically. We find that using higher jet velocities results in more episodic feedback cycles, with higher peaks in the jet power and lower minima. The former is a result of more cold gas feeding stronger feedback, while the latter is a result of stronger shocking or shocking at smaller distances, which leads to more hot gas feeding the BH and reducing its accretion rate. Decreasing the velocity leads to a decrease in the SFR. Note, however, that decreasing the jet velocity at fixed resolution also improves the sampling of feedback (leading to more particles making up the jets and lobes), an effect that might be the main cause of these differences.

In the bottom-left panel of Fig.~\ref{fig:fig8} we show results of varying the feedback efficiency of the kinetic jets. As we can see, the differences are significant. Increasing the feedback efficiency results in fewer and fewer feedback episodes. With an efficiency of $100$ per cent, there is only one initial episode and effectively no star formation. Using an intermediate efficiency leads to two feedback and SFR episodes. It should be noted, however, that all three cases are quenched and thus show negligible star formation as compared to star-forming galaxies. Interestingly, all three simulations show the same minimum in the jet power ($P_\mr{j}\approx3\times10^{42}$ erg s$^{-1}$). This minimum corresponds to hot halo accretion.

In the right-hand panels of Fig.~\ref{fig:fig8} we show the corresponding variations of isotropic feedback. We find that all of the simulations are fairly similar, especially when compared with the variations in the jet case. It should be kept in mind that these simulations are chaotic in nature, so differences in the timing of peaks in the SFR and feedback power may not be very significant. With this in mind, we find that changing the energy type (top right panel) is the variation that has the most significant impact, in the form of changing the periodicity of the feedback events -- the purely thermal case appears to have the longest period between feedback events. It also reaches the lowest value in the feedback power and SFR at $t\approx6$ Gyr. Regardless of these small differences, the typical powers and SFRs are still similar. 

The similarity of thermal and kinetic isotropic feedback, as implemented here, has bearing for cosmological simulations. In particular, the EAGLE simulations (\citealt{Schaye2015}) used a thermal isotropic AGN feedback implementation, whereas in IllustrisTNG (\citealt{Nelson2019}), feedback by AGN is mostly done through kinetic isotropic winds. The results shown here imply that these two feedback implementations are quite similar in their effects (and both of them quite different from kinetic jets). A caveat to this is that our simulations are of idealized clusters. In reality, feedback is expected to occur in various contexts, such as during and after galaxy mergers (see e.g.~\citealt{Gao2020} for observational evidence or \citealt{McAlpine2018} for indications of the same in cosmological simulations) or triggered by disc instabilities (e.g.~\citealt{Menci2014}). In these situations the effects of AGN feedback might be more sensitive to the various parameters and choices we have discussed.

In the middle panel of Fig.~\ref{fig:fig8} we show the results of varying the heating temperature in the thermal isotropic case. The results are again very similar, although the two higher-temperature cases ($\Delta T \geq10^9$ K) show a somewhat lower recurrent peak in the feedback power and SFR at $t\approx6-7$ Gyr. This result may be due to stochastic noise, rather than an indication of an actual trend. In the bottom right-hand panel, we see the results of varying the feedback efficiency. The highest-efficiency case has both lower maxima and higher minima in the power and SFR (more easily visible in the latter), which is likely due to the BH reacting more quickly to the development of a cooling flow: the maxima reached are lower because the feedback can shut off the cooling flow before too much cooling occurs, while the minima are higher because the feedback is then not as explosive.

\subsubsection{Entropy profiles}

\begin{figure*}
\includegraphics[width=1.01\textwidth, trim = 0 10 0 0]{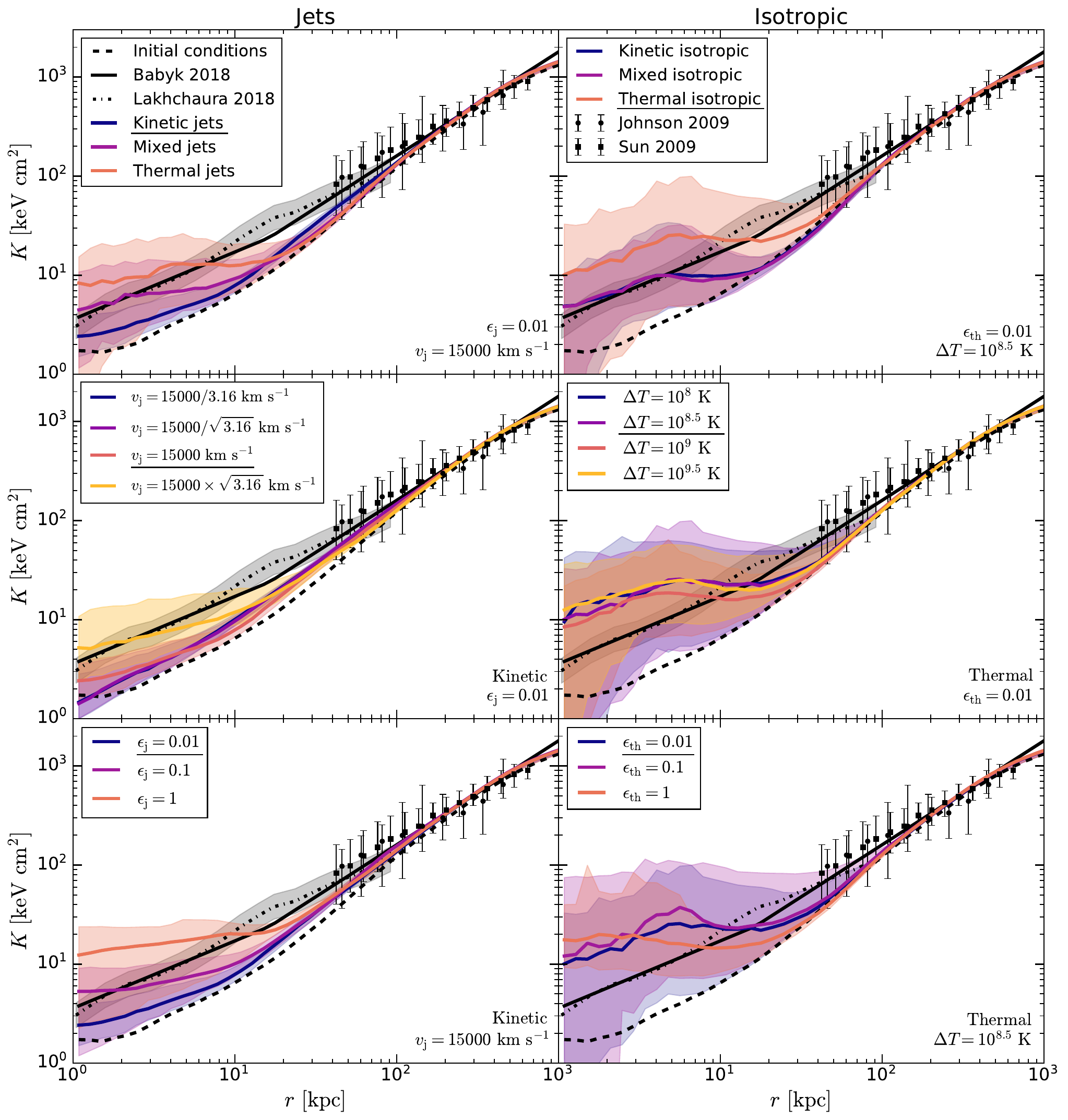}
\caption{Impact of parameter variations on the gas entropy profiles (volume-weighted) of the ICM using both jet (left) and isotropic (right) feedback, for the low-mass galaxy cluster ($M_\mathrm{200}=10^{14}$ $\mathrm{M}_\odot$). The solid lines are medians calculated using 160 snapshots between $t=0$ and $t=8$ Gyr, while the shadings indicate the $16^\mathrm{th}-84^\mathrm{th}$ percentile ranges. The fiducial parameters (underlined and printed in each panel), as well as the parameters being varied, are the same as shown in Fig.~\ref{fig:fig8} and described in its caption. The observational data sets are described in \S~\ref{sec:obs}. 
}
\label{fig:fig9}
\end{figure*}%

Finally, in Fig.~\ref{fig:fig9} we show the entropy profiles for the variations discussed above. From the top-left panel we see that increasing the fraction of kinetic energy in the jets leads to steeper inner entropy profiles, which are in closer agreement to observed ones (in terms of the slope). From the middle-left panel we see that decreasing the jet velocity can also bring the entropy profiles into closer agreement with observations. This may appear counterintuitive considering that real AGN jets are relativistic, and thus high-velocity (see e.g.~review by \citealt{Blandford2019}). However, one should keep in mind that the bulk of the jet material (or energy associated with the jets) is not necessarily relativistic on all scales; the jets are often mostly transrelativistic\footnote{Velocities at which relativistic effects begin to become important, $v_\mr{j}\approx0.1-0.5c$.} (e.g.~\citealt{Jetha2006} and \citealt{Mullin2009}) or subrelativistic (e.g.~\citealt{Shulevski2019}) on kpc scales, the ones we are simulating in this paper. The subrelativistic launching velocities ($v_\mathrm{j}<0.05c$) favoured by these simulations may be indicative of observed jets experiencing significant amounts of entrainment on subgrid scales relative to what we are resolving here (i.e.~below $\approx300$ pc). We find that the two lower-velocity cases shown in the panel have an almost identical entropy profile, indicating that the profiles converge to the same one as the velocity is decreased. From the two higher-velocity cases, we see that increasing the velocity leads to differences in the profiles: the central entropies are higher, and the slope is changed. In addition, the scatter between the different snapshots is increased. Overall, these results indicate that increasing the velocity leads to entropy profiles that are progressively more similar to those found with thermal isotropic feedback, likely due to shock heating (thermalisation) of the jets and inflation of lobes/bubbles at smaller radii.

In the bottom-left panel we show variations of the feedback efficiency. These results indicate that higher efficiencies lead to entropies that are too flat in the centre. The CC/NCC dichotomy could thus partially or wholly be a result of the BH population differing in BH spin -- the low spin ones having lower feedback efficiencies and therefore lower central entropies, whereas the higher-spin ones would be the opposite in this picture.

In the right-hand panels of Fig.~\ref{fig:fig9} we show the same variations for the isotropic case. Overall these are very similar to each other, with the energy type variations (top right-hand panel) being the only ones that show appreciable differences. In particular, the mixed or purely kinetic isotropic wind cases have lower entropies than the purely thermal one, by roughly a factor of two. However, the overall shape of the entropy profile is still the same, and it still disagrees with the observed profiles in terms of the slope.

\subsubsection{Comparison with previous simulations}

We will compare our variations of the feedback implementation with previous work, mostly on idealized galaxy clusters and mainly for the low-mass cluster case ($M_\mr{200}=10^{14}$ $\mathrm{M}_\odot$), since other studies have largely focused on such haloes. We discuss specifically papers that have implemented more than one AGN feedback variant, or that have varied parameters that also correspond to our simulations. 

\cite{Barai2016} performed SPH simulations and compared several implementations of kinetic feedback, as well as one thermal variation. Their feedback implementation is intermediate to our isotropic and jet feedback, since it is bipolar in nature, but with a large opening angle ($45\degree$). They found that using kinetic feedback leads to less star formation than if thermal feedback is used, in agreement with our findings. However, their entropy profiles with thermal feedback are lower than the ones with kinetic feedback, a conclusion opposite to ours (both for the isotropic and jet cases). This is likely a result of their thermal feedback being implemented as a ‘thermal dump' (see footnote \ref{footnote1}), which likely resulted in numerical overcooling. They find that lower-velocity feedback leads to higher central entropies, again in disagreement with our finding. This could be due to differences in the hydrodynamics schemes (GADGET-3 vs. SPHENIX). \cite{Weinberger2022} compared kinetic jets with the kinetic wind implemented in IllustrisTNG; they found that jets are slightly more efficient at quenching star formation, in agreement with our results. They also found that the feedback powers are less time-variable in the jet case than in the kinetic wind case, which is again in agreement with our results. Their interpretation of this is that jets act more on the strongly cooling (but not yet star-forming gas), while the wind acts on the star-forming ISM, including in the vicinity of the BH.

The remaining simulations we compare with were performed using grid-based codes. \cite{Gaspari2014} compared mechanical (kinetic) jet feedback and  thermal isotropic feedback across a range of halo masses ($10^{13}-10^{15}$ $\mr{M}_\odot$), finding that the former leads to cooler cores, in agreement with our results. However, they implemented kinetic jets as self-regulated (with the accretion rate determined from the properties of gas), while their thermal feedback was implemented as a blast with a fixed power (heating all gas near the BH), which is not a fair comparison. \cite{Meece2017} compared different feedback models in a massive halo ($M_\mr{200}\approx10^{15}$ $\mathrm{M}_\odot$). They found that purely thermal jets are less efficient at preventing cooling flows from developing than either mixed or purely kinetic ones, in agreement with our findings. However, in disagreement with our results, they find lower central entropies with thermal feedback, similar to \cite{Barai2016} (this is, again, probably a result of using low heating temperatures as part of a ‘thermal dump' that likely led to too much numerical overcooling). \cite{Ehlert2023} compared dense (i.e.~low-velocity) and light (high-velocity) jets, finding that the results are relatively similar. However, it should be pointed out that the majority of these papers, including the last one, perform their simulations for a relatively short time (usually $1-2$ Gyr or less). This is of order the length of the typical cycle of activity (cooling and feedback) we find in our simulations. Thus, most of these papers may be biasing their results to the first episode of high-activity.

Finally, while we found that the choice of heating temperature used for thermal isotropic feedback has little effect on our results, especially for the entropy profiles, this is in disagreement with previous work in a cosmological context (e.g.~\citealt{LeBrun2014}, \citealt{Hahn2017}). Those studies found that the choice of heating temperature affects both the total mass of the ICM (the gas fraction) as well as its distribution and properties (the thermodynamical profiles). This difference between our results and cosmological studies is likely due to our simulations focusing on isolated and self-regulated systems (assumptions that break down for realistic haloes).

\section{Summary and conclusions}
\label{sec:conclusions}

Using the SWIFT simulation code (\citealt{Schaller2023}) and the SPHENIX SPH implementation (\citealt{Borrow2022}), we have compared different prescriptions of AGN feedback. For this purpose we used a well-tested set-up of idealized galaxy groups and clusters (\citealt{Nobels2022}) with virial masses $M_\mr{200}=10^{13}$, $10^{14}$ and $10^{15}$ $\mathrm{M}_\odot$, which we initialized in a relatively cool-core state. We focused on comparing thermal isotropic (\citealt{Booth2009}) and kinetic jet feedback (\citealt{Husko2022_spin_driven}) -- the former representing the effects of radiatively-driven wind (i.e.~quasar) feedback, and the latter the effects of feedback by relativistic jets. 

We first tested these AGN feedback implementations in unison with a BH spin evolution model based on equations describing subgrid accretion discs. This model gives variable feedback efficiencies, as well as variable jet directions. We assumed that thermal isotropic feedback occurs at high normalized accretion rates (Eddington ratios $\dot{m}>0.01$), when the disc is thin and radiatively efficient, whereas kinetic jets are assumed to be launched at low Eddington ratios ($\dot{m}<0.01$), when the disc is thick and advection-dominated. We compared this hybrid model with one where the disc is always thin and launching isotropic winds, as well one where the disc is always thick and launching jets. We then simplified the set-up by assuming constant feedback efficiencies and fixing the jet direction. In this simplified set-up, we further varied the feedback efficiency, the energy per feedback event, as well as the type of energy being used for feedback (thermal vs. kinetic) for both the isotropic and jet cases. From the simulations performed and the analysis presented in this paper, we find the following:

\begin{itemize}
    \item Kinetic jet feedback leads to more efficient quenching of star formation in the central galaxies than thermal isotropic (wind) feedback. This applies across the whole mass scale range we have tested. It is true in simulations using detailed models of BH spin evolution (resulting in variable feedback efficiencies/jet directions), as well as ones without (using constant feedback efficiencies/jet directions). A larger fraction of the feedback energy couples to large radii in the jet case, resulting in overall more energy being injected in that case in order to quench cooling flows.
    \item Due to a smaller fraction of the feedback energy coupling to the intracluster gas at smaller radii, and a larger fraction at larger radii, the central gas entropies are significantly lower with kinetic jet feedback than with thermal isotropic feedback. They are also in closer agreement with observations in terms of the inner slope. In addition to the median central entropies being lower, median central densities are higher and median central temperatures lower, despite cooling flows being weaker and/or shorter-lived.
    \item We find that isotropic feedback is largely insensitive to the choice of feedback efficiency and energy per feedback event. By varying the type of energy being injected (kinetic, mixed and thermal), we find that the thermal isotropic case has a somewhat higher central entropy and a feedback cycle with the longest periodicity. However, all of these isotropic feedback implementations are still more similar to each other than any of them is to kinetic jet feedback. This may indicate that the isotropic kinetic feedback employed in some cosmological simulations (e.g.~IllustrisTNG) is quite similar in its effects to the isotropic thermal feedback employed in other simulations (e.g.~EAGLE). However, all of our isotropic feedback is energy-dominated, so the conclusions may change somewhat for momentum-dominated winds.
    \item Jet feedback is sensitive to all of the choices mentioned in the previous point. High feedback efficiencies can prevent any cooling flows from developing, leading to higher central entropies. Increasing the jet velocity leads to more frequent cooling flows (and more star formation), but it also leads to higher mean central entropies with shallower slopes, due to strong shocks and heating at small radii. In other words, kinetic jet feedback is progressively more similar to thermal isotropic feedback as the jet velocity is increased. Jet feedback is most efficient if it is kinetic, rather than thermal or mixed. The jet direction is unimportant, as long as it does not change more frequently than every $\approx1$ Gyr, which it is unlikely to do in galaxy clusters with realistic BH spin evolution. Constant jet efficiencies lead to highly periodic cooling flows, unlike in the variable-efficiency cases.
    \item In order to recover the observed entropy profiles across a large range of masses (galaxy group to rich cluster scales), it may be necessary to choose jet velocities carefully. In particular, low velocities may be required in galaxy groups/low-mass clusters in order to yield steep entropy profiles, while high jet velocities may be required to reproduce cored entropy profiles and X-ray cavities in rich galaxy clusters. Alternatively, variable jet efficiencies from a BH spin evolution model, in conjunction with different accretion/merger histories, might naturally lead to some of these differences. We find that a hybrid model with both thermal isotropic and kinetic jet feedback (depending on the BH accretion rate) has the lowest central entropies, and may thus be the most promising. On the other hand, our jet-only model is disfavoured on account of excessive BH mass growth. This growth is due to strong jet-induced spindown of BHs, leading to very low BH spins (and therefore jet efficiencies of order $0.1$ per cent).
    \item The differences between simulated entropy profiles with varying AGN feedback implementations are similar in magnitude to differences that arise if the numerical details are varied (e.g.~the artificial conductivity and viscosity of the hydrodynamics code). This means that these physical and numerical variations are somewhat degenerate with regards to entropy profiles. Bringing different numerical codes in agreement would thus significantly improve the potential of simulations to discriminate between different AGN feedback implementations.
\end{itemize}

We caution that these conclusions may only be valid for isolated systems such as the ones studied in this paper. Thus, some of them may not fully apply in the context of cosmological simulations. Despite this caveat, the results presented in this paper should be valuable when considering different implementations of AGN feedback in cosmological simulations of galaxy formation and evolution. This is particularly true as the AGN feedback implementations are becoming more complicated and realistic.

\section*{Acknowledgements}
The research in this paper made use of the SWIFT open-source simulation code (\url{http://www.swiftsim.com}, \citealt{Schaller2023})
version 0.9.0. The swiftsimio Python library was used to analyze and visualize the data from the simulations (\citealt{Borrow2020_swiftsimio}, \citealt{Borrow_2021_swiftsimio}). The work has been performed under the Project HPC-EUROPA3 (INFRAIA-2016-1-730897), with the support of the EC Research Innovation Action under the Horizon 2020 Framework Programme of the European Union (H2020); in particular, FH gratefully acknowledges the support of the Leiden Observatory and the computer resources and technical support provided by SURFsara, the Dutch national HPC facility. This project was also funded by the AHEAD2020 Programme under Grant Agreement 871158. This project has received funding from the Netherlands Organization for Scientific Research (NWO) through research
programme Athena 184.034.002. FH would like to acknowledge support from the Science Technology Facilities Council through a CDT studentship (ST/P006744/1), and CGL acknowledges support from STFC consolidated grants ST/T000244/1 and ST/X001075/1. This work used the DiRAC@Durham facility managed by the Institute for Computational Cosmology on behalf of the STFC DiRAC HPC Facility (www.dirac.ac.uk). The equipment was funded by BEIS capital funding via STFC capital grants ST/K00042X/1, ST/P002293/1, ST/R002371/1 and ST/S002502/1, Durham University and STFC operations grant ST/R000832/1. DiRAC is part of the National e-Infrastructure.

\section*{Data availability}

The data underlying this article will be provided upon reasonable request to the corresponding author. The code and initial conditions used to generate the data are available online: \url{https://gitlab.cosma.dur.ac.uk/swift/swiftsim}.

\bibliographystyle{mnras}
\bibliography{jet_bibliography} 

\appendix

\section{Effects of jet redirection and precession}
\label{sec:app1}

\begin{figure*}
\includegraphics[width=1.01\textwidth, trim = 0 10 0 0]{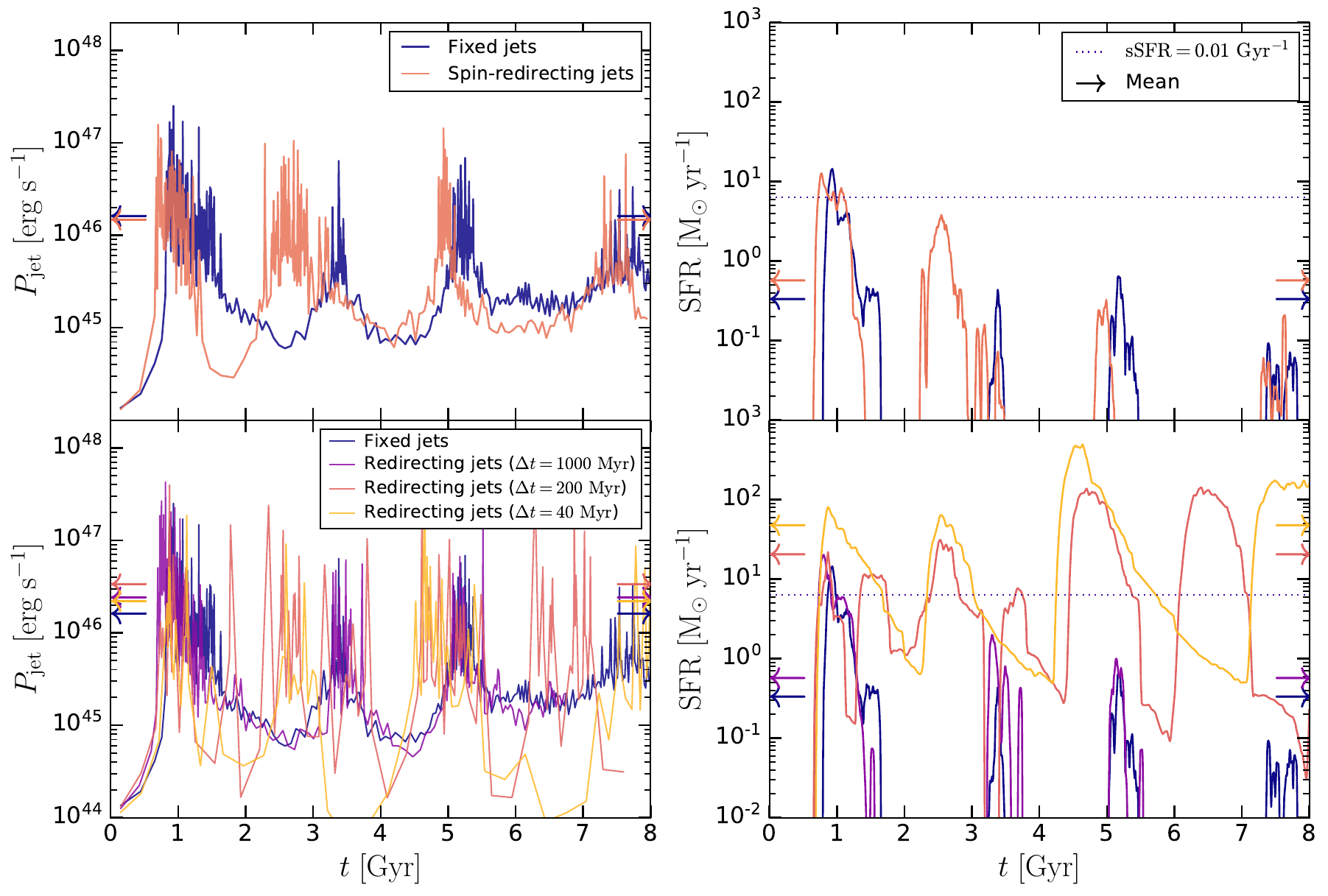}
\caption{Time dependence of the jet powers (left) and star formation rates (right) in the high-mass galaxy cluster ($M_\mathrm{200}=10^{15}$ $\mathrm{M}_\odot$) simulations that feature jet redirection, compared with a case that has a fixed jet direction (along the $z-$axis). In the top panels we compare with a case that features jet redirection using the spin evolution model. The bottom panels show several cases where the jets are held fixed in a direction that randomly changes, with the period of these redirections shown in the legend. The arrows indicate averages over the 8 Gyr simulation run time. The parameters of the simulations correspond to the sixth row of Table \ref{tab:tab2}. }
\label{fig:figA1}
\end{figure*}%

\begin{figure*}
\includegraphics[width=1.01\textwidth, trim = 0 10 0 0]{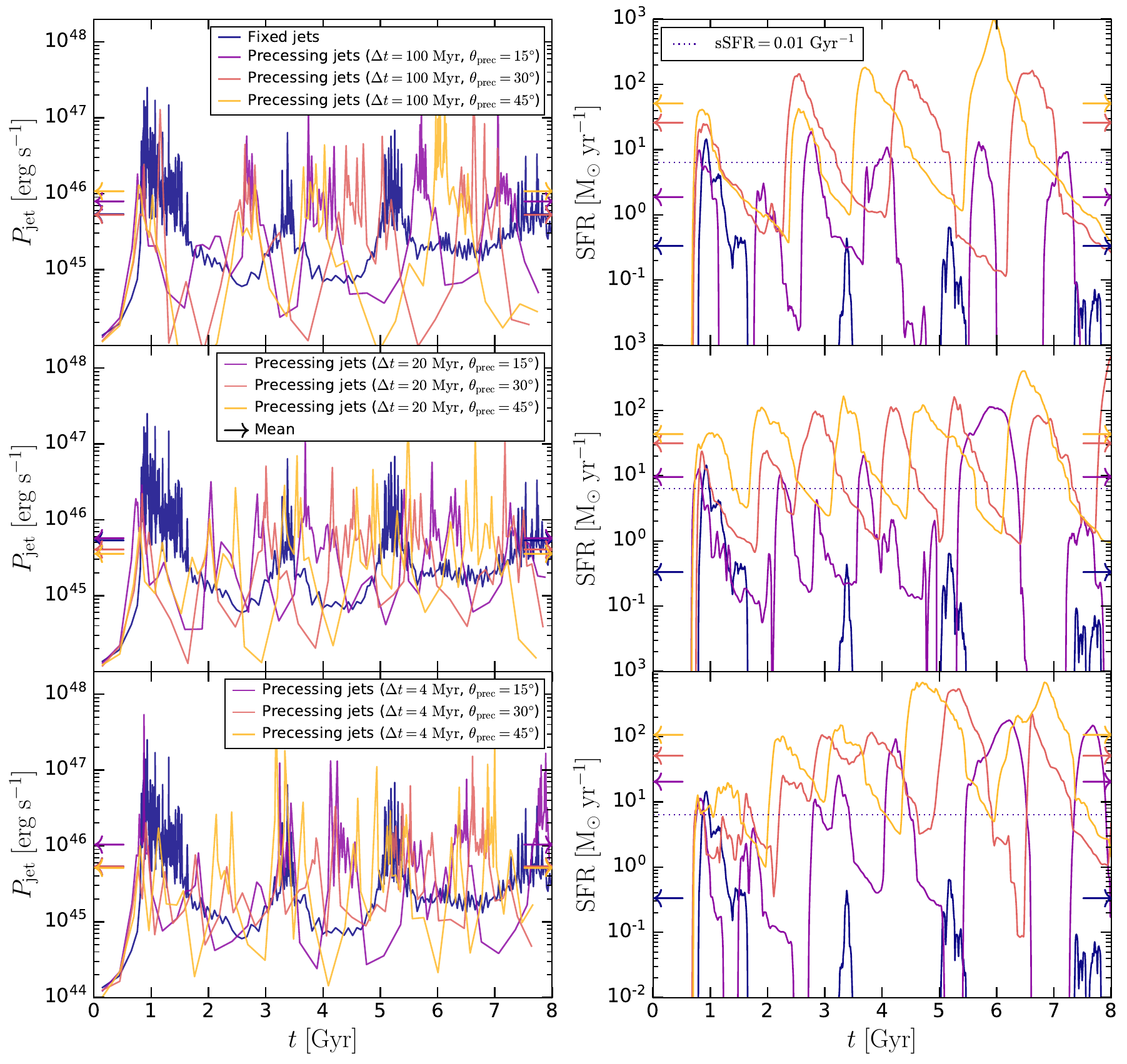}
\caption{Same as Fig.~\ref{fig:figA1}, but showing cases with precessing rather than redirecting jets. From top to bottom we show several cases with different periods of jet precession. In each row (i.e.~at every fixed precession period) we also vary the precession angle. }
\label{fig:figA2}
\end{figure*}%

For the purposes of the main results in this paper, we fixed the jets to be along the $z-$axis when considering simplified feedback without BH spin evolution. This immediately leads to the following questions: how justified is this assumption, and how important is the change of the jet direction for the effects of feedback? We ran some additional simulations in order to answer these questions. These simulations employed either manually redirecting or precessing jets. There are many ways in which both of these processes could be implemented. We used a fairly simple implementation, since these results are meant to be illustrative. We tested these cases in our fiducial high-mass galaxy cluster set-up ($M_\mathrm{200}=10^{15}$ $\mathrm{M}_\odot$), since we found redirection to be more likely for this halo mass (if BH spin evolution is used).

In the redirecting case, the jets were initially directed along the $z-$axis. With a period of $\Delta t$, they were then instantaneously redirected to another, randomly chosen axis. We tested three periods: $\Delta t=1000$ Myr, $\Delta t=200$ Myr, and $\Delta t=40$ Myr. These are compared with the fixed-direction case in Fig.~\ref{fig:figA1}, alongside a case that has spin-driven jet redirection, but a constant jet efficiency ($\epsilon_\mr{j}=0.01$, as in all of these simulations). The spin-redirecting case appears to show similar behaviour as the fixed-axis case, despite the jets being redirected during each of the cooling episodes (for a total of 4 times, i.e.~once per each cooling episode, although this is not shown here). The case with manual redirection every $\Delta t=1000$ Myr is again very similar to the fixed case, and therefore to the spin-redirecting case. However, if redirection is done more often ($\Delta t \leq200$Myr), the jet powers are more variable and the SFRs are more similar to the thermal isotropic case. We interpret this to be a result of the redirection time-scale being similar to (of the same order of magnitude as) the typical duration of a jet episode, which can be up to $100$ Myr. We speculate that this may be due to jets often being redirected while they are in the process of inflating a pair of lobes, or otherwise moving to large radii (where effective heating seems to be necessary in order for cooling flows to be shut off effectively).

In the precessing cases, we manually precessed the jets with a period of $\Delta t$ about the $z-$axis, with a precession angle of $\theta_\mr{prec}$. We did not nutate the jets as well, i.e.~they did not ‘cover' the region between the $z-$axis and the circle on which they were precessing. Note that the effects of jet precession are probably quite similar to the effects of using a larger opening angle. We tested three values of $\Delta t$: $\Delta t=100$ Myr, $\Delta t=20$ Myr, and $\Delta t=4$ Myr. These are relatively shorter than in the redirecting case, because we expect that the BH spin vector can change in direction by small values (e.g.~$15\degree$) with very little mass accretion, which is not true for full redirection. For each of the precession time-scales, we tested three precession angles: $\theta_\mr{prec}=15\degree$, $\theta_\mr{prec}=30\degree$ and $\theta_\mr{prec}=45\degree$. The results of these tests are shown in Fig.~\ref{fig:figA2}. It appears that jet precession leads to significant differences in all cases shown here. The only combination(s) that result in fairly low SFRs are those with $\theta_\mr{prec}=15\degree$ and $\Delta t\geq20$ Myr. However, even these cases show higher SFRs than the fixed-axis case. Cases with larger precession angles appear quite similar in their effects to thermal isotropic feedback. The precession time-scale does not appear to have a large impact.

\section{Mass flux associated with accretion and feedback}
\label{sec:app2}

\begin{figure*}
\includegraphics[width=1.01\textwidth, trim = 0 10 0 0]{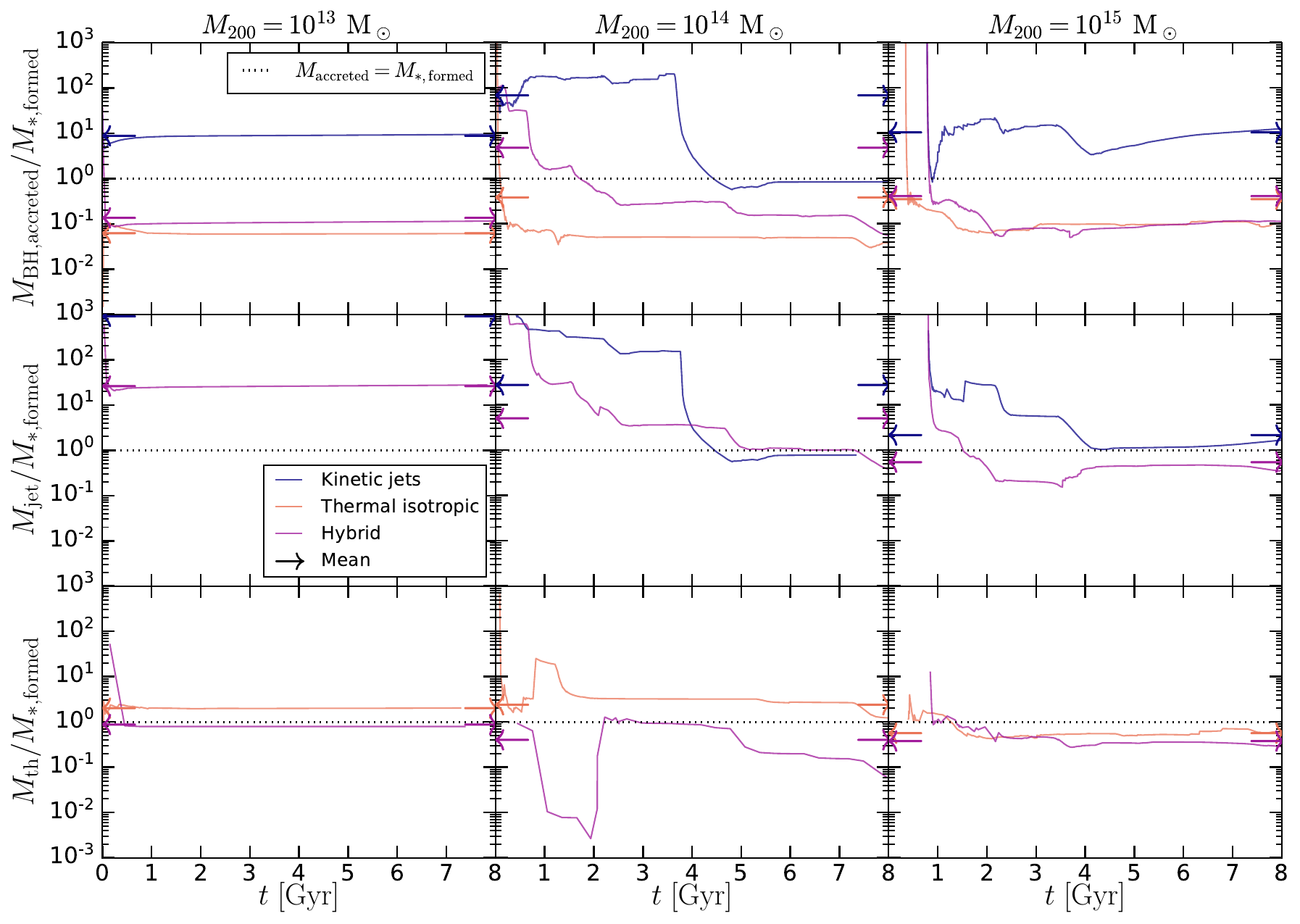}
\caption{Ratio of total mass accreted (top row), launched into the jets (middle row) and heated (bottom row) by the BH, and the total stellar mass formed. The arrows indicate averages over the 8 Gyr simulation run time. This figure is an extension of Fig.~\ref{fig:fig1}.}
\label{fig:figB1}
\end{figure*}%

In Fig.~\ref{fig:figB1} we show some additional quantities from the same simulations as in Fig.~\ref{fig:fig1}, namely: the ratio of total mass accreted, launched into the jets and heated by the BH, to the total stellar mass formed. These ratios are plotted for our simulations with spin evolution spanning the galaxy group ($M_\mathrm{200}=10^{13}$ $\mathrm{M}_\odot$) to high-mass cluster scale ($M_\mathrm{200}=10^{15}$ $\mathrm{M}_\odot$). We plot these quantities in order to glean information on whether BH accretion and feedback are directly interfering with star formation (by depriving it of cool gas either by accreting, kicking or heating it), or indirectly by e.g.~causing outflows of the same gas or shutting off cooling flows that supply this gas. These ratios should be treated as meaningful only once the amount of stars that have formed is appreciable, and not very low due to feedback being effective. For this reason, the results from the left-hand column should not be considered too meaningful (since very little star formation occurs in the galaxy group case), while for the galaxy cluster cases, they become meaningful only at $t=2-4$ Gyr, depending on the case.

From the galaxy cluster cases we see that the amount of mass accreted by the BH is significant in all cases, with the mass flux associated with feedback even more significant (for at least one of the feedback channels in the given simulation). This is true even for the high-mass cluster, which is the most star-forming of the systems we study, and where we find that the combined mass of all the heated and kicked particles in the hybrid case (as an example) to be roughly as large as the total mass of all stars formed ($\approx3\times10^{11}$ $\mathrm{M}_\odot$). Overall, these plots indicate that feedback mechanisms in simulations may directly interfere with the formation of stars (by depriving it of cold gas), even when the rate of star formation is relatively high. This effect may, however, still be subdominant to the indirect effects of feedback. These plots also show that BHs in the kinetic jet-only case cannot self-regulate their growth, due to low jet efficiencies as a result of spindown.

\section{Fraction of black hole growth at low vs. high Eddington ratios}
\label{sec:app2p1}

\begin{figure*}
\includegraphics[width=1.01\textwidth, trim = 0 10 0 0]{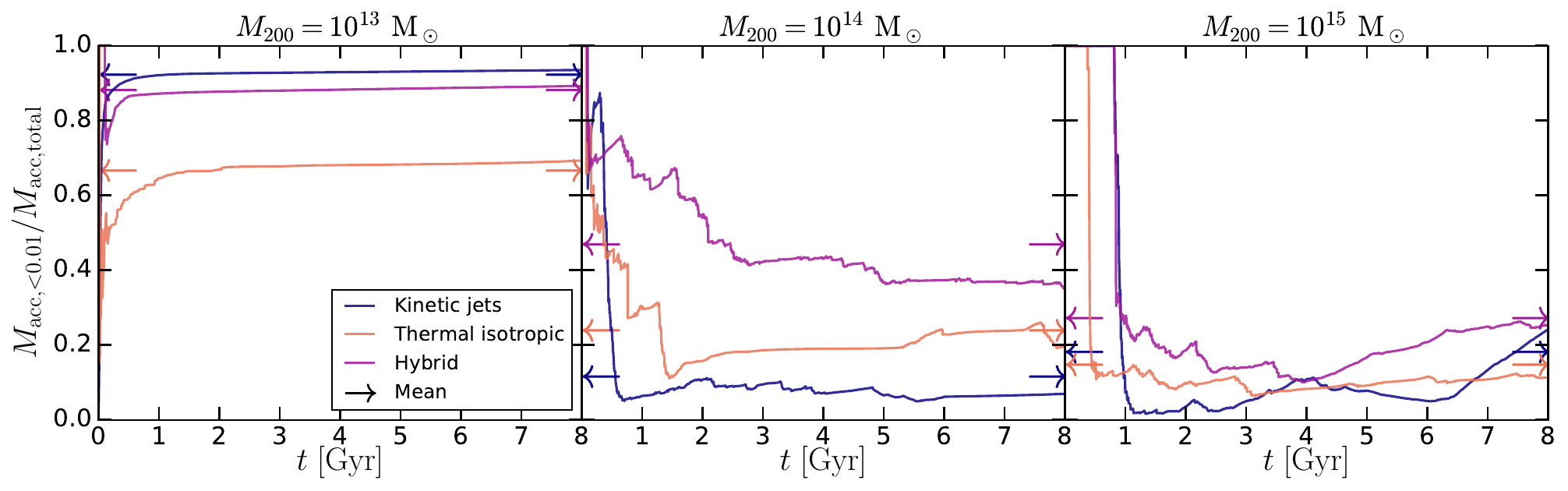}
\caption{BH mass growth that occurs at low Eddington ratios ($\dot{m}<0.01$) as a fraction of the total mass growth, for our simulations with BH spin evolution. This figure is an extension of Fig.~\ref{fig:fig3}.}
\label{fig:figB2}
\end{figure*}%

In Fig.~\ref{fig:figB2} we show the cumulative fraction of mass accreted when $\dot{m}<\dot{m}_\mathrm{crit}=0.01$ (corresponding to the thick disc in the hybrid and kinetic-only case) as a function of time, for all 9 simulations discussed in \S~\ref{sec:res_spin}. This figure is an extension of Fig.~\ref{fig:fig3}. We see that in the galaxy group case, most of the growth is at low Eddington ratios, except at the very beginning. However, this reflects the fact that there is an initial burst of high accretion rate growth at the beginning of the simulation, after which the system is fully quenched. In the galaxy cluster cases, we see that most of the BH growth occurs when $\dot{m}>0.01$, despite the fact that this condition is not fulfilled most of the time. The growth at low Eddington ratios, however, is by no means negligible.

\section{Dimensionless entropy profiles}
\label{sec:app2p2}

\begin{figure}
\includegraphics[width=1.01\columnwidth, trim = 0 10 0 0]{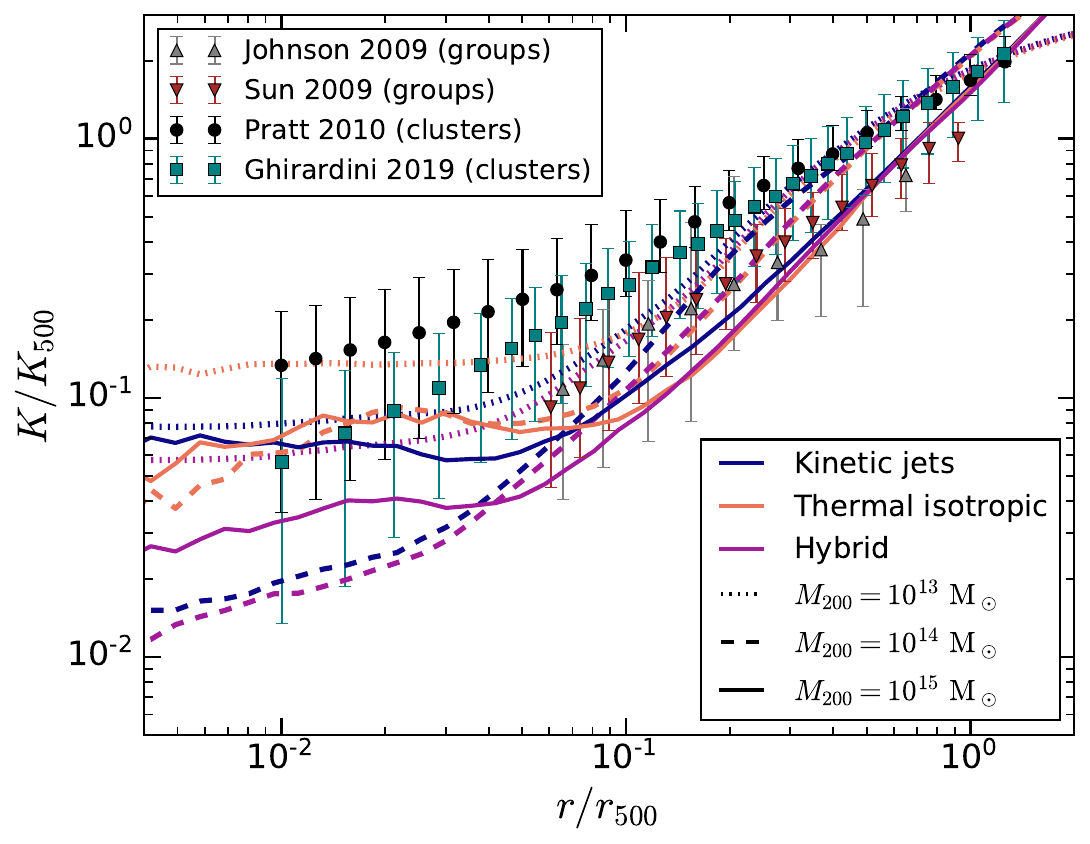}
\caption{Dimensionless entropy profiles for our simulations with BH spin evolution. This figure may be considered an alternative to Fig.~\ref{fig:fig4}. Note that we define the entropy $K_{500}$ using the actual baryon fractions of our simulated haloes, rather than the cosmic baryon fraction (the usual choice). We find that this leads to better agreement between the profiles at large distances.}
\label{fig:figB3}
\end{figure}%

In Fig.~\ref{fig:figB3} we show the dimensionless entropy profiles $K/K_{500}$ as a function of the scaled radius $r/r_{500}$, for the 9 simulations with BH spin evolution, discussed in \S~\ref{sec:entropy_profiles_spin} (this plot may be considered an alternative way of showing the data in Fig.~\ref{fig:fig4}). We define the entropy $K_{500}$ as $k_\mathrm{B}T_{500}/n_\mathrm{e,500}^{2/3}$, where $T_{500}=GM_{500}\mu m_\mathrm{p}/2r_{500}$ and $n_\mathrm{e,500}=500f_\mathrm{b,0}\rho_\mathrm{c}/\mu_\mathrm{e} m_\mathrm{p}$. Here, $\rho_\mathrm{c}$ is the critical density, $\mu_\mathrm{e}=1.14$ the mean molecular weight per free electron and $f_\mathrm{b,0}\approx0.16$ the cosmic baryon fraction (e.g.~\citealt{Barnes2017}). Overall, from Fig.~\ref{fig:figB3} we see that all of the simulated dimensionless profiles are similar, although there is some disagreement in normalisation at large radii. This can be avoided (and the profiles made even more similar) if the cosmic baryon fraction $f_\mathrm{b,0}$ in the definition of $K_{500}$ is replaced by the actual baryon fraction of each of the haloes, $f_\mathrm{b,500}$, although we do not show those profiles here. We find that the low-mass clusters ($M_{200}=10^{14}$ $\mathrm{M}_\odot$) with jets show the lowest dimensionless entropy. This result may not be significant, however, given that these are single realizations of idealized and isolated clusters. Most of our profiles lie below the observations shown in the figure, although this is by construction (we simulate relatively CC systems).

\section{Periodicity between jet episodes}
\label{sec:app3}

\begin{figure*}
\includegraphics[width=1.01\textwidth, trim = 0 10 0 0]{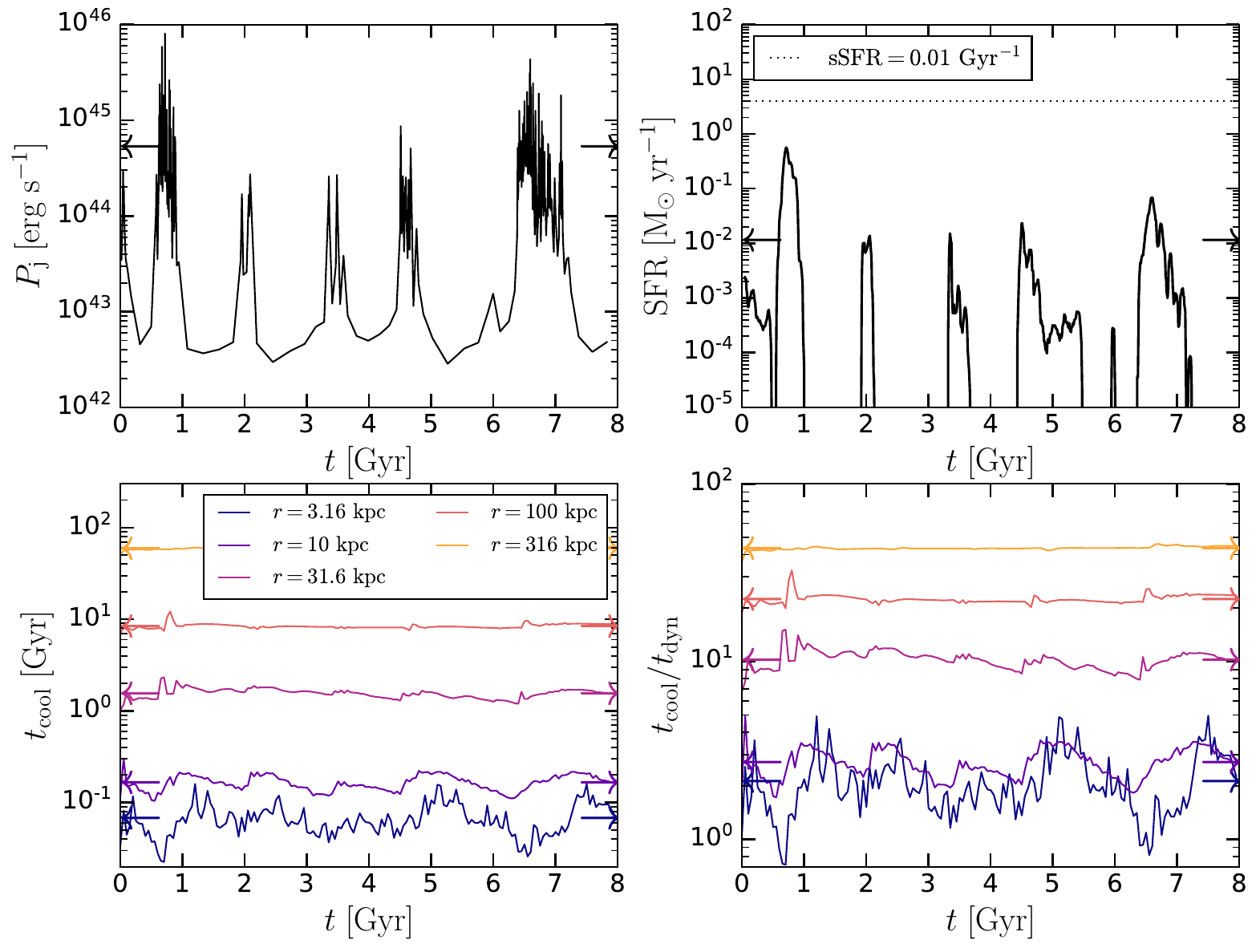}
\caption{Time dependence of quantities related to the quenching/feedback process in our fiducial $M_\mathrm{200}=10^{14}$ $\mathrm{M}_\odot$ simulation with fixed jet feedback (the parameters are given in the third row of Table \ref{tab:tab2}). From top left to bottom right we show the jet power, SFR, cooling time at several radii (see legend) and cooling time to dynamical time ratio for the same radii. These plots show that the periodicity between jet/cooling episodes is set by the cooling time of the gas with $t_\mathrm{cool}/t_\mathrm{dyn}\approx10$, due to all gas with that ratio below $10$ undergoing cooling.}
\label{fig:figC1}
\end{figure*}%

In Fig.~\ref{fig:figC1} we show the approximate periodicity in relevant quantities related to the feedback cycle for our fiducial simulation (with a fixed feedback efficiency $\epsilon_\mathrm{j}=0.01$ and the jets directed along the $z-$axis) of the low-mass galaxy cluster ($M_\mathrm{200}=10^{14}$ $\mathrm{M}_\odot$). The top left-hand panel shows the jet power: it has 5 peaks that appear to be roughly equally separated in time, while the top right-hand panel shows the same for the star formation rate. The peaks in jet power and SFR roughly coincide. The bottom row shows the cooling time and the cooling time to dynamical time ratio ratio at several radii from $r=3.16$ kpc to $r=316$ kpc. According to the hypothesis of \cite{Nobels2022} (and references therein), all gas with $t_\mr{cool}/t_\mr{dyn}<10$ will cool effectively and contribute to the cooling flow. From the right-hand panel we see that the gas at $r=31.6$ kpc has a roughly constant value of the ratio, $t_\mr{cool}/t_\mr{dyn}\approx10$. If we then look at the left-hand panel, that same gas has a roughly constant cooling time of $\approx1.5$ Gyr. This is also roughly the period between the cooling flow episodes. Our results are thus in agreement with the above-mentioned hypothesis. While we have shown only this one case, we find that the same holds true across all our simulations.

\bsp	
\label{lastpage}
\end{document}